\documentclass[]{aa}

\usepackage{graphicx,subcaption}
\usepackage{xcolor}
\usepackage{tikz}
\usepackage{xspace}
\usepackage[hidelinks]{hyperref}

\usepackage{txfonts}
\newcommand{\kms}{\rm km\,s^{-1}}
\newcommand{\Msun}{\rm M_\odot}
\newcommand{\zphot}{z_{\rm phot}}
\newcommand{\zspec}{z_{\rm prism}}

\newcommand{\micron}{\rm \mu m}
\newcommand{\Oiii}{[O\,{\sc iii}]\xspace}
\newcommand{\Nii}{[N\,{\sc ii}]\xspace}
\newcommand{\Hb}{{\rm H}$\beta$\xspace}
\newcommand{\Ha}{{\rm H}$\alpha$\xspace}

\begin{document}

   \title{RUBIES: a complete census of the bright and red distant Universe with JWST/NIRSpec  }

   \author{Anna de Graaff\inst{\ref{i1}}\thanks{degraaff@mpia.de}
        \and Gabriel Brammer\inst{\ref{i2},\ref{i_NBI}}
        \and Andrea Weibel \inst{\ref{i_ge}}
        \and Zach Lewis\inst{\ref{i3}}\thanks{NSF Graduate Research Fellow}
        \and Michael V. Maseda\inst{\ref{i3}}
        \and Pascal A. Oesch \inst{\ref{i_ge},\ref{i2},\ref{i_NBI}}
        \and Rachel Bezanson \inst{\ref{i_pitt}}
        \and Leindert A. Boogaard\inst{\ref{i1},\ref{i_strw}}
        \and Nikko J. Cleri \inst{\ref{i4}, \ref{i5}, \ref{i6}}
        \and Olivia R. Cooper \inst{\ref{utaustin},\ref{i2}}\thanks{NSF Graduate Research Fellow}
        \and Rashmi Gottumukkala \inst{\ref{i2}, \ref{i_NBI}}
        \and Jenny E. Greene \inst{\ref{i_princeton}}
        \and Michaela Hirschmann \inst{\ref{epfl}}
        \and Raphael E. Hviding \inst{\ref{i1}}
        \and Harley Katz \inst{\ref{i_chicago}}
        \and Ivo Labb\'e \inst{\ref{i_swin}}
        \and Joel Leja \inst{\ref{i4}, \ref{i5}, \ref{i6}}
        \and Jorryt Matthee \inst{\ref{i_ista}}
        \and Ian McConachie\inst{\ref{i3}}
        \and Tim B. Miller \inst{\ref{i_ciera}}
        \and Rohan P. Naidu \inst{\ref{i_mit}}\thanks{NASA Hubble Fellow}
        \and Sedona H. Price \inst{\ref{i_pitt}}
        \and Hans-Walter Rix \inst{\ref{i1}}
        \and David J. Setton \inst{\ref{i_princeton}}\thanks{Brinson Prize Fellow}
        \and Katherine A. Suess \inst{\ref{i_cu}}
        \and Bingjie Wang \inst{\ref{i4}, \ref{i5}, \ref{i6}}
        \and Katherine E. Whitaker\inst{\ref{i_umass},\ref{i2}}
        \and Christina C. Williams \inst{\ref{NOIRLab}}
}

   \institute{Max-Planck-Institut f\"ur Astronomie, K\"onigstuhl 17, D-69117 Heidelberg, Germany\label{i1}
    \and 
    Cosmic Dawn Center (DAWN), Copenhagen, Denmark\label{i2} 
    \and 
    Niels Bohr Institute, University of Copenhagen, Jagtvej 128, Copenhagen, Denmark\label{i_NBI} 
    \and 
    Department of Astronomy, University of Geneva, Chemin Pegasi 51, 1290 Versoix, Switzerland\label{i_ge}    
    \and 
    Department of Astronomy, University of Wisconsin-Madison, 475 N. Charter St., Madison, WI 53706 USA\label{i3}  
    \and 
    Department of Physics and Astronomy and PITT PACC, University of Pittsburgh, Pittsburgh, PA 15260, USA\label{i_pitt}
    \and 
    Leiden Observatory, Leiden University, PO Box 9513, NL-2300 RA Leiden, The Netherlands\label{i_strw}
    \and 
    Department of Astronomy \& Astrophysics, The Pennsylvania State University, University Park, PA 16802, USA\label{i4}
    \and 
    Institute for Computational \& Data Sciences, The Pennsylvania State University, University Park, PA 16802, USA\label{i5}
    \and 
    Institute for Gravitation and the Cosmos, The Pennsylvania State University, University Park, PA 16802, USA\label{i6}
    \and 
    Department of Astronomy, The University of Texas at Austin, Austin, TX, USA\label{utaustin}
    \and 
    Department of Astrophysical Sciences, Princeton University, 4 Ivy Lane, Princeton, NJ 08544, USA\label{i_princeton}
    \and 
    Institute of Physics, Laboratory for Galaxy Evolution, Ecole Polytechnique Federale de Lausanne, Observatoire de Sauverny, Chemin Pegasi 51, 1290 Versoix, Switzerland\label{epfl}
    \and 
    Department of Astronomy \& Astrophysics, University of Chicago, 5640 S Ellis Avenue, Chicago, IL 60637, USA\label{i_chicago}
    \and
    Centre for Astrophysics and Supercomputing, Swinburne University of Technology, Melbourne, VIC 3122, Australia\label{i_swin}
    \and 
    Institute of Science and Technology Austria (ISTA), Am Campus 1, 3400 Klosterneuburg, Austria\label{i_ista}
    \and 
    Center for Interdisciplinary Exploration and Research in Astrophysics (CIERA), Northwestern University,1800 Sherman Ave, Evanston, IL 60201, USA\label{i_ciera}
    \and 
    MIT Kavli Institute for Astrophysics and Space Research, 77 Massachusetts Ave., Cambridge, MA 02139, USA\label{i_mit}
    \and 
    Department for Astrophysical \& Planetary Science, University of Colorado, Boulder, CO 80309, USA\label{i_cu}
    \and
    Department of Astronomy, University of Massachusetts, Amherst, MA 01003, USA\label{i_umass}
    \and 
    NSF’s National Optical-Infrared Astronomy Research Laboratory, 950 North Cherry Avenue, Tucson, AZ 85719, USA\label{NOIRLab}
    }


\titlerunning{RUBIES Overview}
\authorrunning{de Graaff et al.}

  \abstract{
   We present the \emph{Red Unknowns: Bright Infrared Extragalactic Survey} (RUBIES), providing JWST/NIRSpec spectroscopy of red sources selected across $\sim150$\,arcmin$^2$ from public JWST/NIRCam imaging in the UDS and EGS fields. RUBIES novel observing strategy offers a well-quantified selection function: the survey is optimised to reach high ($>70\%$) spectroscopic completeness for bright and red ($\mathrm{F150W-F444W}>2$) sources that are very rare. To place these rare sources in context, we simultaneously observe a reference sample of the $2<z<7$ galaxy population, sampling sources at a rate that is inversely proportional to their number density in the 3D parameter space of F444W magnitude, $\mathrm{F150W-F444W}$ colour, and photometric redshift. In total, RUBIES observes $\sim3000$ targets across $1<z_{\rm phot}<10$ with both the PRISM and G395M dispersers, and $\sim1500$ targets at $z_\mathrm{phot}>3$ using only the G395M disperser. The RUBIES data reveal a highly diverse population of red sources that span a broad redshift range ($z_\mathrm{spec}\sim1-9$), with photometric redshift scatter and outlier fraction that are 3 times higher than for similarly bright sources that are less red. This diversity is not apparent from the photometric spectral energy distributions (SEDs). Only spectroscopy reveals that the SEDs encompass a mixture of galaxies with dust-obscured star formation, extreme line emission, a lack of star formation indicating early quenching, and luminous active galactic nuclei. As a first demonstration of our broader selection function we compare the stellar masses and rest-frame $U-V$ colours of the red sources and our reference sample: red sources are typically more massive ($M_*\sim10^{10-11.5}\,\Msun$) across all redshifts. However, we also find that the most massive systems span a wide range in $U-V$ colour. We describe our data reduction procedure and data quality, and publicly release the reduced RUBIES data and vetted spectroscopic redshifts of the first half of the survey through the DAWN JWST Archive.
  }


\maketitle

\clearpage
\section{Introduction}

The first cycle of observations with the James Webb Space Telescope (JWST; \citealt{Gardner2023}) delivered extraordinary near-infrared imaging of the best-studied extragalactic deep fields. Among a wealth of discoveries in the high-redshift Universe \citep{ISSI2024}, perhaps the most surprising finding has been the great abundance of very red sources that were previously undetected with the Hubble Space Telescope (HST), and unresolved or undetected with the Spitzer Space Telescope. These new sources are likely to be at high redshift, and many are suggested to be substantially more luminous and more massive than expected from previous observations and theoretical models. These results raise major questions: How did the brightest galaxies assemble their stellar mass on extremely short timescales? What evolutionary phases have been missing from existing studies due to incompleteness at excessively red colours?

Although unified by having red colours over $\sim1-4\,\micron$, the new sources discovered with the NIRCam instrument onboard JWST \citep{Rieke2023} have highly heterogeneous morphologies \citep[e.g.][]{Nelson2023,PerezGonzalez2023,Labbe2023b}, suggestive of multiple classes of objects with different formation paths. At the highest redshifts, a red colour (with respect to HST) typically reflects the Lyman break in the rest-frame UV {(specifically, the spectral break at the Lyman limit of $912\AA$ at $z\lesssim5$, and the Lyman-$\alpha$ break due to absorption by the neutral intergalactic medium at $z\gtrsim5$; e.g. \citealt{Madau_1996,Steidel_1996,Giavalisco2002,Steidel_2003})} coupled with either strong emission lines or possibly a Balmer break at rest-frame optical wavelengths \citep[e.g.][]{Eyles2005,Labbe2010}. Searches designed for these types of spectral energy distributions (SED) have rapidly yielded a vast number of candidate galaxies at $z>7$, and beyond $z=10$ \citep[e.g.][]{Castellano2022,Finkelstein_2023,Atek_2022,Donnan_2022,Adams_2022,Naidu2022}. Several of these candidates were significantly brighter than anticipated and suggested to be extremely massive galaxies, reaching stellar masses of $M_*\approx10^{11}\,\rm M_\odot$ before the Universe is 800 Myr old \citep{Labbe2023}. The abundance and masses of these systems have sparked a debate whether these findings are consistent with the standard cosmological model \citep{MBK2023} {, and whether the redshift and mass estimates themselves are correct \citep[e.g.][]{Endsley2023,Kocevski2023}}.

At redshifts $z\sim1-6$, red sources that are not detected at $\sim1\,\micron$ but luminous at $\sim4\,\micron$ may be strongly obscured by dust. Mid- and far-infrared missions as well as ground-based sub-millimetre facilities had previously uncovered a population of sources that are extremely luminous at such long wavelengths, but often faint or undetected with HST, especially at higher redshifts \citep[e.g.][]{Franco2018,Wang2019,Casey2019,Williams2019,Manning2022}: these sub-millimetre galaxies (SMGs) are typically at a redshift of $z\sim1-5$, and are thought to be massive galaxies with extremely high star formation rates \citep[for a review, see][]{Casey2014,Hodge2020}. Thanks to the improved sensitivity of JWST, we are now able to detect near-infrared emission from such SMGs out to $z\sim5$ \citep{Herard2023,Sun2024}, and also extend this population to lower luminosities \citep{Price2023}. The newly discovered population of extremely red sources likely contributes significantly to the stellar mass budget of the high-redshift Universe \citep[e.g.][]{Nelson2023,Fudamoto2022,Barrufet2023,Gottumukkala2024,PerezGonzalez2023,Xiao2023,Weibel2024,Williams2024}.

In contrast with this population of highly star-forming galaxies, a large number of photometric candidate massive quiescent galaxies have been identified out to $z\sim5$ \citep{Carnall2023,Long2024,Valentino2023,PerezGonzalez2023}. The broad-band SEDs of these systems are consistent with strong Balmer breaks, resulting in a red colour ($\rm F150W-F444W\gtrsim2$). Indeed, spectroscopic follow-up with JWST/NIRSpec has now confirmed the presence of old stellar populations in several of these systems at $z\sim4.5-5$ \citep[e.g.][]{Carnall2023b,deGraaff2024c}, with the highest redshift massive quiescent galaxy discovered at $z=7.3$ \citep[][]{Weibel2024b}. The existence of such objects is surprising: the formation of massive ($\gtrsim 10^{10}\,\Msun$) galaxies at $z\gtrsim4-5$ simultaneously requires rapid mass assembly in the first Gyr, and cessation of star formation in an epoch where the star formation activity in galaxies is typically only increasing. The great abundance of massive quiescent galaxies at these high redshifts would pose a challenge for many galaxy formation models \citep{Valentino2023}.

Lastly, a mysterious sample of extremely compact red sources is ill-described by all of the above classes of objects. An apparently characteristic feature is a `v-shaped' SED \citep[e.g.][]{Furtak2023,Barro2024,Labbe2023b}, i.e. a blue rest-frame UV continuum and red rest-frame optical continuum, which has proved challenging to model with many standard SED fitting codes \citep[e.g.][]{Killi2023,Wang2024a}. With photometric redshifts ranging from $z\sim0$ to $z\sim9$, the nature of these sources remains highly debated. Some are likely to be cool dwarf stars in the Milky Way \citep[e.g.][]{Burgasser2024,Hainline2024,Holwerda2024}, while the first spectroscopic measurements for others have unveiled broad Balmer lines suggestive of accreting black holes \citep[e.g.][]{Kocevski2023,Harikane2023,Matthee2024,Greene2024}. If the photometric redshifts are correct and the emission originates from active galactic nuclei (AGN), then these sources may reflect the early formation of massive black holes in high-redshift galaxies, challenging models of black hole growth \citep{Greene2024}. However, if the SED is instead dominated by stars, some of these objects may represent the most massive systems in the high-redshift Universe, forming the likely progenitors of early-type galaxies at $z\sim0$ \citep{Labbe2023,Baggen2023,Akins2024,Wang2024b}.

The limiting factor in understanding the nature of these different bright and red sources in the early Universe is the coarse wavelength sampling from broadband photometry alone. Spectroscopy at near-infrared wavelengths is crucial to characterise the intrinsic shape of the SEDs and the presence of strong emission lines that can be degenerate with continuum breaks in broadband photometry. 
Multi-object spectroscopy with the NIRSpec instrument \citep{Jakobsen2022,Boeker2023} has proved extremely powerful thus far: 
even at modest depths, early spectroscopic programmes have confirmed the redshifts of over a dozen $z>8$ galaxies, revealed a great abundance of emission lines, as well as continuum emission in galaxies out to $z\sim10$ {\citep[e.g.][]{Curti2023,CurtisLake2023,RobertsBorsani2023,Fujimoto2023,Fujimoto2023b,Wang2023,AHaro2023a,AHaro2023b}.} 

However, a great difficulty with spectroscopic programmes with the NIRSpec microshutter array (MSA; \citealt{Ferruit2022}) is the target selection. In the mask design process, sources that are designated to be high priority have a high probability of being observed, but this probability drops rapidly for lower priority classes as more sources are placed on the mask \citep{Bonaventura2023}. The definition of `high' and `low' priority depends entirely on the science programme and can be difficult to quantify, if provided at all. 

Large spectroscopic programmes in Cycle 1 have predominantly prioritised the search for the highest redshift galaxies, which typically are selected to have SEDs consistent with Lyman breaks with blue UV slopes (e.g. the NIRSpec guaranteed time observations (GTO) programmes and early release science programmes; \citealt{Eisenstein2023,Maseda2024,Finkelstein_2023,Treu2022}). Moreover, many of these spectroscopic targets were selected from HST imaging catalogues. Because very red sources were not present in the photometric catalogues or not prioritised in the mask design procedure, the total number of such sources with follow-up spectroscopic observations has been extremely limited thus far.

In this paper we present the \emph{Red Unknowns: Bright Infrared Extragalactic Survey} (RUBIES), a {$\sim60$} hour Cycle 2 spectroscopic follow-up programme with the NIRSpec/MSA of sources selected from public NIRCam imaging obtained in Cycle 1. With 18 pointings spread across two legacy extragalactic deep fields ($\sim150$\,arcmin$^2$), RUBIES is currently the largest JWST/NIRSpec survey in terms of both area and number of targets outside of the GTO programmes. The motivation for RUBIES is twofold. First, with both low- and medium-resolution spectroscopy over a wide area we are able to uncover the nature of a large sample of $\sim100$ extremely rare, red sources ($\rm F150W-F444W>3$) at high redshifts. Second, we wish to place these rare sources into a cosmological context, by observing a census sample of the $z\sim2-7$ galaxy population. Unique to RUBIES is the fact that this census sample follows a well-quantified selection function (thus yielding well-defined spectroscopic completeness) based on only three measurements: the NIRCam $\rm F150W-F444W$ colour, F444W magnitude and photometric redshift.

We present the novel observing strategy developed to achieve our target selection in Section~\ref{sec:observations}, as well as the resulting completeness in the 3D parameter space of $\rm F150W-F444W$ colour, $\rm F444W$ magnitude and photometric redshift. In Section~\ref{sec:reduction} we describe the data reduction procedure, which includes the derivation of custom calibration products. We also assess the data quality by comparing the relative flux and wavelength calibration between the low-resolution and medium-resolution spectra. We present an overview of major science goals and initial scientific results in Section~\ref{sec:science}, and provide a summary in Section~\ref{sec:summary}. Throughout we specify magnitudes using the AB system \citep{ABmags}. Where relevant, we assume a flat $\Lambda$CDM cosmology with $\Omega_{\rm m}=0.3$ and $h=0.7$.

\section{Observing strategy}\label{sec:observations}

\subsection{Image data}\label{sec:imaging}

\begin{figure*}
    \centering
    \begin{subfigure}{0.85\linewidth}
    \includegraphics[width=\linewidth]{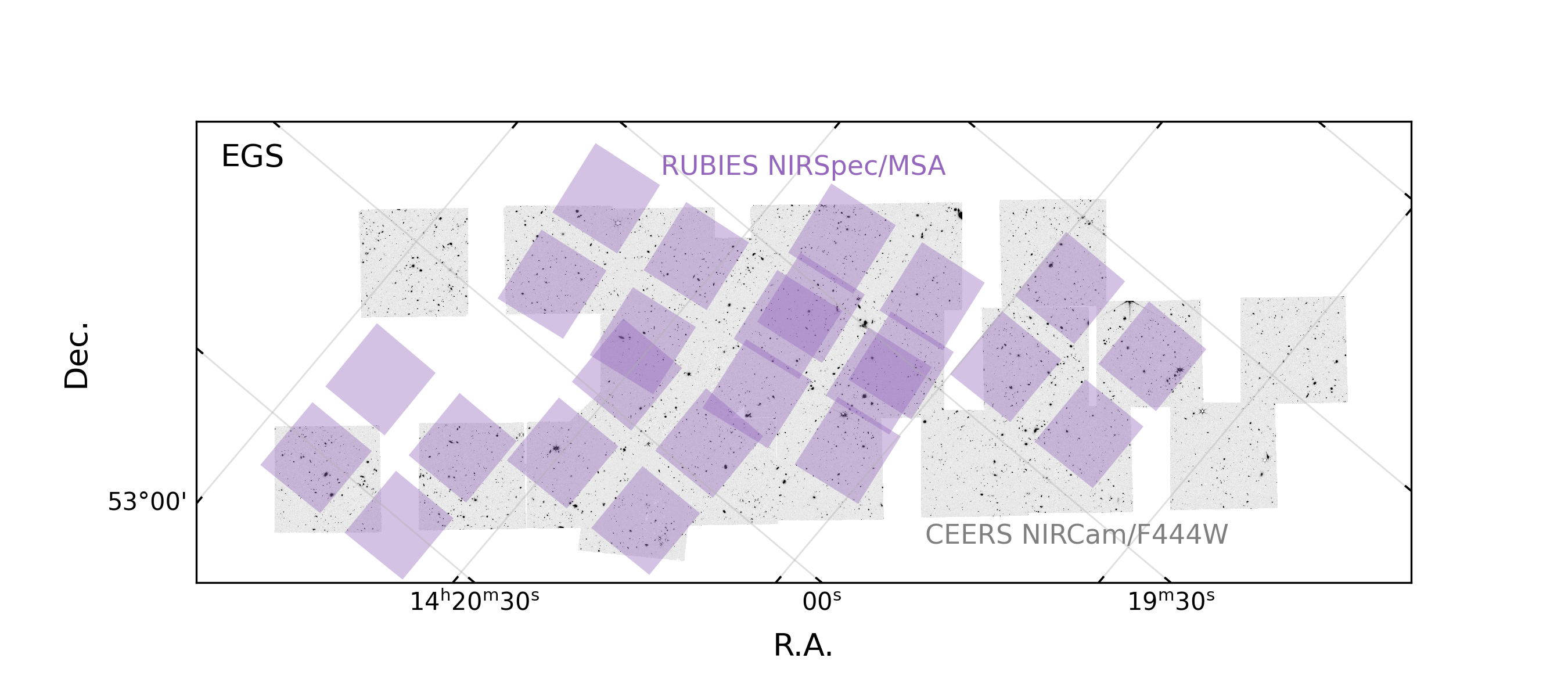}

    \end{subfigure}\vspace{-0.2cm}
    \begin{subfigure}{0.85\linewidth}
    \includegraphics[width=\linewidth]{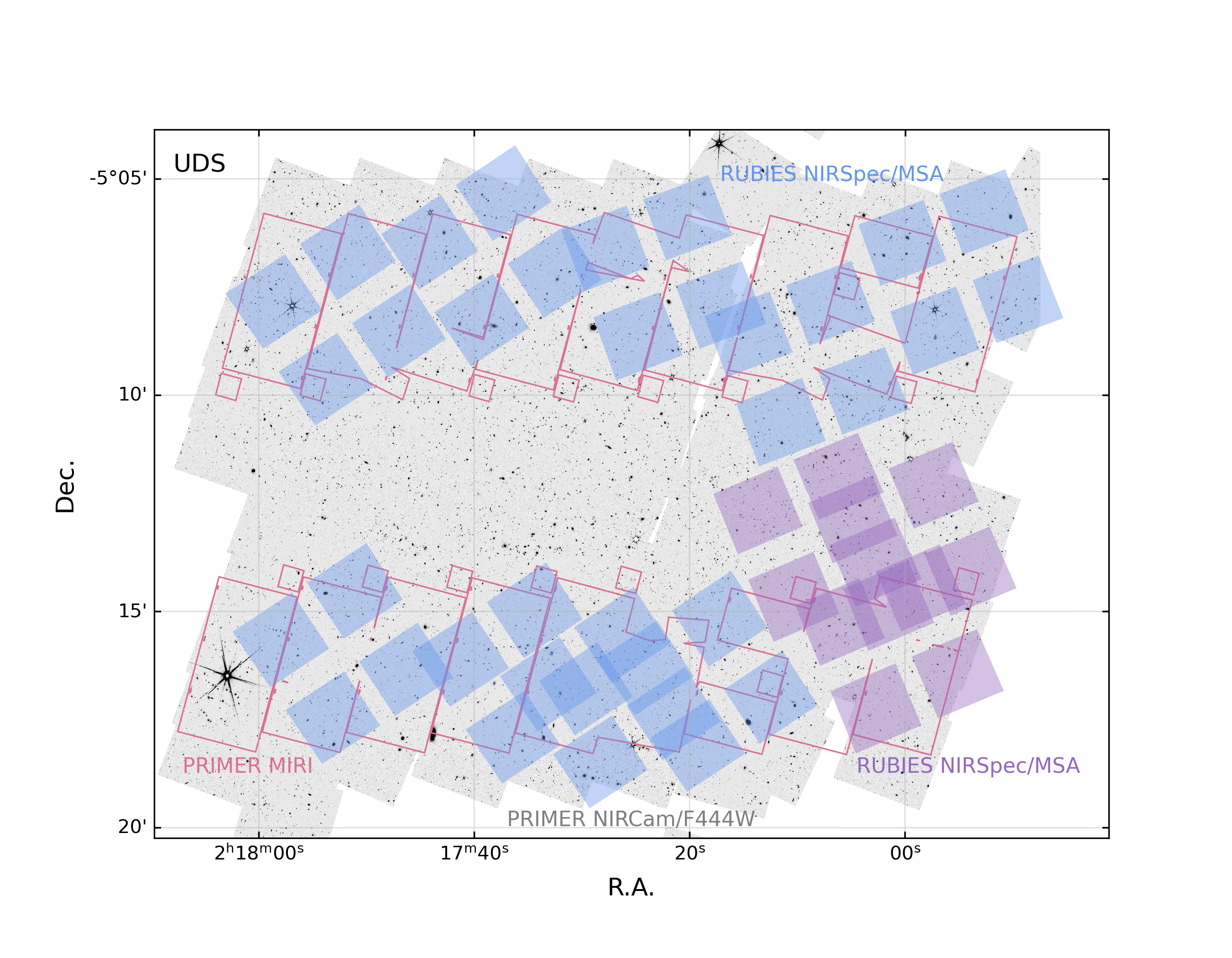}

    \end{subfigure}
    
    \caption{RUBIES footprint of 18 NIRSpec/MSA pointings in the UDS and EGS fields. Purple pointings correspond to the first half of observations in January-March 2024 and form the focus of the current data release. Background images show the NIRCam F444W image mosaics, primarily constructed from public imaging of the CEERS and PRIMER surveys. For the UDS we also show the outline of the PRIMER MIRI imaging footprint in pink.}
    \label{fig:footprint}
\end{figure*}

RUBIES (programme ID 4233; PIs: de Graaff \& Brammer) targets two extragalactic legacy fields: the Ultra-deep Survey (UDS) and the Extended Groth Strip (EGS). Both fields were previously observed with HST as part of the Cosmic Assembly Near-infrared Deep Extragalactic Legacy Survey (CANDELS; \citealt{Grogin2011,Koekemoer2011}) and 3D-HST Survey \citep{Brammer2012,Skelton2014}, which combined provide imaging in F606W, F814W, F125W, F140W, and F160W filters. Moreover, both fields have extensive ancillary data ranging from X-ray to radio wavelengths.

In JWST Cycle 1, the EGS formed the focus of the Cosmic Evolution Early Release Science Survey {(CEERS; PID 1345, PI: Finkelstein; \citealt{Finkelstein2025})}. The NIRCam imaging obtained in this programme spans 7 filters (F115W, F150W, F200W, F277W, F356W, F410M and F444W), reaching a $5\sigma$ point source depth (in a circular aperture of radius $0.1\arcsec$) of 28.6 mag in the F444W filter over an area of approximately 80\,arcmin$^2$ \citep{Bagley2023,Finkelstein_2023}. In addition, smaller sets of imaging data from programmes 2279 (PI: Naidu), 2514 (PIs: Williams \& Oesch; {\citealt{Williams2024:panoramic}}) and 2750 (PI: Arrabal Haro) add depth and area to some of the above filters. Finally, although not publicly available at the time of target selection, F090W imaging from programme 2234 (PI: Ba\~nados; Khusanova et al. in prep.) covers the full CEERS footprint, which we use for flux calibration of the RUBIES spectra.

The UDS is one of two legacy fields targeted by the Public Release IMaging for Extragalactic Research (PRIMER) Survey (PID 1837; PI: Dunlop). The PRIMER NIRCam imaging in the UDS covers a wide area of 224 arcmin$^2$ in 8 different filters (F090W, F115W, F150W, F200W, F277W, F356W, F410M and F444W), reaching an average image depth of 27.9 mag in the F444W filter for apertures of radius $0.15\arcsec$ \citep{Donnan2024}. Imaging from pure parallel programmes 2514 (PIs: Williams \& Oesch) and 3990 (PI: Morishita) further increase depth and area in various parts of the UDS field.

Finally, we note that the PRIMER Survey was designed as a coordinated parallel programme, such that MIRI \citep{Wright2023} imaging was obtained simultaneously with NIRCam imaging. This provides mid-infrared imaging in the F770W and F1800W filters over an area of approximately 125 arcmin$^2$, close to half of the NIRCam footprint. In Figure~\ref{fig:footprint} we show the NIRCam F444W imaging in the EGS and UDS fields, together with the outline of the MIRI footprint in the UDS. In the near future MIRI imaging will also be publicly available across the entire CEERS area in the EGS from programme 3794 (PI: Kirkpatrick).

All publicly available imaging data were reduced using \texttt{grizli} \citep{grizli}, described in detail in \citet{Valentino2023}. We use image mosaics from the DAWN JWST Archive (DJA) version 7.2\footnote{For the first three RUBIES masks observed in January 2024 we used version 7.0 for the target selection. The main difference between these two versions is an improved treatment of hot pixels in the long wavelength filters. }, which have a pixel scale of $0.04\arcsec$.

\subsection{Spectroscopic observations}\label{sec:spec_obs}

The RUBIES observations consist of 18 NIRSpec MSA pointings, with 12 pointings located in the UDS and 6 in the EGS (Table~\ref{tab:pointings}). We show the footprint of the survey in Figure~\ref{fig:footprint}, with pointings that were observed between January-March 2024 in purple, forming the primary focus of this paper and data release. Pointings shown in blue were observed very recently (August 2024) or are still scheduled. The total area spanned by the NIRSpec MSA quadrants is approximately $150$ arcmin$^2$ after accounting for small overlaps between pointings. The choice for these precise locations is described in Section~\ref{sec:masks}, although we note here that for the UDS we aimed for a strong overlap with the PRIMER MIRI footprint. 

We observe each pointing with two different disperser/filter combinations: the low-resolution ($R\sim100$) PRISM/Clear and the medium-resolution ($R\sim1000$) G395M/F290LP modes, covering $0.6-5.3\,\micron$ and $2.9-5.3\,\micron$, respectively. For each target on the mask we open 3 microshutters to construct a slit. A 3-point nodding strategy is used, with an integration time of 963\,s  per exposure (65 groups using the NRSIRS2RAPID readout pattern). The total exposure time per source is 48\,min for each disperser/filter combination. A small number of sources ($\sim 1\%$) were observed in two separate pointings and therefore have double this exposure time. 

The PRISM and G395M observations are taken consecutively and at exactly the same location, but do not use the same masks. After obtaining the PRISM data, we reconfigure the MSA to place extra sources onto the mask before observing with the G395M disperser. Because the spectral traces from the G395M disperser are long (spanning approximately the length of one detector), this leads to a large number of overlapping traces on the detector. However, because the background is much lower at medium resolution than for the PRISM observations and the majority of sources are faint (Section~\ref{sec:catalog}), we can allow for large numbers of overlapping traces (typically up to $\approx 5$) without significant sacrifice to the data quality (a strategy that was also used in \citealt{Maseda2023}). Only in rare cases do we detect continuum emission from very bright sources at the depth of RUBIES and contamination therefore forms a problem, but we find that the large number of PRISM observations allow us to disentangle such overlapping traces. This approach uses the NIRSpec MSA in a similar fashion to the NIRCam grism mode, with the key difference being that the majority of MSA shutters remain closed and the background is therefore substantially reduced. We describe the selection of the `grating-only' targets in Section~\ref{sec:masks}.

\begin{table}
\caption{Observed RUBIES pointings.}
\begin{center}
 \begin{tabular}{ccccc}\hline\hline
  Visit & RA & Dec & APA & Obs. date\\
     & (J2000) & (J2000) & (deg) & \\\hline
  \multicolumn{5}{c}{UDS}\\\hline
  1:1 &  02:17:01  & -05:15:51 & 203.00 & 2024-01-16 \\   
  1:2 &  02:16:59  & -05:13:29 & 203.00 & 2024-01-18 \\   
  1:3 &  02:17:08  & -05:13:17 & 203.00 & 2024-01-19 \\  
  2:1 &  02:16:55  & -05:07:09 & {200.75} & {2024-12-19} \\  
  2:2 &  02:17:09  & -05:09:16 & {200.74} & {2024-12-19} \\  
  2:3 &  02:17:23  & -05:07:17 & {200.74} & {2024-12-19}\\  
  3:1 &  02:17:38  & -05:06:47 & 33.56 & 2024-07-25 \\  
  3:2 &  02:17:53  & -05:08:09 & 33.55 & 2024-07-25 \\  
  3:3 &  02:17:52  & -05:15:59 & 33.55 & 2024-07-25\\  
  4:1 &  02:17:35  & -05:16:26 & 33.59 & 2024-08-08 \\  
  4:2 &  02:17:27  & -05:17:00 & 33.59 & 2024-08-09 \\  
  4:3 &  02:17:18  & -05:16:38 & 33.60 & 2024-08-09\\  \hline
  \multicolumn{5}{c}{EGS}\\\hline
  5:1 &  14:20:24  &  52:57:40 & 0.88 & 2024-03-20 \\
  5:2 &  14:20:02  &  52:53:55 & 0.80 & 2024-03-20 \\
  5:3 &  14:19:15  &  52:48:10 & 0.65 & 2024-03-20 \\
  6:1 &  14:19:45  &  52:56:25 & 7.84 & 2024-03-13 \\
  6:2 &  14:19:29  &  52:52:13 & 7.79 & 2024-03-13 \\
  6:3 &  14:19:38  &  52:51:49 & 7.82 & 2024-03-13\\\hline  
 \end{tabular}
\tablefoot{The aperture position angle (APA), is the angle of the NIRSpec microshutters as projected onto the sky, and differs from the position angle of the telescope itself. }
\label{tab:pointings}
\end{center}
\end{table}

\subsection{Parent catalogue}\label{sec:catalog}

The parent catalogue of RUBIES was largely constructed from the source catalogues (version 7.2) that are publicly available on the DJA. These catalogues were created by performing source detection on an inverse variance weighted stack of the NIRCam F277W, F356W and F444W mosaics with \texttt{SEP} \citep{Barbary2016}, a Python implementation of \texttt{SourceExtractor} \citep{Bertin1996}, and subsequently measuring photometry for the detected sources in circular apertures of radius $0.25\arcsec$ for all available bands. {Uncertainties on these aperture fluxes were measured from the weight images, by summing the pixel variances in the same apertures. We note that we chose an aperture of radius $0.25\arcsec$, as this matches the effective radii of galaxies at the median photometric redshift of our survey ($z\sim4$; e.g. \citealt{Kartaltepe2023,Ormerod2024,Sun2024:morph}). } The aperture fluxes were rescaled to `total' fluxes using the ratio between the Kron aperture flux and the circular aperture flux measured from the detection image. Importantly, these measurements do not account for variations in the point spread function (PSF) as a function of wavelength. 

Photometric redshifts were estimated using \texttt{eazy} \citep{Brammer2008} with the \texttt{agn\_blue\_sfhz\_13} template set and without any priors. This template set is optimised for a broad redshift range, and includes templates of emission line dominated sources, as well as an empirical template of a compact red AGN designed to roughly match the source from \cite{Killi2023}. {The redshift fits were run using an iterative estimation of zero-point offsets for each filter, as described in \citet{Whitaker2011} and \citet{Skelton2014}. Briefly, this algorithm computes the residual between the observed and best-fit model photometry for each filter (keeping F277W as a reference point); the average zero point offset (per filter) is then computed from all objects in the catalogue. The redshift fitting is repeated with these new zero point estimates, in order to iteratively minimise the residuals for all filters. Although the absolute flux calibration of NIRCam \citep{Gordon2022} is currently better than $<1-2\%$ for the filters used here\footnote{\url{https://jwst-docs.stsci.edu/jwst-calibration-status/nircam-calibration-status/nircam-imaging-calibration-status}}, we find that the DJA aperture photometry and PSF-matched aperture photometry of W24 differ by an approximately constant offset (of up to $\approx 0.1$ mag for short wavelengths), with secondary scatter due to source morphology and colour gradients. These empirical zero-point offsets therefore effectively apply an average correction that (partially) compensates for the different PSFs of the images that were not convolved to a common PSF before the aperture photometry was performed. }

To test this catalogue, we compared the identified sources and photometric redshifts to the catalogues of \citet[][hereafter W24]{Weibel2024}. This second set of catalogues was created using the same software (\texttt{SourceExtractor}, \texttt{eazy}), but with the critical difference that empirical PSF models were used to smooth the mosaics to match the PSF of the F444W mosaics. We find that in general the two catalogues agree very well: the overlap in sources is large (96\% of sources in the DJA catalogue are also in the W24 catalogue), and the aperture photometry agrees to within $<0.1$ mag even for the bluest filters. The photometric redshifts also agree well, with an outlier fraction ($\Delta \zphot / (1+\zphot)>0.2$) of 0.12, many of which are faint sources ($\rm F444W>27$ mag). 

Although we found that the two catalogues agree well overall, we opted to use the DJA catalogue (i.e. without PSF matched photometry) as our primary catalogue. First, the DJA catalogues were (at the time) available for all fields, providing a homogeneous catalogue from the start. Second, the source detection of W24 was optimised for the detection of high-redshift sources. Upon visual inspection we found that many sources at $z\sim1-3$ were deblended into multiple components, whereas they would constitute a single source detection in the DJA catalogue. Because a large fraction of RUBIES targets are at intermediate redshifts, the latter scenario is preferred for the purposes of designing our spectroscopic follow-up programme.

However, we applied some modifications to the DJA catalogue to form the final RUBIES parent sample. Most importantly, we supplemented the DJA catalogue with high-fidelity high-redshift ($z>6.5$) targets from the catalogues of W24, which we describe in further detail below. We also made use of the quality flags available in the W24 catalogues to filter out artefacts, bright stars and diffraction spikes through a cross-matching (with separation $<0.2\arcsec$) between the two catalogues. Finally, we visually inspected all (yes \textit{all}) bright sources in the catalogue ($\mathrm{F150W}<20$ regardless of redshift; $\mathrm{F150W}<24$ or $\mathrm{F444W}<24$  for $\zphot>3$) to weed out diffraction spikes and stars that escaped the quality flags of W24. The final catalogue contains approximately 200,000 sources with good-quality photometry (64,311 in the EGS, 137,049 in the UDS).

\subsection{Target prioritisation}\label{sec:targets}

\begin{figure*}
    \centering
    \includegraphics[width=\linewidth]{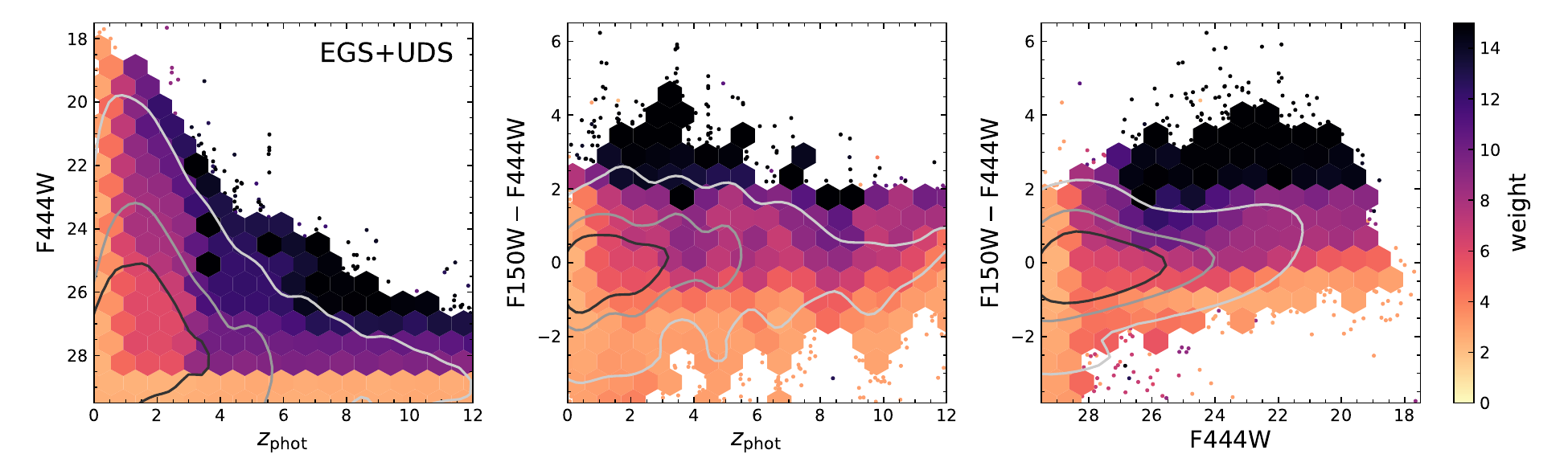}
    \caption{Distribution of targets in the RUBIES parent catalogue in different projections of the 3D parameter space of $\rm F150W-F444W$ colour, total $\rm F444W$ magnitude and best-fit photometric redshift. Black, grey and white contours enclose the 50th, 75th and 95th percentiles of all sources in the catalogue, respectively. The colour coding shows the average weight of sources in a bin; for bins containing fewer than 10 objects we show individual data points. Weights are computed for targets according to their number density in this 3D parameter space, such that the rarest sources receive the highest weight (see Section~\ref{sec:targets}). }
    \label{fig:target_weights}
\end{figure*}

The target prioritisation is split in two categories according to the scientific motivation for RUBIES: the highest priority red sources (`Rubies') themselves, and the census sample of the high-redshift galaxy population. 

High-priority Rubies were selected in two different ways:
\begin{itemize}
    \item \textit{Red sources.} We selected all sources with $\rm F150W-F444W>2$ and $\rm F444W<27$, where the colour was measured in a circular aperture of radius $0.25\arcsec$ and a $1\sigma$ upper limit was used in case of non-detection in the F150W filter. {The upper limit on the magnitude ($\rm F444W<27$) was chosen to yield a marginal detection of continuum emission within 48\,min of PRISM observations ($S/N\sim1-2\,$pix$^{-1}$), which we estimated based on extensive simulations with the JWST Exposure Time Calculator using realistic galaxy size distributions.} As we may expect template fitting to fail for very rare sources, we did not use any photometric redshift information for the selection (and we indeed find a high outlier fraction for these sources, as discussed in Section~\ref{sec:spec_zs}). Image cutouts of all sources in the DJA catalogue meeting these criteria were visually inspected, regardless of photometric quality flag, to check whether the sources are real or artefacts. This yielded 1269 sources across both fields.
    
    \item \textit{Bright high-redshift sources.} We selected sources with $\zphot>6.5$, probability $P(\zphot >6)>0.5$ and $\rm F444W<27$ from either the DJA or the W24 catalogues (i.e. duplicate sources only need to meet these criteria in one catalogue to be selected). We further required that sources are covered by all available JWST broad filters, have a signal-to-noise ratio $S/N>8.5$ in the stacked long wavelength mosaic and are undetected ($S/N<3$) in filters below $1\,\micron$ (F435W, F606W, and F814W or F090W where available). For the W24 catalogues we used {(PSF-matched)} photometry measured from smaller circular apertures of radius $0.16\arcsec$. The SEDs and image cutouts were visually inspected for all sources to assess whether the source is (i) real or an artefact and (ii) likely to be at high redshift. Sources were inspected by three reviewers (AdG, AW, PO) and selected as a good high-redshift candidate through a simple majority, resulting in a total of 868 objects.
\end{itemize}
There are 39 sources which met the criteria for both of the above selections. We further subdivided each class of targets into Priority 1 and 2 classes: extremely red sources with $\rm F150W-F444W>3$ (317), and high-redshift candidates with $\rm F444W<26.5$ (442) were assigned Priority 1; all other sources were assigned Priority 2.

Next, we assigned priority to the sources that form the census survey using a simple selection function based on three quantities: the $\rm F150W-F444W$ colour, total $\rm F444W$ magnitude and best-fit photometric redshift. As an estimate of the number density of each source, we computed the distance $d_8$ to its 8th nearest neighbour in this 3D parameter space. We then assigned a weight $W$ to each source that is inversely proportional to the natural logarithm of this number density: $W = -3 \ln(d_8)$, with a maximum of $W=15.8$. In this way, sources in the extremes of the colour-magnitude-redshift space receive the highest weight, while sources that are very common are assigned a low weight. We note that, for the purposes of the mask design procedure (Section~\ref{sec:masks}), we broadly subdivided the census sample in two priority classes, split by photometric redshift ($\zphot=3$) with the higher redshift group having higher priority. In practice, this ensures that high-redshift targets are always placed on the mask before low-redshift sources, regardless of the computed weight. Finally, Priority 1 and 2 sources were also given a weight following this strategy, on average resulting in $W({\rm P1})\approx15.8$ and $W({\rm P2})\approx14.5$\,, respectively.

In Figure~\ref{fig:target_weights} we show the distribution of the full parent catalogue in the three projections of the $\rm F150W-F444W$, $\rm F444W$ and photometric redshift parameter space. Contours enclose the 50th, 75th and 95th percentiles of the parent catalogue: unsurprisingly, the vast majority of sources are at low redshifts, faint and relatively blue. The oscillatory features in the $\rm F150W-F444W$ vs. $\zphot$ plane result from the Balmer break and Lyman break shifting in and out of the $\rm F150W$ filter. We show the average weight across the parameter space in colour, demonstrating that the reddest, brightest and highest-redshift sources receive the highest weights.

\subsection{Mask design}\label{sec:masks}

To create masks for the NIRSpec MSA we allocated shutters to sources according to their weight and priority class. The `best' mask can then be defined as the one that reaches the highest combined weight of all allocated targets. The difficulty in designing a survey is that we not only wish to optimise individual masks, but also optimise the total weight of the survey across all 18 pointings. Currently, no software exists to tackle this problem: both the default MSA Planning Tool (MPT) and eMPT software \citep{Bonaventura2023} were designed to optimise single masks. We therefore used a combination of existing and custom tools to design the RUBIES masks.

\subsubsection{Pointing locations}\label{sec:pointings}

Our aim was to find the optimal set of pointings that maximises the number of observed Priority 1 Rubies across the full survey area. To do so, we leveraged the initial pointing algorithm (IPA) of the eMPT software, which can very efficiently search for pointing locations that contain a large number of Priority 1 sources. We ran the IPA for the PRISM disperser for a large number of starting points (a grid of points separated by $\approx 1.5\arcmin$) and used a search box of $1\arcmin$. For each IPA run, this returns multiple groups of pointing locations containing typically $\approx 10$ Priority 1 sources for which PRISM spectra can be obtained while avoiding overlapping traces. Collecting all pointing groups, this yielded hundreds of possible pointing locations. 

Many of these pointing locations are redundant as they target (largely) the same sets of high-priority sources. We therefore pruned the list of pointing locations iteratively. We first computed the total number of unique Priority 1 sources covered by the full list of pointings, and checked which pointing contributes the lowest number of unique sources. We then removed this pointing from the list and repeated this process until we reached a list of 25 pointing locations. With a manageable number of 25 pointings, we could then perform a brute force computation of the number of unique Priority 1 sources for every possible combination of $N$ pointings within the set of 25 pointings, where the number of pointings $N$ depends on the field. 

This approach typically yielded a small number ($\sim 5$) of sets of pointing locations with an equal number of high-priority sources. We decided between these equivalent sets of pointings based on other priorities: in the UDS we aimed for strong overlap with the footprint of the existing MIRI imaging. We also selected a few ($\sim 10$) `Priority 0' targets with weight $W=100$, which for example include the brightest sources of \citet{Labbe2023} (published in \citealt{Wang2024b}) and the $z\approx7$ massive quiescent galaxy of \citet{Weibel2024b}. These sources helped decide between otherwise degenerate pointings, but we stress that, because there are very few such Priority 0 sources, their selection does not bias the overall selection function.

The mask designs of eMPT are conservative in the sense that only shutters that result in complete traces are opened for the PRISM disperser: shutters with traces that would be partially truncated by the detector chip gap or the edge of the detector are censored \citep{Bonaventura2023}. For RUBIES we decided to allow for such truncated traces, as in practice we have found this to significantly affect only a very small number of sources. The optimal set of pointings found with the above workaround for the eMPT therefore merely served as a starting point: we used a search radius of $30\arcsec$ around the location of these pointings to search for pointings with (i) the highest combined target weight (Section~\ref{sec:targets}) and (ii) the least overlap between pointings, which results in the final pointing locations shown in Figure~\ref{fig:footprint}.

\begin{figure*}
    \centering
    \includegraphics[width=0.8\linewidth]{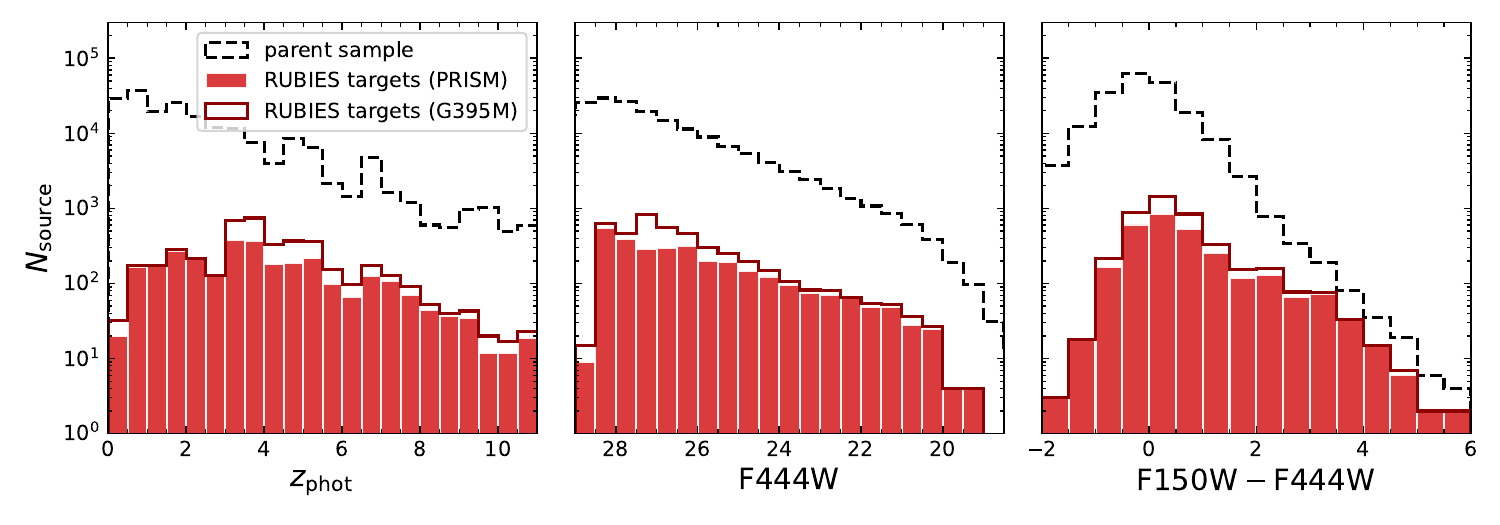}
    \includegraphics[width=0.8\linewidth]{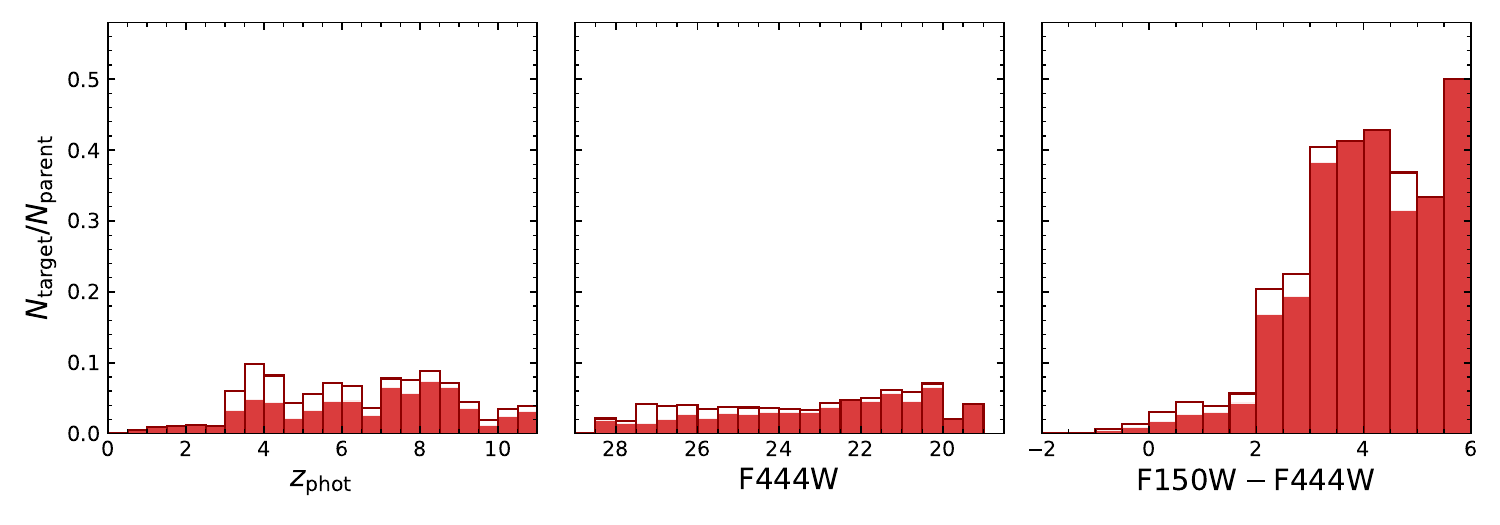}
    \caption{Distribution of the RUBIES parent sample (spanning the full EGS and UDS NIRCam area of $\sim 300\,$arcmin$^2$) and of the targets selected for spectroscopic follow-up, shown in the space of parameters used to define the selection function: photometric redshift, F444W magnitude, and $\rm F150W-F444W$ colour. Top panels show absolute source counts, while bottom panels show the fraction of observed targets with respect to the full parent sample. The RUBIES selection is strongly biased (as intended) toward red sources and preferentially targets brighter sources at $\zphot>3$, although the majority of the observed targets are still faint ($\rm F444W>26$) and relatively blue ($\rm F150W-F444W \sim 0$). }
    \label{fig:sample_hists}
\end{figure*}

\subsubsection{MSA configuration}\label{sec:msa_hacks}

We have developed a custom algorithm to configure the MSA, described as follows. Using the latest version of APT (at the time of observing), we export the file containing information on operable, failed closed and failed open shutters. For a given pointing location and aperture position angle (APA), we then construct a list of sources that fall in open shutters. We generally use a lenient definition of `open', such that source centroids that fall on the walls between shutters are also included (we choose to do so, because many sources are spatially extended). Only for the Priority 0 and 1 sources do we use a stricter definition -- the true open shutter area -- to minimise slit losses.

Next, we place sources on the mask one at a time (opening slitlets of 1x3 shutters per source), moving down the parent catalogue ordered by the priority class and source weight, starting with the highest weight and priority. We construct an empirical trace model for the NIRSpec PRISM spectra that reverse engineers the trace calibration implemented in the full STScI JWST pipeline using spectroscopic observations from the CEERS survey. We first fit a quadratic polynomial for each trace in extracted CEERS PRISM spectra, i.e. the detector $y$ cross-dispersion location of the trace as a function of the $x$ detector axis, and then approximate the full PRISM trace model by fitting a 2D cubic polynomial to these trace coefficients as a function of the MSA shutter row and columns (separated by MSA quadrant). For each source to be placed on the mask, we use this trace model to assess whether its 3-shutter trace overlaps with those of already allocated shutters to decide whether or not the triplet of shutters can be opened; if there is overlap, the algorithm moves further down the list. The combined weight of the mask is then simply the sum of the weights of all allocated targets. We compute this combined weight for all points on a finely spaced grid in the vicinity of the pointing location from eMPT (Section~\ref{sec:pointings}) to determine the final pointing location and MSA configuration.  We note that for a particular specified spacecraft pointing, set of valid MSA shutters, and a catalogue of source positions and weights, our shutter allocation procedure is deterministic and optimal for the \textit{weight-ordered} sources but is not necessarily optimal for the \textit{total} weight (e.g., swapping two sources $j$ and $k$ that would overlap with source $i$ but not with each other and where $w_i > \mathrm{max}(w_j, w_k)$ and $w_i < w_j + w_k$).

This procedure typically allocates $\sim 170$ (and up to 200) targets per PRISM mask. As a last step for the PRISM masks, we open blank sky shutters in areas where there is sufficient space left on the detector. This typically results in $\sim 30-40$ background shutters per mask. These background shutters are extremely valuable for calibration and reduction purposes, as discussed in Section~\ref{sec:reduction} and Appendix~\ref{sec:sky_appendix}.

For the G395M masks, we begin at the same location and by allocating the same targets as for the PRISM mask, such that all sources observed with the PRISM disperser are also observed with the G395M disperser. No restriction is imposed against sources whose 3-shutter G395M spectra overlap (see Section~\ref{sec:spec_obs}), and because of strongly overlapping spectra we do not open any background shutters. We do add further sources to the mask, provided that their best-fit photometric redshift $\zphot>3.0$ ($\zphot>3.3$ for the EGS) where we may expect to observe the H$\alpha$ line in the G395M data. This increases the number of sources by approximately 50\% ($\sim 250$ sources per mask), i.e. one third of all sources targeted in the survey have only a G395M observation.

Finally, we thoroughly inspect all open shutters (science, background, and failed open) in the Astronomer's Proposal Tool, to check that no bright stars have entered the mask. As the UDS contains multiple bright ($G<14$) stars, in a few cases this led to repeating the search for optimal pointings.

\subsubsection{Spectroscopic completeness}

\begin{figure}
    \centering
    \includegraphics[width=\linewidth]{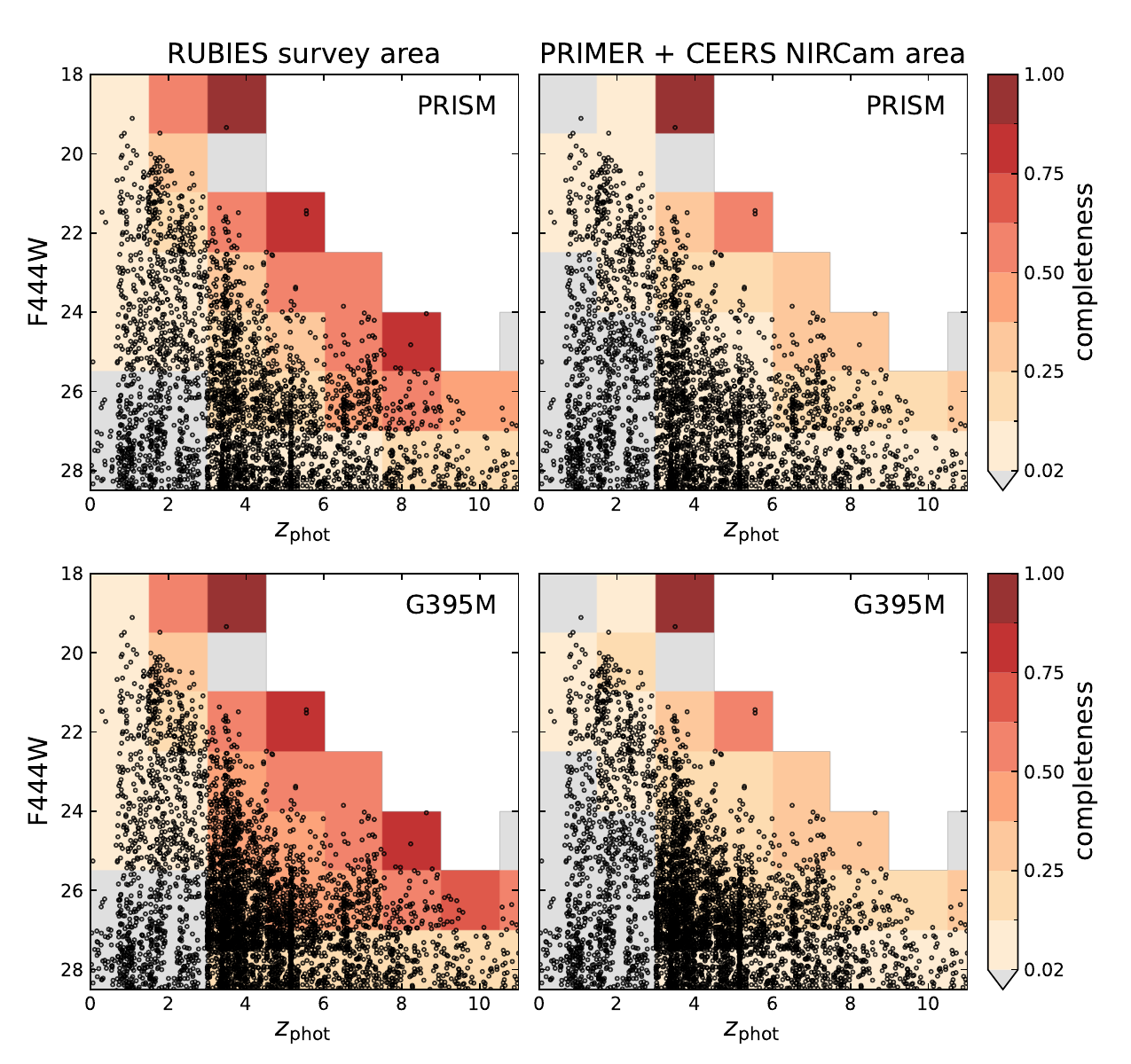}
    \caption{Distribution of photometric redshifts and F444W magnitudes of RUBIES targets for the PRISM (top) and G395M (bottom) observations. Colour coding shows the spectroscopic completeness in each bin: on the left this is computed as the fraction of targets in the RUBIES NIRSpec footprint that are observed. On the right this is calculated as the fraction of observed targets from the full parent catalogue (i.e. the total PRIMER and CEERS area, approximately double the area covered by RUBIES). The RUBIES selection function achieves high ($>50\%$) spectroscopic targeting completeness for bright, high-redshift sources, even reaching $>70\%$ in the extremes of the parameter space. }
    \label{fig:zmag_completeness}
\end{figure}

\begin{figure*}[!t]
    \centering
    \includegraphics[width=0.9\linewidth]{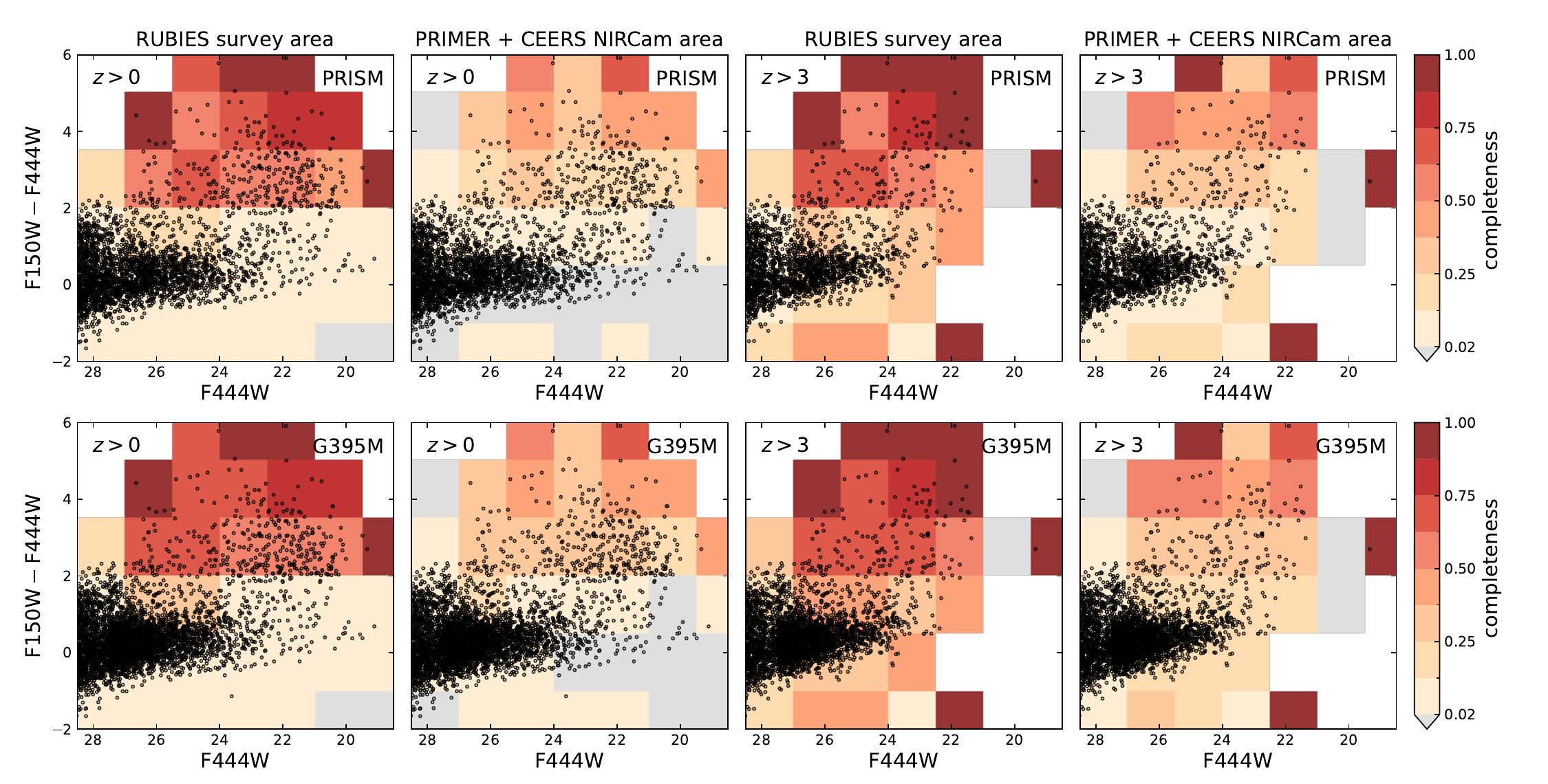}
    \caption{Distribution of F444W magnitudes and $\rm F150W-F444W$ colours of RUBIES targets for the PRISM (top) and G395M (bottom) observations. Symbols and colour scale are the same as in Figure~\ref{fig:zmag_completeness}: the colour coding indicates the spectroscopic completeness computed for the RUBIES footprint alone or the full NIRCam area of the parent catalogue. The two sets of panels on the left show all sources, whereas those on the right only show sources with $\zphot>3$. RUBIES reaches very high completeness for red sources, especially at $\zphot>3$. Comparison of the two different measures of completeness shows that RUBIES is biased toward red sources, which is the result of our pointing location optimisation (Section~\ref{sec:pointings}). Nevertheless, because the vast majority of sources in the parent sample are faint and blue (Figure~\ref{fig:sample_hists}), the majority of RUBIES sources are also faint and blue, forming a critical comparison sample to place the rare, red sources in the context of the broader galaxy population. }
    \label{fig:clr_mag_completeness}
\end{figure*}

In total, RUBIES targets {2901} sources in both the PRISM and G395M observations. Approximately 300 of these targets are red, and 200 are bright high-redshift candidates (as defined in Section~\ref{sec:targets}). An additional $\sim1500$ sources are observed with only the G395M disperser. Figure~\ref{fig:sample_hists} shows the redshift, F444W and $\rm F150-F444W$ distributions of the selected targets in comparison to the parent sample, as well as the ratio between the target sample and parent sample. RUBIES is clearly strongly biased toward red sources, demonstrating the success of our mask design procedure. Despite what the survey name suggests, however, the majority of RUBIES targets are relatively blue: this reflects the fact that the vast majority of sources in the parent sample are blue, and these sources form the census sample that is critical to place the rare red sources into the context of the full galaxy population. 

We further evaluate the selection function of the survey by computing the achieved completeness in the 3D parameter space used for target prioritisation. Here, we define completeness as the ratio of targets that are observed and the targets that could have been observed. The latter depends on the area used: i.e., whether we only consider sources in the area covered by the NIRSpec quadrants ($\sim 150\,$arcmin$^2$), or the full RUBIES parent catalogue ($\sim 300\,$arcmin$^2$). 

In Figure~\ref{fig:zmag_completeness} we show the completeness as a function of photometric redshift and F444W magnitude: left panels show the completeness computed for the RUBIES NIRSpec footprint, right panels the completeness using the full NIRCam area of PRIMER and CEERS. Black points show the selected RUBIES targets, and the colour coding indicates the completeness in a given bin. We separately show the PRISM (top) and G395M (bottom) masks, which highlights the fact that the G395M masks predominantly target sources at $\zphot>3$.

We reach high completeness for the brightest sources at $\zphot>3$ (typically $>50\%$, and $>70\%$ for the extremes). Comparing the completeness within the survey area vs. the full available NIRCam area, we see that RUBIES is biased (as intended) toward bright, high-redshift sources: although the full NIRCam area is a factor $\approx 2$ larger than the RUBIES area, the completeness does not differ by a simple factor 2 between the left and right panels. For sources that are common the completeness is low ($<10\%$), but we still sample many such sources: the bulk of RUBIES targets are fainter sources at $\zphot\sim3-5$. Because we sample many of these sources and have a well-defined weight for each observed target, we can correct for incompleteness in future studies of e.g. scaling relations or mass functions.

Similarly, in Figure~\ref{fig:clr_mag_completeness} we show the completeness as a function of $\rm F150W-F444W$ colour and $\rm F444W$ magnitude. As before, we distinguish between the PRISM and G395M masks, as well as the completeness computed for the RUBIES area and full NIRCam area. We further differentiate between the full survey and sources with $\zphot>3$. These figures demonstrate that RUBIES reaches very high completeness ($\gtrsim 70\%$) for the reddest sources, especially for sources with $\zphot>3$, even though the photometric redshift was not explicitly included in the target prioritisation for the reddest sources (Section~\ref{sec:targets}). As in Figure~\ref{fig:zmag_completeness}, comparing the two different measures of completeness shows that RUBIES is biased toward redder sources as the result of our pointing location optimisation {(which maximises the number of red sources observed across the survey, see Section~\ref{sec:pointings})}. 

\begin{figure}
    \centering
    \includegraphics[width=\linewidth]{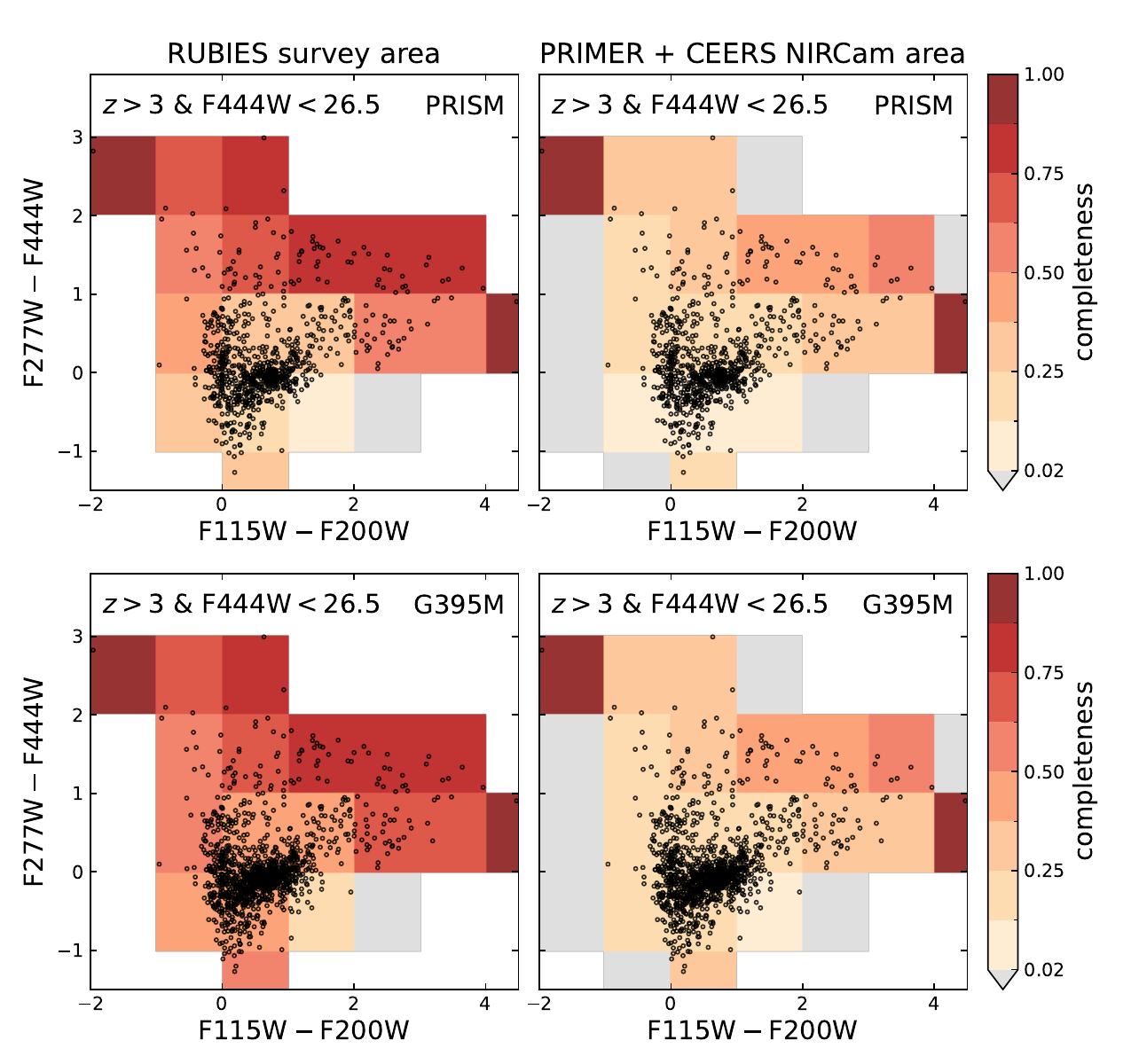}
    \caption{Distribution of RUBIES targets for the PRISM (top) and G395M (bottom) observations in the colour-colour space of the NIRCam broad filters $\rm F115W-F200W$ and $\rm F277W-F444W$. Symbols and colour scale are the same as in Figure~\ref{fig:zmag_completeness}; we only show targets with $\zphot>3$ and $\rm F444W<26.5$. Although RUBIES targets were not selected in this parameter space, the RUBIES selection function samples the (broad) distribution very well. The completeness increase along both colour axes and reaches very high values ($>80\%$) in the extremes.}
    \label{fig:clr_clr_completeness}
\end{figure}

Finally, we show the same completeness measurements in colour-colour space spanned by four NIRCam broad bands ($\rm F115W-F200W$ vs. $\rm F277W-F444W$) in Figure~\ref{fig:clr_clr_completeness}. 
RUBIES targets were not selected in this parameter space, and the colour distribution therefore provides valuable insight into the consequences of our selection function. We plot only relatively bright sources with $\rm F444W<26.5$ and $\zphot >3$. The majority of sources are blue in both $\rm F115W-F200W$ and $\rm F277W-F444W$ colours, but the tails of the distribution span a very wide range in colour ($\approx 4\,$mag). The completeness increases along both colour axes, reaching very high values ($>80\%$) in the extremes of the distribution.

\section{Data processing}\label{sec:reduction}

All RUBIES spectra are reduced with the latest version of \texttt{msaexp}\footnote{\url{https://github.com/gbrammer/msaexp}} \citep{msaexp}. A previous version was described in \citet{Heintz2024}, and corresponds to version 2 of NIRSpec data released on the DAWN JWST Archive\footnote{\url{https://dawn-cph.github.io/dja}}. In this Section and Appendix~\ref{sec:sky_appendix}, we provide a brief description of the reduction pipeline and primarily focus on the changes with respect to the description in \citet{Heintz2024}. Notably, these changes include a major improvement to the absolute flux calibration, and reductions with two different background subtraction strategies (local and global).

Moreover, we use the multiple observations of the RUBIES targets (i.e. both PRISM and G395M observations) to assess the relative flux and wavelength calibration of our dataset. A similar analysis was previously performed by the NIRSpec GTO team \citep{Bunker2023,JADES_DR3} for data from the JWST Advanced Deep Extragalactic Survey \citep[JADES][]{Eisenstein2023} reduced with the NIRSpec GTO pipeline. The large number of RUBIES targets and our multi-disperser observing strategy now enables such a characterisation also for GO data and the public pipeline \texttt{msaexp}.

The reduction and data quality as described here correspond to the new version 3 of NIRSpec data on the DJA\footnote{\url{https://s3.amazonaws.com/msaexp-nirspec/extractions/nirspec_graded_v3.html}}. We publicly release all reduced RUBIES spectra from the first half of observations (January-March 2024) through the DJA. We also provide visually vetted spectroscopic redshifts through this database, as described in Section~\ref{sec:spec_zs}.

\subsection{Data reduction}

We begin by running the uncalibrated (\texttt{uncal}) exposures downloaded from the Mikulski Archive for Space Telescopes (MAST) through the \href{https://jwst-pipeline.readthedocs.io/en/latest/jwst/pipeline/calwebb_detector1.html}{Detector1Pipeline} steps of the standard \texttt{jwst} pipeline\footnote{Pipeline version 1.14.0 with calibration files \texttt{jwst\_1225.pmap} from the Calibration Reference Data System (CRDS) } after inserting a mask for large cosmic-ray snowball events \citep{RigbyJWST} calculated with \texttt{snowblind} \citep{snowblind} before the ramp-fit step.  We compute a correction for the $1/f$ striping in the count-rate (\texttt{rate}) exposure products.  We compute a pedestal offset of the science extension and a multiplicative scaling of the read noise extension from un-illuminated portions of the detector arrays and run the modified products through the \href{https://jwst-pipeline.readthedocs.io/en/latest/jwst/pipeline/calwebb_spec2.html}{Spec2Pipeline} steps of the standard pipeline up to the photometric calibration.

With flat-fielded, flux-calibrated, 2D spectra of each source on a mask saved to individual files, it is here that further \texttt{msaexp} processing deviates from the standard \texttt{jwst} pipeline. We begin by applying updated corrections for the vignetting of the MSA bars to each 2D spectrum derived as described in Appendix~\ref{sec:barshadow}. The sky background of each source spectrum can be effectively removed by taking straight differences of the 2D spectra obtained at the three spacecraft nod offset positions, i.e., $S^\prime_{A} = S_A - (S_B + S_C)/2$ and $V^\prime_A = S_A + (S_B + S_C)/2$ for the 2D $S$ science and $V$ variance arrays. We have also implemented a global sky subtraction approach for the PRISM spectra that is described in Appendix~\ref{sec:skybackground}. The global sky subtraction is recommended especially for bright extended sources. We find that the nodded background subtraction still performs better for compact sources with low $S/N$ due to occasional small residual artefacts in the global sky subtraction strategy. In the remainder of this paper, we specify explicitly which background subtraction is used.

The three offset exposures are combined in a rectified pixel grid with perpendicular cross-dispersion and wavelength axes using a 2D histogram that is analogous to the ``drizzle'' algorithm \citep{Drizzle} in the cross-dispersion axis but where the pixel independence is preserved and correlated noise is eliminated along the wavelength axis.  The wavelength grids are fixed for all PRISM and G395M spectra with sampling close to the that of the native detector pixels.

\begin{figure*}[!t]
    \centering
    \includegraphics[width=\linewidth]{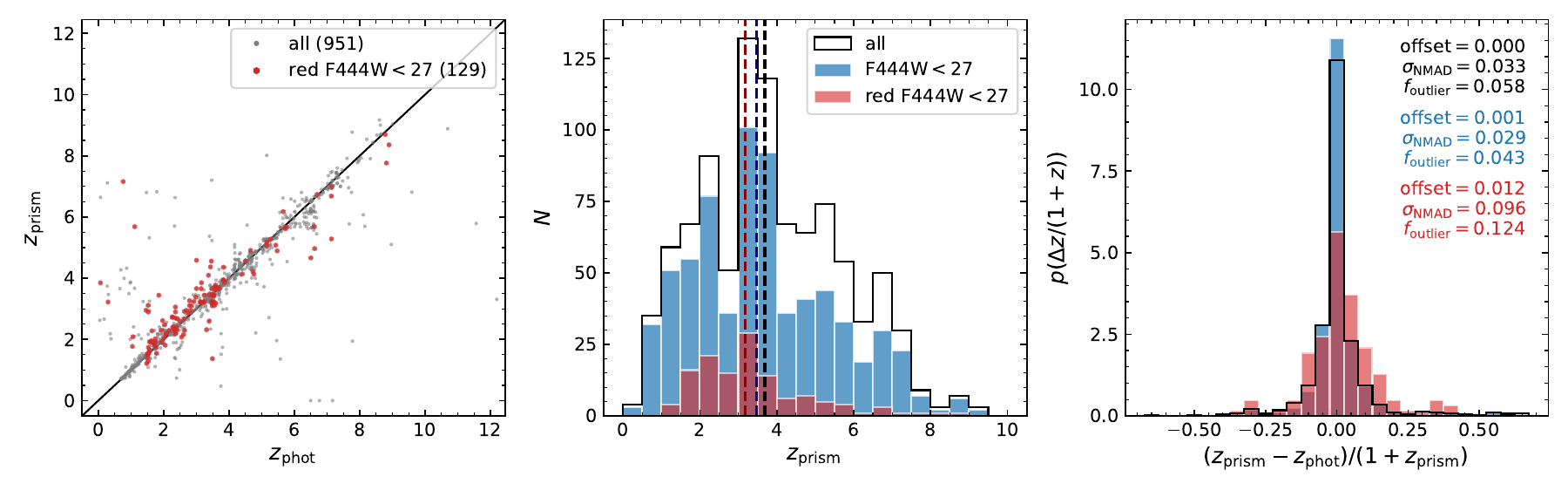}
    \caption{Robust spectroscopic redshifts from the first half of RUBIES PRISM observations (obtained between January-March 2024). Left: comparison between the best-fit photometric redshifts used for target selection vs. {the best-fit spectroscopic redshift ($\zspec$)}. Middle: spectroscopic redshift distribution for all targets and the subset of red targets, defined as $\rm F150W-F444W>2$; dashed lines show the median redshifts. Right: Differences between the photometric and spectroscopic redshifts. Overall there is good agreement between the photometric and spectroscopic redshifts. However, for red sources the photometric redshift scatter is a factor 3 higher than for the census sample, with an outlier fraction that is factor 3 higher than for similarly bright sources that are less red. This illustrates the need for spectroscopy, in particular for red sources.}
    \label{fig:zphot_zspec}    
\end{figure*}

The source location along the slitlet must be known by any strategy used to extract a 1D spectrum from the rectified 2D combination.  This location is provided by the \texttt{jwst} pipeline \href{https://jwst-pipeline.readthedocs.io/en/latest/jwst/assign_wcs/index.html#assign-wcs-step}{AssignWcsStep} using the spacecraft pointing telemetry and the catalogue positions used to generate the MSA mask plan.  While the combined precision of the spacecraft pointing after the MSA target acquisition and the catalogue astrometry is clearly sufficient such that sources fall within the planned opened shutters, catalogue errors of just 20~mas (1/5 pixel) in the astrometry of individual sources would result in easily detectable offsets along the slitlet relative to the nominal position. We fit a cross-dispersion profile for each source in the frame of the curved 2D traces in the detector cutouts with parameters for a spatial offset and a scalar Gaussian width that is added in quadrature to a Gaussian approximation to the wavelength-dependent PSF.  This 2D profile model is rectified and combined in the same way as the science data and used for a final optimally-weighted \citep{Horne86} 1D extraction.  Finally, we derive an effective extended-source path-loss correction for light outside of the slitlet for each source using the a priori position within the shutter and assuming an azimuthally-symmetric Gaussian profile with the fitted width.

\subsection{Spectroscopic redshifts}\label{sec:spec_zs}

We used the {least squares template fitting method} implemented in \texttt{msaexp} to estimate spectroscopic redshifts for the reduced PRISM spectra (with global background subtraction) with the same template set that was used for the photometric redshifts.  This omits the sources that were only observed with the G395M disperser, which we defer to a future data release paper. All PRISM spectra and template fits were visually inspected to (i) assess the redshift fit and (ii) check for major data quality issues. In this visual inspection process the best-fit redshift from \texttt{msaexp} can be manually updated to a different redshift by the inspector, although for the majority of sources the best-fit redshift matches the inspected redshift.

The spectra were graded as follows:
\begin{itemize}
    \item grade 0: major data quality issue
    \item grade 1: no features apparent in the spectrum
    \item grade 2: ambiguous redshift (e.g. a single line detection)
    \item grade 3: robust redshift
\end{itemize}

For the data obtained in the first half of the survey (January-March 2024) there are 951 sources with grade 3, which corresponds to an overall redshift success rate of 65\%. When also including grade 2 sources (77) this increases to 70\%. For the highest priority (1 and 2) sources we achieve an even higher success rate: we obtain robust (grade 3) spectroscopic redshifts for 90\% of red sources, and 82\% of bright high-redshift sources. Sources that we could not establish a redshift for are typically faint, or at very low redshift ($z<0.5$) where there are few discernible features in the near-infrared. Redshifts for these sources may be recovered by including photometric information, which we plan to incorporate in future. Data quality issues (grade 0) affect only a small fraction of targets ($<1\%$).

We compare the best-fit photometric redshifts and PRISM redshifts (hereafter $\zspec$) in Figure~\ref{fig:zphot_zspec}. Overall we find good agreement between the photometric and spectroscopic redshifts: the scatter, computed as the normalised median absolute deviation, is small with $\sigma(\Delta z/(1+z)) = 0.033$, and the outlier fraction is low $f_{\rm outlier}=0.06$, defined as the fraction of sources for which $\Delta z/(1+z)>0.15$. However, we find that the same is not true for the reddest sources (i.e. the red Priority 1 and 2 Rubies), as the scatter is a factor 3 larger, with a larger outlier fraction of 0.12. We find that for 27\% (12\%) of these red targets $|\zphot-\zspec|>0.5$ ($>1.0$), with some extreme discrepancies where $\zphot\sim 1$ but $\zspec>5$. These sources typically have extremely red, smoothly rising broad-band SEDs; the template fitting here fails with photometry alone due to a lack of strong features or ill-fitting templates, but the spectra show (in some cases strong) emission lines.

We also show the redshift distribution of the full sample of sources with robust redshifts, and the subsets of moderately bright ($\rm F444W<27$) sources and red sources. We find a median redshift of $\zspec=3.7$, with the highest redshift being at $\zspec=9.3$. The reddest sources, targeted without any selection on photometric redshift, tend to have slightly lower redshifts with a median of $\zspec\approx 3.2$, although the redshift distribution has a long tail extending to $\zspec=8.7$.

\subsection{Wavelength and flux calibration}\label{sec:calibration}

\begin{figure*}
    \centering
    \includegraphics[width=0.8\linewidth]{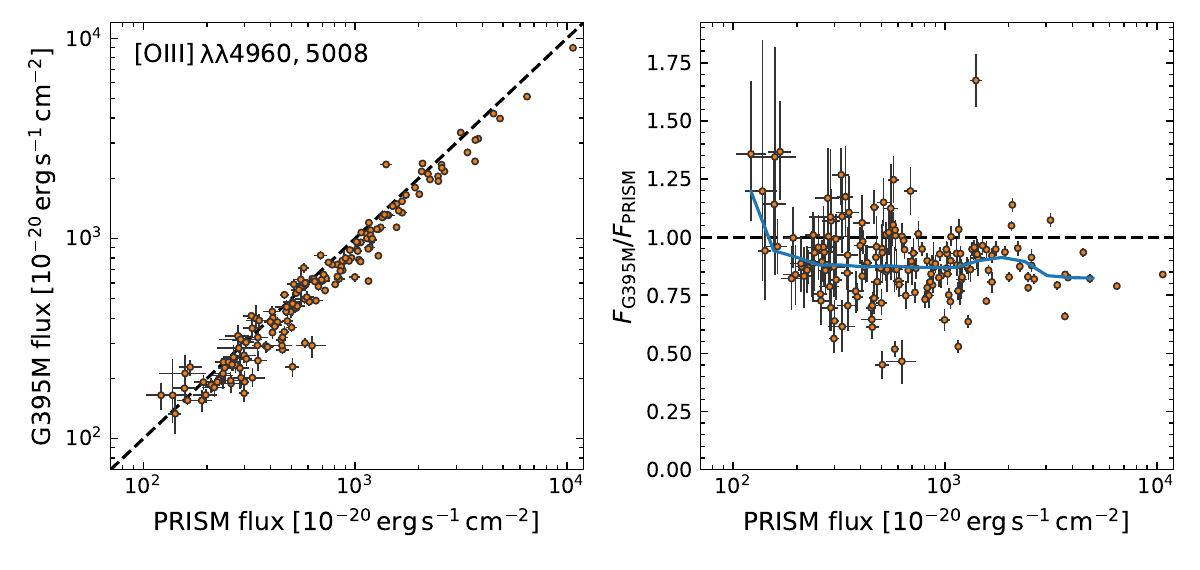}
    \caption{Comparison between the fluxes measured from the PRISM and G395M spectra, using the \Oiii$\lambda\lambda 4960,5008$ emission line doublet. The blue solid line shows the running median. We find a systematic offset between the two gratings, despite the fact that the slit losses and spectral extraction method used are identical. The PRISM fluxes are brighter by approximately $10-15$\%, and this offset does not depend significantly on the line flux itself, and likely points to a calibration issue.  }
    \label{fig:flux_check}
\end{figure*}

\begin{figure*}
    \centering
    \includegraphics[width=0.95\linewidth]{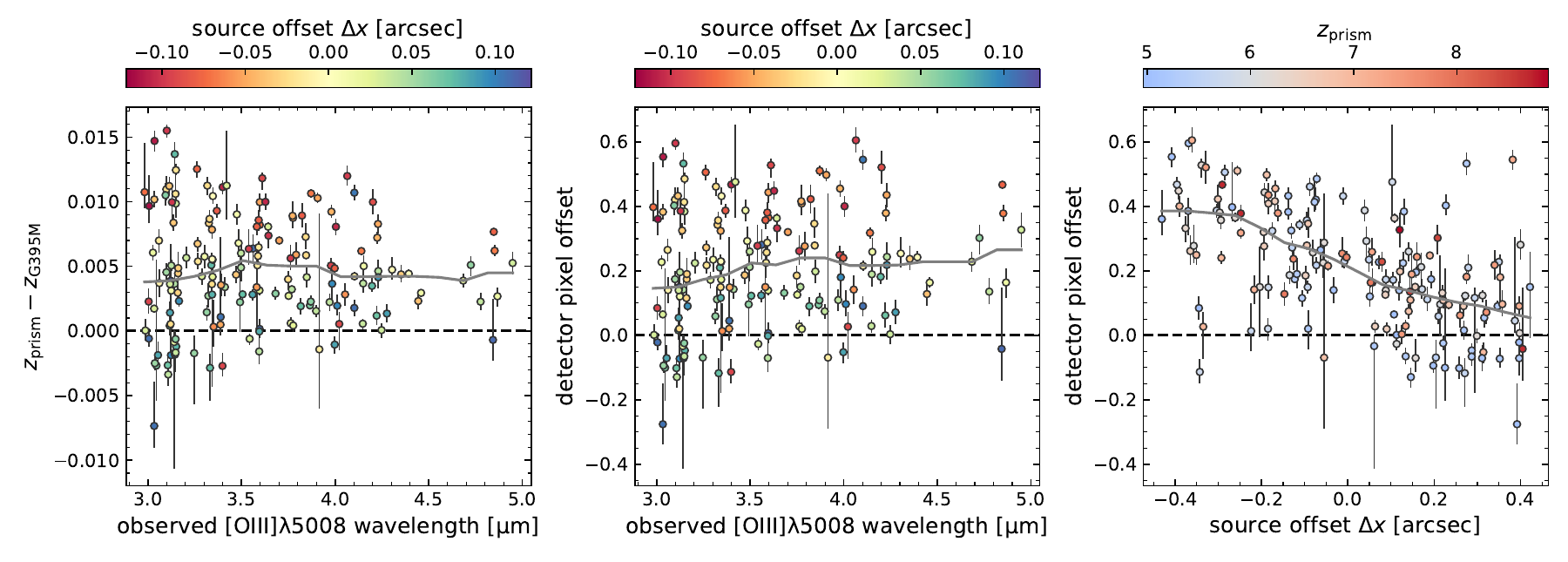}
    \caption{Redshift and wavelength offset between the observed \Oiii$\lambda 5008$ emission lines measured from the PRISM and G395M spectra. Taking the G395M spectrum as `truth', we find a systematic offset of $\Delta z\sim 0.0044$ or $\sim0.25$ detector pixel for the PRISM spectrum, which does not appear to depend significantly on wavelength (grey solid lines show the running median). The scatter can be partially explained by the larger uncertainty for fainter emission lines. In addition, the intrashutter position of the source (i.e. the spatial offset in the dispersion direction) also introduces wavelength offsets of up to 1 pixel, if the source is point-like and located at the edge of the shutter. In practice, high-redshift sources are (moderately) spatially extended, resulting in smaller offsets. We indeed find a correlation between the source position in the slit and the wavelength offset.}
    \label{fig:wave_offset}
\end{figure*}

We use the PRISM and G395M spectra to test the flux and wavelength calibration of the spectra. A similar exercise was performed by \citet{Bunker2023} and \citet{JADES_DR3} for JADES, where substantial offsets were found between fluxes and wavelengths measured from the same emission lines in different dispersers. However, the spectra in these papers were reduced using the NIRSpec GTO pipeline, which differs in significant ways from \texttt{msaexp}. 

Starting from the robust redshifts measured in the previous section (\ref{sec:spec_zs}), we select sources for which the \Hb line and \Oiii doublet fall in the wavelength range of the G395M disperser, $\zspec > 4.96$. Next, we fit the three emission lines using a custom emission line fitting software that accounts for both the broadening of emission lines by the line spread function (LSF) and the undersampling of the LSF by the NIRSpec detectors. {The latter is critical, as the NIRSpec LSF for a point source has a width of only $\sim 1-1.5$ pixel \citep{deGraaff2024a}, and fitting, e.g., a Gaussian profile to such an undersampled line could severely over- or underestimate the line flux: because the flux density profile is highly non-linear across the pixel, the integrated flux of the profile across the pixel differs from the flux obtained by simply evaluating the profile at the mid-point of the pixel. In order to robustly fit a Gaussian line profile to the undersampled NIRSpec data, we therefore first construct the Gaussian emission line model on a fine wavelength grid (a factor 5 higher than the NIRSpec wavelength sampling) and subsequently integrate the model with a Riemann sum. }

We assume a single Gaussian line profile for the \Hb and \Oiii doublet, and assume the same kinematics for both line species by fitting for a single velocity dispersion parameter $\sigma_{\rm gas}$. For a small number of sources with strong outflows this may be inaccurate, but on average we find that these assumptions do not lead to significant residuals. We do not fix the flux ratio of the \Oiii doublet in order to verify that our measurements retrieve the expected theoretical ratio. After constructing the model, we convolve the emission lines with the LSF of an idealised point source \citep[see Appendix A of][]{deGraaff2024a}. This LSF is likely too narrow for many of the spatially-extended sources in the sample, and we therefore do not consider the velocity dispersion measurements themselves to be physically meaningful. We note that we do not use the LSF curves provided in the JWST User Documentation (JDox), as we find these to be too broad for many of our sources and would therefore yield incorrect fluxes.  

We fit the wavelength range around the \Hb and \Oiii emission line complex ($\pm 0.35\,\micron$), and approximate the continuum using a 1st order polynomial. Because we find that the uncertainties are typically underestimated by the data reduction pipeline, we use the continuum flux around the emission lines to compare the scatter of the continuum to the median value of the error spectrum in the same wavelength range, and subsequently rescale the error spectrum by the ratio of the two. The fitting itself is performed using the Markov Chain Monte Carlo (MCMC) sampling method implemented in the \texttt{emcee} package \citep{emcee}. 

These fits were run for both the PRISM and G395M spectra (using the nodded/local background subtraction for both dispersers) and yield realistic error bars for the measured fluxes. We select sources with $S/N >3$ for the \Oiii$\lambda 5008$ line measured from the G395M spectrum and $F_{[\rm O\,{III}]\,\lambda 5008}>1\times10^{-18}\,\rm erg\,s^{-1}\,cm^{-2}$, and remove a few (5) bad fits, leaving a sample of 186 objects. We compare the combined \Oiii$\lambda\lambda 4960,5008$ flux in Figure~\ref{fig:flux_check} and find a good correlation, but with a systematic offset of approximately 10\%, as the PRISM fluxes are higher.

Because the spectra were taken consecutively and at the exact same location in the sky, the slit losses are the same for both dispersers. We have also used the same background subtraction for the reduction of both sets of spectra, and used identical extraction profiles. The offset therefore likely reflects a systematic calibration issue, the source of which is unclear. \citet{JADES_DR3} report a similar offset of $\sim 10-15\%$ between the PRISM and G395M dispersers, albeit in the \textit{opposite} direction. We note that if we use the same (older) calibration files as used for the JADES data release (corresponding to version 2 of the DJA, which used \texttt{jwst\_1180.pmap} from the CRDS), we obtain the same result as \citet{JADES_DR3}. We therefore conclude that the absolute flux calibration of NIRSpec MSA spectroscopy remains uncertain at the $10-20\%$ level.

Next, we test the relative wavelength calibration. We compute the redshift difference between the PRISM and G395M fits and convert this to a wavelength offset: we assume the G395M redshift is the `true' value, and calculate the wavelength offset of the \Oiii$\lambda 5008$ emission line in the PRISM spectrum. We convert this wavelength offset to the offset in detector pixels, using the dispersion curves provided on the JDox\footnote{\url{https://jwst-docs.stsci.edu/jwst-near-infrared-spectrograph/nirspec-instrumentation/nirspec-dispersers-and-filters}}. This neglects the fact that the PRISM traces vary slightly in length across the detector, although this is a secondary effect.

Figure~\ref{fig:wave_offset} shows there is a systematic offset in the wavelength solution between the PRISM and G395M spectra that does not depend significantly on wavelength (or redshift) itself. The offset is approximately 0.25 pixel, which for the low-resolution PRISM quickly translates to a large redshift and velocity offset (ranging from $\sim 100-1100\,\kms$; or a redshift offset of $\Delta z\sim0.0044$). Our finding is in good agreement with the results reported by \citet{Bunker2023} and \citet{JADES_DR3}, and points to either a calibration issue or an error in both reduction pipelines.

The scatter seen in Figure~\ref{fig:wave_offset} in part can be explained by the measurement uncertainty for spectra with lower S/N as well as the fact that we have not accounted for variation in the length of the PRISM traces in converting the wavelength offsets to pixels. However, the scatter is also partially a physical effect. The reduction pipeline assumes that a source is in the centre of the shutter or that the slit is illuminated uniformly to compute the wavelength solution. However, extragalactic sources observed with the NIRSpec MSA rarely satisfy either of these conditions. The spatial offset in the dispersion direction of the centroid of the source with respect the shutter centre therefore translates into a wavelength offset, the magnitude of which depends on the spectral resolution of the disperser. The width of the slit is approximately two detector pixels: a maximum offset of $\pm 1$ pixel can therefore be expected if a source is point-like and located on the edge of the shutter. In practice, sources are typically (moderately) spatially extended, resulting in wavelength offsets that are difficult to estimate and correct for. The right-hand panel of Figure~\ref{fig:wave_offset} indeed shows a correlation between the centroid offset in the dispersion direction and the wavelength offset.

\section{Science objectives}\label{sec:science}

With {4444} targeted sources spread over a wide area of $\sim150$ arcmin$^2$, RUBIES is among the largest spectroscopic programmes performed with JWST/NIRSpec thus far. The combination of low- and medium-resolution spectroscopy allows for a detailed characterisation of the stellar population properties, properties of the interstellar medium, dust and active black holes. The broad range in colour, magnitude and redshift spanned by the RUBIES targets opens up a wealth of opportunities to investigate the growth of galaxies and black holes in the early Universe. 

\subsection{Nature of the reddest and brightest high-redshift sources}

RUBIES provides the first statistical samples of rare red and bright objects: in total the survey targets approximately 300 sources redder than $\rm F150W-F444W>2$\,, 120 of which are even more extreme with $\rm F150W-F444W>3$. Similarly, of approximately 200 high-redshift ($\zphot\gtrsim7$) candidates, we observe 12 (64) sources brighter than $\rm F444W<25$ ($\rm F444W<26$). As discussed in Section~\ref{sec:spec_zs}, we obtain high-quality spectra and robust redshifts for nearly 90\% of these high-priority targets. Collecting such a large sample is critical: we find that the population of red and bright sources is highly heterogeneous. 

The red and bright sources span a wide range in redshift ($\zspec\sim1-9$), and have diverse spectral properties. Broadly, we can identify four groups of objects, although we also find great diversity within each group. 
Figure~\ref{fig:red_spectra} demonstrates these different types, showing colour images (created from the F150W, F277W and F444W NIRCam images), the full low-resolution PRISM spectrum, and a selected wavelength range of the medium-resolution G395M spectrum. From top to bottom, we can distinguish:

\begin{itemize}
    \item {\bf Dust-obscured star-forming galaxies.} The RUBIES colour selection yields a large sample of objects with bright continuum emission that continues rising from the rest-frame UV to the rest-frame near-infrared, consistent with strong attenuation by dust. Previous work based on photometry alone has shown that these red sources  were not detected by HST, but likely contribute significantly to the stellar mass and star formation rate density of the high-redshift Universe \citep[e.g.][]{Nelson2023,Barrufet2023}. 
    
    We detect emission lines in many of these sources, allowing for a precise redshift determination that is difficult to obtain from broad-band photometry alone. We typically find strong \Ha and Paschen line emission indicative of high star formation rates \citep[typically at $z\sim2-4$; similar to the findings of][]{Barrufet2024}, and a suite of forbidden emission lines in a subset of sources. In conjunction with existing far-infrared and sub-mm constraints from Herschel and ALMA, these measurements place unique constraints on the ISM and dust properties of massive star-forming galaxies at cosmic noon (a first exploration of which is presented in {\citealt{Cooper2024}}). Moreover, by leveraging the full continuum SED, we are able to constrain the stellar population properties and trace the stellar mass growth of the most massive galaxies at $z>2$ (Gottumukkala et al. in prep.). 

\begin{figure*}[!h]
    \centering
    \includegraphics[width=\linewidth]{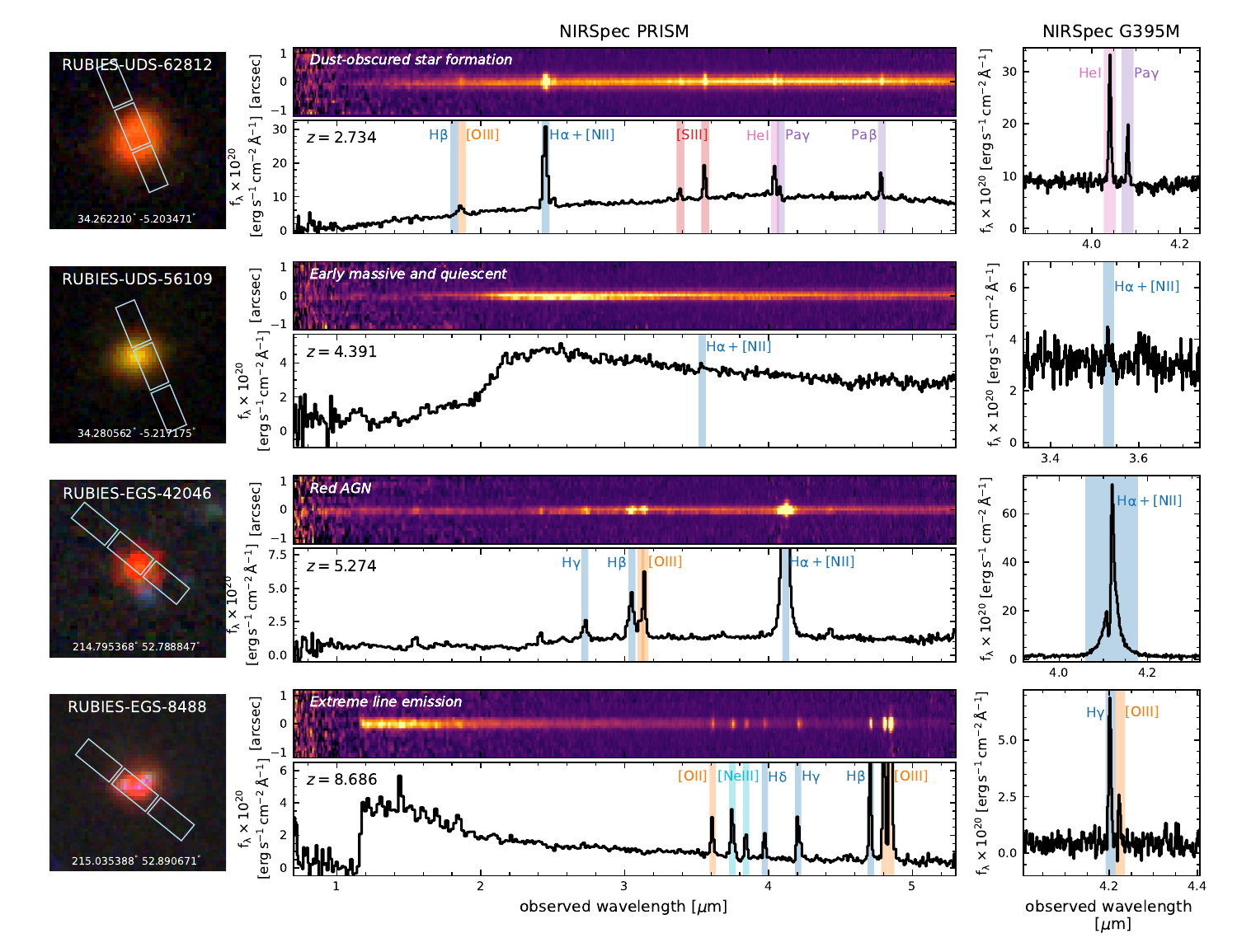}
    \caption{Example spectra of high-priority `Rubies'. False colour images are constructed from the NIRCam F150W, F277W and F444W filters, and show the location of the NIRSpec microshutters. The low-resolution PRISM spectra (with global background subtraction, see Section~\ref{sec:reduction}) reveal a great diversity in spectral shapes, both in terms of continuum features (Balmer breaks and jumps) and the presence of different emission lines. The medium-resolution G395M data deblend lines that are difficult to interpret from the PRISM spectroscopy alone, and provide strong constraints on the ionised gas kinematics. We broadly identify four types of SEDs, from top to bottom: galaxies with strongly dust-obscured star formation, massive quiescent galaxies, extremely red AGN, and extreme emission line galaxies. Note that medium-resolution NIRSpec data of RUBIES-EGS-8488 was also presented by \citet{Larson2023} with ID CEERS\_1019. }
    \label{fig:red_spectra}
\end{figure*}

    \item {\bf Massive quiescent galaxies at $z>4$.} The red colour selection also picks up bright sources with remarkably strong Balmer breaks at $z>4$, which -- unlike the dusty star-forming systems -- are not red at rest-frame optical wavelengths and show no or weak line emission. This implies that these galaxies formed a large amount of stellar mass ($M_*>10^{10}\,\Msun$) within only $\sim 1$\,Gyr and subsequently ceased forming stars. Although massive quiescent galaxies are common at $z<1$ \citep[e.g.][]{Muzzin2013}, at $z>4$ the star formation activity in the vast majority of galaxies is very high \citep[e.g.][]{Speagle2014}, and the finding of such massive quiescent galaxies therefore challenges current models of galaxy formation \citep{Valentino2023}. 
    
    RUBIES has discovered two of the most extreme such systems to date: an extremely massive ($M_*\approx10^{11}\,\Msun$) quiescent galaxy at $z=4.9$ that formed and quenched in the epoch of reionisation \citep{deGraaff2024c}; and a massive ($M_*\approx10^{10.2}\,\Msun$) quiescent galaxy at $z=7.3$ that is a likely progenitor of the massive quiescent galaxies seen at $z<4$ \citep{Weibel2024b}. The analysis of the full population of high-redshift massive quiescent galaxies in RUBIES will set a benchmark for future galaxy formation models and their uncertain models for physical processes such as AGN feedback.

    \item {\bf Red AGN, ultra-massive galaxies and `little red dots'.} Among the most debated sources found with JWST are the extremely compact red objects dubbed little red dots (LRDs). This term was originally coined for sources that show symmetric broad Balmer lines \citep{Matthee2024}, which, combined with narrow forbidden lines, suggest that the broadening arises in gravitational motions around a supermassive BH, rather than outflows or supernovae \citep[e.g.][]{Furtak2023,Greene2024,Killi2023,Kocevski2023}. However, since then many other explanations have been proposed, as the SEDs of photometrically-selected objects with similar colours and morphologies may also be consistent with compact star-forming galaxies \citep[e.g.][]{PerezGonzalez2023,Williams2024}. For some of the highest redshift sources ($z>7$), the latter interpretation implies extremely high stellar masses in tension with the standard cosmological model \citep{Labbe2023}. 

\begin{figure*}
    \centering
    \begin{subfigure}{0.4\linewidth}
        \includegraphics[width=\linewidth]{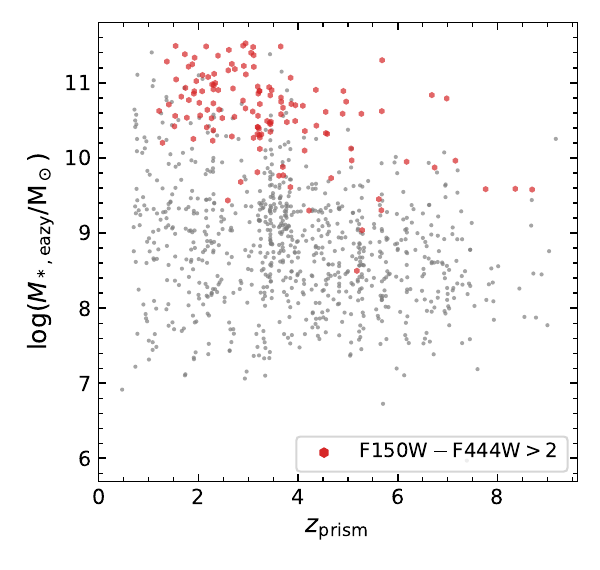}
    \end{subfigure}
    \begin{subfigure}{0.4\linewidth}
        \includegraphics[width=\linewidth]{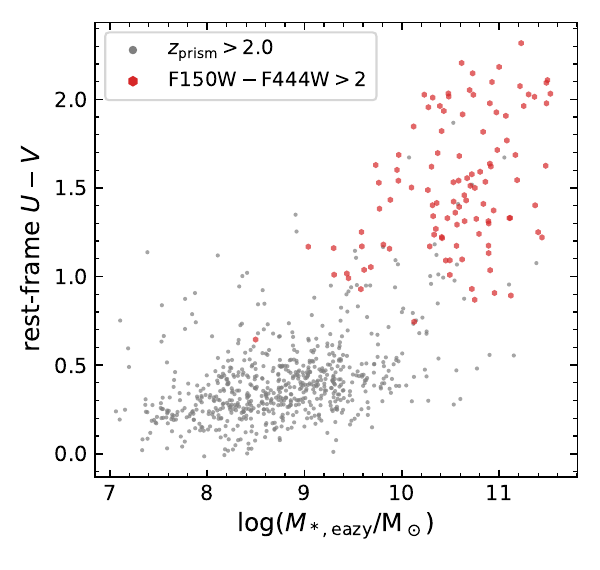}
    \end{subfigure}
    \caption{Stellar mass vs. spectroscopic redshift and rest-frame $U-V$ colour, estimated from \texttt{eazy} fitting to PSF-matched photometry of \citet{Weibel2024} for targets with robust spectroscopic redshifts. Sources that were selected with high priority based on their $\rm F150W-F444W$ colour are marked in red. The RUBIES sources span a wide range in redshift, stellar mass and rest-frame colour, with the census sample being predominantly blue and of lower stellar mass. The red sources tend to be massive ($M_*\gtrsim10^{10}\,\Msun$), although there is large diversity in the rest-frame $U-V$ colour at the highest stellar masses. We identify a redshift clustering of approximately 15 massive ($M_*>10^{10}\,\Msun$) red galaxies at $\zspec\approx3.2$, which also corresponds to a spatial clustering in the UDS field and coincides with the early massive quiescent galaxy of \citet{Glazebrook2024}. }
    \label{fig:sample_clr_mass}
\end{figure*}

    The very first {`Ruby' (Priority 1 target)} observed, RUBIES-BLAGN-1 at $z=3.1$ \citep{Wang2024a}, is one of the brightest LRDs discovered thus far, and one of very few LRDs with a MIRI detection in the rest-frame mid-infrared. This single source already demonstrated the complex nature of these systems: broad Balmer ($\rm FWHM\sim 4000\,\kms$) lines are consistent with an actively accreting black hole, but the relatively faint rest-frame mid-infrared emission strongly disfavours the presence of a hot dusty torus that would typically be expected from AGN. RUBIES follow-up of the sources suggested to be in tension with $\Lambda$CDM \citep{Labbe2023} revealed a similarly complex puzzle: broad Balmer lines and narrow \Oiii lines suggest the presence of an AGN; remarkably, however, the SEDs also show Balmer breaks consistent with evolved stellar populations and possibly very high stellar masses at $z\sim7-8$ \citep{Wang2024b}.

    RUBIES has observed many ($\sim 30-50$) sources that -- depending on the definition used -- may be considered to be a little red dot, constituting the largest sample of LRDs with spectroscopic follow-up from JWST/NIRSpec to date. These sources, typically at $\zspec\sim4-8$, show red continua at rest-frame optical wavelengths and broad Balmer lines. We emphasise that the vast majority of this sample is extragalactic: only 4 sources observed thus far turn out to be cool stars. The RUBIES sample provides a unique opportunity to investigate the nature of these LRDs and quantify the fraction of AGN among the population.

    \item {\bf Extreme emission line galaxies. } Some of the brightest (measured in the F444W filter) sources at $z>6$ are dominated by strong, narrow emission lines. Although these objects still show significant rest-frame UV emission, the F444W broad band is boosted by the strong rest-frame optical emission lines, such that the overall colour is red. As opposed to the Balmer breaks found in the other red sources at similarly high redshifts, for some of these objects we detect Balmer jumps indicative of strong hydrogen free-bound nebular continuum emission. 

    These sources have been found to have low metallicities, and the shapes of the SEDs of a subset of the Balmer jump galaxies possibly provide evidence for a top-heavy initial mass function and hot massive stars in the first Gyr \citep{Cameron2024,Katz2024}. In contrast, other studies suggest that these spectra may instead be consistent with the presence of AGN \citep[e.g.][]{Larson2023, Tacchella2024,Li2024}, with some reporting the detection of faint high-ionisation lines that are expected only from AGN \citep{Brinchmann2023,Chisholm2024}. 
    The combination of PRISM spectroscopy, revealing the continuum emission, and G395M spectroscopy, which resolves emission lines such as H$\gamma$ and \Oiii$\lambda4363$ that are blended in the PRISM spectra, in RUBIES provides unique constraints on the ISM conditions of the brightest sources at $z\sim7-9$.

\end{itemize}

\subsection{Census of the $2<z<7$ galaxy population}

\begin{figure*}
    \centering
    \includegraphics[width=0.8\linewidth]{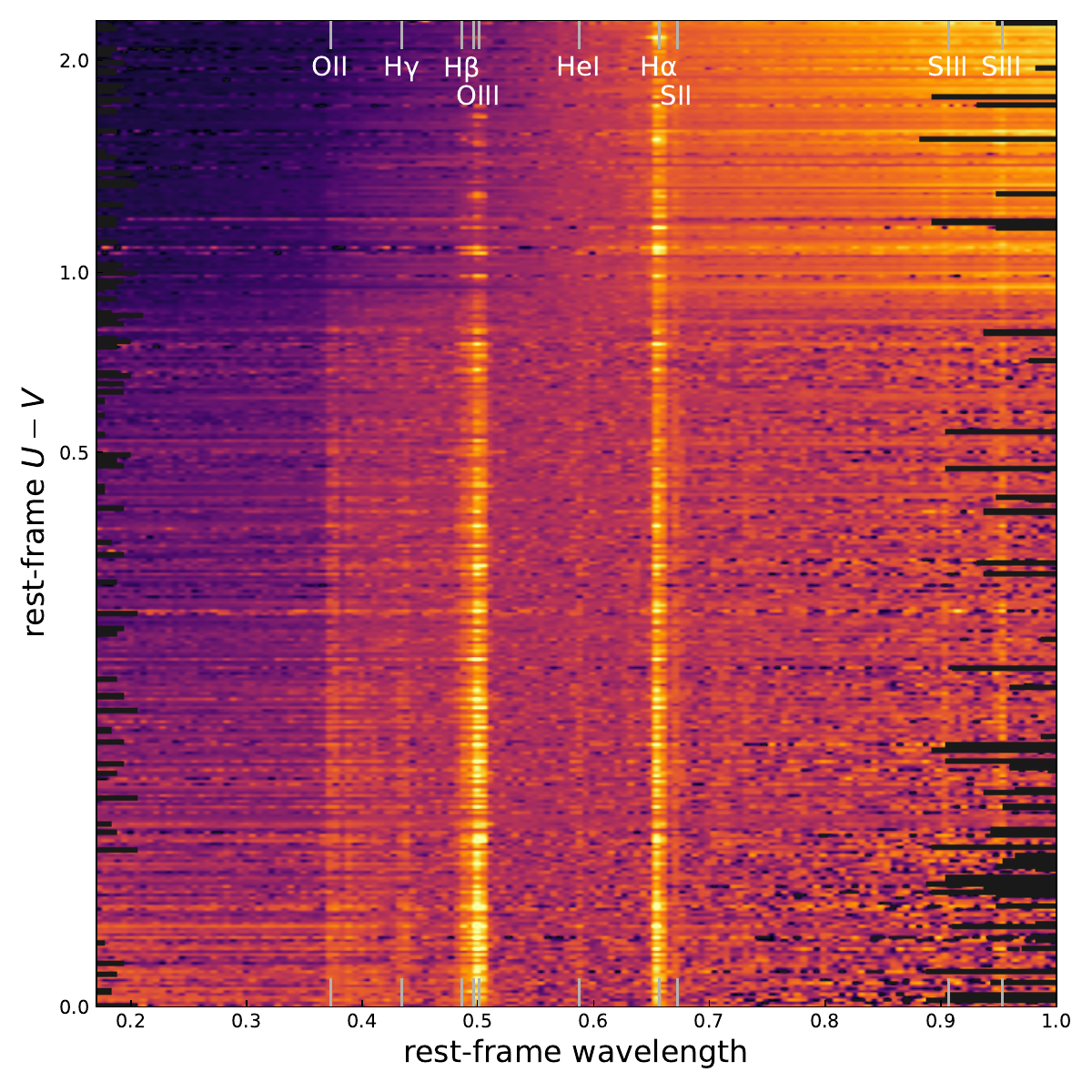}
        \caption{RUBIES spectra at $2<\zspec<5$ and with median continuum $S/N>1$, sorted by rest-frame $U-V$ colour (spectra normalised by the median flux between rest-frame $0.51-0.64\,\micron$). The bluest sources show the strongest emission lines, while the Balmer break becomes more prominent for redder sources. For the very reddest sources (near the top) there is a diversity in spectral shapes and visible emission lines, with a mixture of dusty galaxies, quiescent galaxies and red AGN. } 
    \label{fig:stack_2d}
\end{figure*}

The RUBIES census sample comprises $\sim 4000$ targets, of which $\sim 1000$ are `continuum-bright', defined as PRISM spectra with a median continuum $S/N\gtrsim 3 \rm\,pix^{-1}$. For the remainder we primarily detect (strong) emission lines; these sources typically are fainter than $\rm F444W>26-26.5$ depending on morphology. The census sample was optimised to place the rare Rubies (Section~\ref{sec:targets}) in the context of the broader high-redshift population. However, this broad dataset also offers many ancillary science opportunities.

\begin{itemize}
    \item {\bf Rare sources in context: number densities and properties.} With a well-quantified selection function, RUBIES is uniquely positioned to place strong constraints on the number densities of otherwise rare objects. These measurements are critical to assess whether sources are in tension with cosmological galaxy formation simulations: for example, such simulations make clear predictions for the fraction of quiescent galaxies as a function of stellar mass and redshift. Moreover, having a well-characterised census sample helps to understand which property makes a source `rare', for instance, by studying the occurrence of broad-line AGN or uncommon emission line features as a function of different galaxy stellar population properties.
    
    \item {\bf Star formation histories at $z>2$\,.} For the continuum-bright sample the PRISM spectra encode critical information on the stellar population properties: the spectra lift the degeneracy between spectral breaks and strong emission lines that affect photometric measurements. In addition, for the brightest sources we even detect stellar absorption features at the PRISM resolution \citep[e.g.][]{deGraaff2024c}. Full spectrum fitting will therefore provide crucial insight into the star formation histories of a diverse population of galaxies at $2<z<7$. Moreover, by accounting for the RUBIES selection function, we will also be able to spectroscopically constrain the stellar mass function, which so far has remained restricted to photometric samples at these high redshifts \citep[e.g.][]{Weaver2023,Harvey2024,Weibel2024,Wang2024_massfunction}. 
    
    \item {\bf Star formation and ISM properties across cosmic time.} For the majority of RUBIES targets we observe multiple strong emission lines, both in the PRISM spectra (e.g. \Ha, \Hb, and \Oiii) and in the G395M spectra (which resolve the \Ha and \Nii complex at $z>3.4$). These measurements provide direct constraints on the the star formation, dust and ionising conditions in high-redshift galaxies. 
    
    With the diverse population of galaxies probed in RUBIES, we will be able to extend previous studies of the dust attenuation, metallicity and ionisation conditions of the ISM \citep[e.g.][]{Shapley2023,Sanders2023,Backhaus2024} to larger samples, and, critically, to unexplored regions of parameter space. The medium-resolution data additionally reveal the ionised gas kinematics and will allow for a systematic exploration of outflows from star formation and/or AGN \citep[e.g.][]{Xu2023,Carniani2024}. Furthermore, with the well-quantified selection function of RUBIES we will be able to measure key scaling relations that have so far been limited to relatively small spectroscopic samples or larger photometric samples, such as the star formation main sequence \citep[e.g.][]{Clarke2024,Rinaldi2024}, or the stellar mass-metallicity relation (e.g. \citealt{Nakajima2023,Curti2024}; Lewis et al. in prep.).
    
    \item {\bf Large-scale environment.} The large sample of spectroscopic redshifts enables the investigation of the large scale structure at high redshifts. We can already identify structure in the spectroscopic redshift distribution in Figures~\ref{fig:zphot_zspec} and \ref{fig:sample_clr_mass}, and find that for some peaks this also corresponds to spatial clustering, indicative of an overdensity. Notably, we find an intriguing clustering of sources at $\zspec\approx3.2$ in the UDS. This apparent overdensity contains approximately 15 massive red sources, and corresponds to the same redshift and spatial vicinity of the extremely early, massive quiescent galaxy of \citet{Glazebrook2024}. We explore the properties of this massive overdensity in further detail in McConachie et al. (in prep.). 
\end{itemize}

We provide a first look into the rest-frame properties of this census sample in Figure~\ref{fig:sample_clr_mass}, which shows targets with robust spectroscopic redshifts. The majority of these sources are at $2<z<7$, and span a wide range in stellar mass ($M_*\sim10^{7-11.5}\,\Msun$) and rest-frame $U-V$ colour ($\sim 2.5$ mag). Here, rest-frame properties were estimated by {cross-matching our targets with the PSF-matched photometry from the W24 catalogues and} re-running \texttt{eazy} with redshifts fixed to the (robust) spectroscopic redshifts. 

Although these stellar mass estimates are approximate, particularly for sources that are ill-described by the templates, and do not leverage stellar populations information encoded in the PRISM spectra, two major conclusions can already be drawn. First, the red ($\rm F150W-F444W>2$; red markers) sources are significantly more massive than typical census galaxies (grey points), with $M_*\gtrsim10^{10}\,\Msun$ irrespective of redshift. Second, although red sources are massive, there is no strong trend between the two quantities beyond $M_*\sim10^{9.5}\,\Msun$. Massive galaxies in RUBIES appear to be a diverse population, with rest-frame colours ranging from $U-V\approx0.5$ to $U-V\approx2.5$, and extend the findings of, e.g., \citet{vanDokkum2011} at $z\sim 1$ to higher redshifts. 

Finally, we zoom in on the spectral properties of sources at $2<z<5$ with continuum $S/N>1$ in Figure~\ref{fig:stack_2d}. The spectra are sorted by rest-frame $U-V$ colour, and normalised by the median continuum flux in between the \Oiii and \Ha lines. The data are extremely rich: for the bluest sources we find a wealth of emission lines, whereas the Balmer break becomes more prominent for redder objects. There is a diversity in spectral shapes, with some having a very red continuum (in $f_\nu$) at rest-frame optical wavelengths, while other SEDs flatten beyond the Balmer break. We emphasise that, although the rest-frame $U-V$ colour is a convenient measure for an initial exploration of the galaxy population, it alone cannot separate these different SED types: there is strong scatter in both the SED shapes and emission line ratios even for spectra with identical $U-V$ colours. The $z>2$ RUBIES sources clearly form a multi-dimensional population that remains to be explored in further detail.

\section{Conclusions and data release}\label{sec:summary}

RUBIES is a {60}-hour Cycle 2 programme with JWST/NIRSpec designed to study the red and bright sources that have been newly discovered with JWST/NIRCam in the EGS and UDS fields. With a total of $\sim300$ red objects ($\rm F150W-F444W>2$), $\sim100$ of which are extremely red ($\rm F150W-F444W>3$), RUBIES provides the first statistical samples of rare objects in the early Universe. Crucially, a carefully constructed census sample of $\sim 4000$ sources is able to place these rare sources in the context of the broader galaxy population at $2<z<7$\,: this census sample holds immense legacy value, and is enabled by a custom-developed mask design strategy for NIRSpec that samples the full parent population and relies solely on the measured F444W magnitudes, $\rm F150W-F444W$ colours, and photometric redshifts of sources.

Obtaining such a large spectroscopic sample of red sources is crucial, as this population is highly heterogeneous. We find that the red sources span a wide redshift range, from $1<\zspec<9$, and show diverse spectral properties that do not correlate trivially with redshift, magnitude or colour alone. In comparison to the full galaxy population, the red sources are among the most massive systems at all redshifts and therefore {could possibly} contribute significantly to the stellar mass and star formation rate density in the early Universe. 

A wealth of science questions are still to be explored with the RUBIES dataset. We provide an initial public data release of all RUBIES data obtained between January-March 2024 (i.e., half the survey) through the DAWN JWST Archive\footnote{\url{https://s3.amazonaws.com/msaexp-nirspec/extractions/nirspec_graded_v3.html}}. This release includes reduced PRISM spectra and G395M spectra for all targets, as well as visually-inspected spectroscopic redshifts based on the PRISM spectra. In the future, we will include the remainder of the RUBIES dataset, as well as redshifts measured directly from the G395M spectra. Finally, we note that this release incorporates major recent improvements to the NIRSpec calibration files and data reduction pipeline, although some challenges in the flux and wavelength calibration still remain. Future data releases will incorporate further progress made in the reduction and extraction of the NIRSpec spectra.

\begin{acknowledgements}

We thank the CEERS and PRIMER teams for making their imaging data publicly available immediately. 
This work is based on observations made with the NASA/ESA/CSA James Webb Space Telescope. The data were obtained from the Mikulski Archive for Space Telescopes at the Space Telescope Science Institute, which is operated by the Association of Universities for Research in Astronomy, Inc., under NASA contract NAS 5-03127 for JWST. These observations are associated with programs \#1345, \#1837 \#2234, \#2279, \#2514, \#2750, \#3990 and \#4233.
Support for program \#4233 was provided by NASA through a grant from the Space Telescope Science Institute, which is operated by the Association of Universities for Research in Astronomy, Inc., under NASA contract NAS 5-03127. {REH acknowledges support by the German Aerospace Center (DLR) and the Federal Ministry for Economic Affairs and Energy (BMWi) through program 50OR2403 `RUBIES'.} 
This research was supported by the International Space Science Institute (ISSI) in Bern, through ISSI International Team project \#562.
The Cosmic Dawn Center is funded by the Danish National Research Foundation (DNRF) under grant \#140. This work has received funding from the Swiss State Secretariat for Education, Research and Innovation (SERI) under contract number MB22.00072, as well as from the Swiss National Science Foundation (SNSF) through project grant 200020\_207349. Support for this work for RPN was provided by NASA through the NASA Hubble Fellowship grant HST-HF2-51515.001-A awarded by the Space Telescope Science Institute, which is operated by the Association of Universities for Research in Astronomy, Incorporated, under NASA contract NAS5-26555. 

\end{acknowledgements}

\bibliographystyle{aa}
\bibliography{refs}

\begin{thebibliography}{127}
\expandafter\ifx\csname natexlab\endcsname\relax\def\natexlab#1{#1}\fi

\bibitem[{{Adamo} {et~al.}(2024){Adamo}, {Atek}, {Bagley}, {Ba{\~n}ados},
  {Barrow}, {Berg}, {Bezanson}, {Brada{\v{c}}}, {Brammer}, {Carnall},
  {Chisholm}, {Coe}, {Dayal}, {Eisenstein}, {Eldridge}, {Ferrara}, {Fujimoto},
  {de Graaff}, {Habouzit}, {Hutchison}, {Kartaltepe}, {Kassin}, {Kriek},
  {Labb{\'e}}, {Maiolino}, {Marques-Chaves}, {Maseda}, {Mason}, {Matthee},
  {McQuinn}, {Meynet}, {Naidu}, {Oesch}, {Pentericci},
  {P{\'e}rez-Gonz{\'a}lez}, {Rigby}, {Roberts-Borsani}, {Schaerer}, {Shapley},
  {Stark}, {Stiavelli}, {Strom}, {Vanzella}, {Wang}, {Wilkins}, {Williams},
  {Willott}, {Wylezalek}, \& {Nota}}]{ISSI2024}
{Adamo}, A., {Atek}, H., {Bagley}, M.~B., {et~al.} 2024, arXiv e-prints,
  arXiv:2405.21054

\bibitem[{Adams {et~al.}(2022)Adams, Conselice, Ferreira, Austin, Trussler,
  Juodžbalis, Wilkins, Caruana, Dayal, Verma, \& Vijayan}]{Adams_2022}
Adams, N.~J., Conselice, C.~J., Ferreira, L., {et~al.} 2022, Monthly Notices of
  the Royal Astronomical Society, 518, 4755–4766

\bibitem[{{Akins} {et~al.}(2024){Akins}, {Casey}, {Lambrides}, {Allen},
  {Andika}, {Brinch}, {Champagne}, {Cooper}, {Ding}, {Drakos}, {Faisst},
  {Finkelstein}, {Franco}, {Fujimoto}, {Gentile}, {Gillman}, {Gozaliasl},
  {Harish}, {Hayward}, {Hirschmann}, {Ilbert}, {Kartaltepe}, {Kocevski},
  {Koekemoer}, {Kokorev}, {Liu}, {Long}, {McCracken}, {McKinney}, {Onoue},
  {Paquereau}, {Renzini}, {Rhodes}, {Robertson}, {Shuntov}, {Silverman},
  {Tanaka}, {Toft}, {Trakhtenbrot}, {Valentino}, \& {Zavala}}]{Akins2024}
{Akins}, H.~B., {Casey}, C.~M., {Lambrides}, E., {et~al.} 2024, arXiv e-prints,
  arXiv:2406.10341

\bibitem[{{Arrabal Haro} {et~al.}(2023{\natexlab{a}}){Arrabal Haro},
  {Dickinson}, {Finkelstein}, {Fujimoto}, {Fern{\'a}ndez}, {Kartaltepe},
  {Jung}, {Cole}, {Burgarella}, {Chworowsky}, {Hutchison}, {Morales},
  {Papovich}, {Simons}, {Amor{\'\i}n}, {Backhaus}, {Bagley}, {Bisigello},
  {Calabr{\`o}}, {Castellano}, {Cleri}, {Dav{\'e}}, {Dekel}, {Ferguson},
  {Fontana}, {Gawiser}, {Giavalisco}, {Harish}, {Hathi}, {Hirschmann},
  {Holwerda}, {Huertas-Company}, {Koekemoer}, {Larson}, {Lucas}, {Mobasher},
  {P{\'e}rez-Gonz{\'a}lez}, {Pirzkal}, {Rose}, {Santini}, {Trump}, {de la
  Vega}, {Wang}, {Weiner}, {Wilkins}, {Yang}, {Yung}, \& {Zavala}}]{AHaro2023a}
{Arrabal Haro}, P., {Dickinson}, M., {Finkelstein}, S.~L., {et~al.}
  2023{\natexlab{a}}, \apjl, 951, L22

\bibitem[{{Arrabal Haro} {et~al.}(2023{\natexlab{b}}){Arrabal Haro},
  {Dickinson}, {Finkelstein}, {Kartaltepe}, {Donnan}, {Burgarella}, {Carnall},
  {Cullen}, {Dunlop}, {Fern{\'a}ndez}, {Fujimoto}, {Jung}, {Krips}, {Larson},
  {Papovich}, {P{\'e}rez-Gonz{\'a}lez}, {Amor{\'\i}n}, {Bagley}, {Buat},
  {Casey}, {Chworowsky}, {Cohen}, {Ferguson}, {Giavalisco}, {Huertas-Company},
  {Hutchison}, {Kocevski}, {Koekemoer}, {Lucas}, {McLeod}, {McLure}, {Pirzkal},
  {Seill{\'e}}, {Trump}, {Weiner}, {Wilkins}, \& {Zavala}}]{AHaro2023b}
{Arrabal Haro}, P., {Dickinson}, M., {Finkelstein}, S.~L., {et~al.}
  2023{\natexlab{b}}, \nat, 622, 707

\bibitem[{Atek {et~al.}(2022)Atek, Shuntov, Furtak, Richard, Kneib, Mahler,
  Zitrin, McCracken, Charlot, Chevallard, \& Chemerynska}]{Atek_2022}
Atek, H., Shuntov, M., Furtak, L.~J., {et~al.} 2022, Monthly Notices of the
  Royal Astronomical Society, 519, 1201–1220

\bibitem[{{Backhaus} {et~al.}(2024){Backhaus}, {Trump}, {Pirzkal}, {Barro},
  {Finkelstein}, {Arrabal Haro}, {Simons}, {Wessner}, {Cleri}, {Bagley},
  {Hirschmann}, {Nicholls}, {Dickinson}, {Kartaltepe}, {Papovich}, {Kocevski},
  {Koekemoer}, {Bisigello}, {Jaskot}, {Lucas}, {Jung}, {Wilkins}, {Yung},
  {Ferguson}, {Fontana}, {Grazian}, {Grogin}, {Kewley}, {Kirkpatrick}, {Lotz},
  {Pentericci}, {P{\'e}rez-Gonz{\'a}lez}, {Ravindranath}, {Somerville}, {Yang},
  {Holwerda}, {Kurczynski}, {Hathi}, {Rose}, \& {Davis}}]{Backhaus2024}
{Backhaus}, B.~E., {Trump}, J.~R., {Pirzkal}, N., {et~al.} 2024, \apj, 962, 195

\bibitem[{{Baggen} {et~al.}(2023){Baggen}, {van Dokkum}, {Labb{\'e}},
  {Brammer}, {Miller}, {Bezanson}, {Leja}, {Wang}, {Whitaker}, {Suess}, \&
  {Nelson}}]{Baggen2023}
{Baggen}, J. F.~W., {van Dokkum}, P., {Labb{\'e}}, I., {et~al.} 2023, \apjl,
  955, L12

\bibitem[{{Bagley} {et~al.}(2023){Bagley}, {Finkelstein}, {Koekemoer},
  {Ferguson}, {Arrabal Haro}, {Dickinson}, {Kartaltepe}, {Papovich},
  {P{\'e}rez-Gonz{\'a}lez}, {Pirzkal}, {Somerville}, {Willmer}, {Yang}, {Yung},
  {Fontana}, {Grazian}, {Grogin}, {Hirschmann}, {Kewley}, {Kirkpatrick},
  {Kocevski}, {Lotz}, {Medrano}, {Morales}, {Pentericci}, {Ravindranath},
  {Trump}, {Wilkins}, {Calabr{\`o}}, {Cooper}, {Costantin}, {de la Vega},
  {Hilbert}, {Hutchison}, {Larson}, {Lucas}, {McGrath}, {Ryan}, {Wang}, \&
  {Wuyts}}]{Bagley2023}
{Bagley}, M.~B., {Finkelstein}, S.~L., {Koekemoer}, A.~M., {et~al.} 2023,
  \apjl, 946, L12

\bibitem[{Barbary(2016)}]{Barbary2016}
Barbary, K. 2016, Journal of Open Source Software, 1, 58

\bibitem[{{Barro} {et~al.}(2024){Barro}, {P{\'e}rez-Gonz{\'a}lez}, {Kocevski},
  {McGrath}, {Trump}, {Simons}, {Somerville}, {Yung}, {Arrabal Haro}, {Akins},
  {Bagley}, {Cleri}, {Costantin}, {Davis}, {Dickinson}, {Finkelstein},
  {Giavalisco}, {G{\'o}mez-Guijarro}, {Hathi}, {Hirschmann}, {Holwerda},
  {Huertas-Company}, {Kartaltepe}, {Koekemoer}, {Lucas}, {Papovich}, {Pirzkal},
  {Seill{\'e}}, {Tacchella}, {Wuyts}, {Wilkins}, {de la Vega}, {Yang}, \&
  {Zavala}}]{Barro2024}
{Barro}, G., {P{\'e}rez-Gonz{\'a}lez}, P.~G., {Kocevski}, D.~D., {et~al.} 2024,
  \apj, 963, 128

\bibitem[{{Barrufet} {et~al.}(2025){Barrufet}, {Oesch}, {Marques-Chaves},
  {Arellano-Cordova}, {Baggen}, {Carnall}, {Cullen}, {Dunlop}, {Gottumukkala},
  {Fudamoto}, {Illingworth}, {Magee}, {McLure}, {McLeod}, {Micha{\l}owski},
  {Stefanon}, {van Dokkum}, \& {Weibel}}]{Barrufet2024}
{Barrufet}, L., {Oesch}, P.~A., {Marques-Chaves}, R., {et~al.} 2025, \mnras,
  537, 3453

\bibitem[{{Barrufet} {et~al.}(2023){Barrufet}, {Oesch}, {Weibel}, {Brammer},
  {Bezanson}, {Bouwens}, {Fudamoto}, {Gonzalez}, {Gottumukkala}, {Illingworth},
  {Heintz}, {Holden}, {Labbe}, {Magee}, {Naidu}, {Nelson}, {Stefanon}, {Smit},
  {van Dokkum}, {Weaver}, \& {Williams}}]{Barrufet2023}
{Barrufet}, L., {Oesch}, P.~A., {Weibel}, A., {et~al.} 2023, \mnras, 522, 449

\bibitem[{{Bertin} \& {Arnouts}(1996)}]{Bertin1996}
{Bertin}, E. \& {Arnouts}, S. 1996, \aaps, 117, 393

\bibitem[{{B{\"o}ker} {et~al.}(2023){B{\"o}ker}, {Beck}, {Birkmann},
  {Giardino}, {Keyes}, {Kumari}, {Muzerolle}, {Rawle}, {Zeidler}, {Abul-Huda},
  {Alves de Oliveira}, {Arribas}, {Bechtold}, {Bhatawdekar}, {Bonaventura},
  {Bunker}, {Cameron}, {Carniani}, {Charlot}, {Curti}, {Espinoza}, {Ferruit},
  {Franx}, {Jakobsen}, {Karakla}, {L{\'o}pez-Caniego}, {L{\"u}tzgendorf},
  {Maiolino}, {Manjavacas}, {Marston}, {Moseley}, {Ogle}, {Perna},
  {Pe{\~n}a-Guerrero}, {Pirzkal}, {Plesha}, {Proffitt}, {Rauscher}, {Rix},
  {Rodr{\'\i}guez del Pino}, {Rustamkulov}, {Sabbi}, {Sing}, {Sirianni}, {te
  Plate}, {{\'U}beda}, {Wahlgren}, {Wislowski}, {Wu}, \&
  {Willott}}]{Boeker2023}
{B{\"o}ker}, T., {Beck}, T.~L., {Birkmann}, S.~M., {et~al.} 2023, \pasp, 135,
  038001

\bibitem[{{Bonaventura} {et~al.}(2023){Bonaventura}, {Jakobsen}, {Ferruit},
  {Arribas}, \& {Giardino}}]{Bonaventura2023}
{Bonaventura}, N., {Jakobsen}, P., {Ferruit}, P., {Arribas}, S., \& {Giardino},
  G. 2023, \aap, 672, A40

\bibitem[{{Boylan-Kolchin}(2023)}]{MBK2023}
{Boylan-Kolchin}, M. 2023, Nature Astronomy, 7, 731

\bibitem[{{Brammer}(2023{\natexlab{a}})}]{grizli}
{Brammer}, G. 2023{\natexlab{a}}, {grizli}

\bibitem[{{Brammer}(2023{\natexlab{b}})}]{msaexp}
{Brammer}, G. 2023{\natexlab{b}}, {msaexp: NIRSpec analyis tools}

\bibitem[{{Brammer} {et~al.}(2008){Brammer}, {van Dokkum}, \&
  {Coppi}}]{Brammer2008}
{Brammer}, G.~B., {van Dokkum}, P.~G., \& {Coppi}, P. 2008, \apj, 686, 1503

\bibitem[{{Brammer} {et~al.}(2012){Brammer}, {van Dokkum}, {Franx},
  {Fumagalli}, {Patel}, {Rix}, {Skelton}, {Kriek}, {Nelson}, {Schmidt},
  {Bezanson}, {da Cunha}, {Erb}, {Fan}, {F{\"o}rster Schreiber}, {Illingworth},
  {Labb{\'e}}, {Leja}, {Lundgren}, {Magee}, {Marchesini}, {McCarthy},
  {Momcheva}, {Muzzin}, {Quadri}, {Steidel}, {Tal}, {Wake}, {Whitaker}, \&
  {Williams}}]{Brammer2012}
{Brammer}, G.~B., {van Dokkum}, P.~G., {Franx}, M., {et~al.} 2012, \apjs, 200,
  13

\bibitem[{{Brinchmann}(2023)}]{Brinchmann2023}
{Brinchmann}, J. 2023, \mnras, 525, 2087

\bibitem[{{Bunker} {et~al.}(2024){Bunker}, {Cameron}, {Curtis-Lake},
  {Jakobsen}, {Carniani}, {Curti}, {Witstok}, {Maiolino}, {D'Eugenio},
  {Looser}, {Willott}, {Bonaventura}, {Hainline}, {{\"U}bler}, {Willmer},
  {Saxena}, {Smit}, {Alberts}, {Arribas}, {Baker}, {Baum}, {Bhatawdekar},
  {Bowler}, {Boyett}, {Charlot}, {Chen}, {Chevallard}, {Circosta}, {DeCoursey},
  {de Graaff}, {Egami}, {Eisenstein}, {Endsley}, {Ferruit}, {Giardino},
  {Hausen}, {Helton}, {Hviding}, {Ji}, {Johnson}, {Jones}, {Kumari}, {Laseter},
  {L{\"u}tzgendorf}, {Maseda}, {Nelson}, {Parlanti}, {Perna}, {Rauscher},
  {Rawle}, {Rix}, {Rieke}, {Robertson}, {Rodr{\'\i}guez Del Pino}, {Sandles},
  {Scholtz}, {Sharpe}, {Skarbinski}, {Stark}, {Sun}, {Tacchella}, {Topping},
  {Villanueva}, {Wallace}, {Williams}, \& {Woodrum}}]{Bunker2023}
{Bunker}, A.~J., {Cameron}, A.~J., {Curtis-Lake}, E., {et~al.} 2024, \aap, 690,
  A288

\bibitem[{{Burgasser} {et~al.}(2024){Burgasser}, {Bezanson}, {Labbe},
  {Brammer}, {Cutler}, {Furtak}, {Greene}, {Gerasimov}, {Leja}, {Pan}, {Price},
  {Wang}, {Weaver}, {Whitaker}, {Fujimoto}, {Kokorev}, {Dayal}, {Nanayakkara},
  {Williams}, {Marchesini}, {Zitrin}, \& {van Dokkum}}]{Burgasser2024}
{Burgasser}, A.~J., {Bezanson}, R., {Labbe}, I., {et~al.} 2024, \apj, 962, 177

\bibitem[{{Cameron} {et~al.}(2024){Cameron}, {Katz}, {Witten}, {Saxena},
  {Laporte}, \& {Bunker}}]{Cameron2024}
{Cameron}, A.~J., {Katz}, H., {Witten}, C., {et~al.} 2024, \mnras, 534, 523

\bibitem[{{Carnall} {et~al.}(2023{\natexlab{a}}){Carnall}, {McLeod}, {McLure},
  {Dunlop}, {Begley}, {Cullen}, {Donnan}, {Hamadouche}, {Jewell}, {Jones},
  {Pollock}, \& {Wild}}]{Carnall2023}
{Carnall}, A.~C., {McLeod}, D.~J., {McLure}, R.~J., {et~al.}
  2023{\natexlab{a}}, \mnras, 520, 3974

\bibitem[{{Carnall} {et~al.}(2023{\natexlab{b}}){Carnall}, {McLure}, {Dunlop},
  {McLeod}, {Wild}, {Cullen}, {Magee}, {Begley}, {Cimatti}, {Donnan},
  {Hamadouche}, {Jewell}, \& {Walker}}]{Carnall2023b}
{Carnall}, A.~C., {McLure}, R.~J., {Dunlop}, J.~S., {et~al.}
  2023{\natexlab{b}}, \nat, 619, 716

\bibitem[{{Carniani} {et~al.}(2024){Carniani}, {Venturi}, {Parlanti}, {de
  Graaff}, {Maiolino}, {Arribas}, {Bonaventura}, {Boyett}, {Bunker}, {Cameron},
  {Charlot}, {Chevallard}, {Curti}, {Curtis-Lake}, {Eisenstein}, {Giardino},
  {Hausen}, {Kumari}, {Maseda}, {Nelson}, {Perna}, {Rix}, {Robertson}, {Del
  Pino}, {Sandles}, {Scholtz}, {Simmonds}, {Smit}, {Tacchella}, {{\"U}bler},
  {Williams}, {Willott}, \& {Witstok}}]{Carniani2024}
{Carniani}, S., {Venturi}, G., {Parlanti}, E., {et~al.} 2024, \aap, 685, A99

\bibitem[{{Casey} {et~al.}(2014){Casey}, {Narayanan}, \& {Cooray}}]{Casey2014}
{Casey}, C.~M., {Narayanan}, D., \& {Cooray}, A. 2014, \physrep, 541, 45

\bibitem[{{Casey} {et~al.}(2019){Casey}, {Zavala}, {Aravena}, {B{\'e}thermin},
  {Caputi}, {Champagne}, {Clements}, {da Cunha}, {Drew}, {Finkelstein},
  {Hayward}, {Kartaltepe}, {Knudsen}, {Koekemoer}, {Magdis}, {Man}, {Manning},
  {Scoville}, {Sheth}, {Spilker}, {Staguhn}, {Talia}, {Taniguchi}, {Toft},
  {Treister}, \& {Yun}}]{Casey2019}
{Casey}, C.~M., {Zavala}, J.~A., {Aravena}, M., {et~al.} 2019, \apj, 887, 55

\bibitem[{{Castellano} {et~al.}(2022){Castellano}, {Fontana}, {Treu},
  {Santini}, {Merlin}, {Leethochawalit}, {Trenti}, {Vanzella}, {Mestric},
  {Bonchi}, {Belfiori}, {Nonino}, {Paris}, {Polenta}, {Roberts-Borsani},
  {Boyett}, {Brada{\v{c}}}, {Calabr{\`o}}, {Glazebrook}, {Grillo}, {Mascia},
  {Mason}, {Mercurio}, {Morishita}, {Nanayakkara}, {Pentericci}, {Rosati},
  {Vulcani}, {Wang}, \& {Yang}}]{Castellano2022}
{Castellano}, M., {Fontana}, A., {Treu}, T., {et~al.} 2022, \apjl, 938, L15

\bibitem[{{Chisholm} {et~al.}(2024){Chisholm}, {Berg}, {Endsley}, {Gazagnes},
  {Richardson}, {Lambrides}, {Greene}, {Finkelstein}, {Flury}, {Guseva},
  {Henry}, {Hutchison}, {Izotov}, {Marques-Chaves}, {Oesch}, {Papovich},
  {Saldana-Lopez}, {Schaerer}, \& {Stephenson}}]{Chisholm2024}
{Chisholm}, J., {Berg}, D.~A., {Endsley}, R., {et~al.} 2024, \mnras, 534, 2633

\bibitem[{{Clarke} {et~al.}(2024){Clarke}, {Shapley}, {Sanders}, {Topping},
  {Brammer}, {Bento}, {Reddy}, \& {Kehoe}}]{Clarke2024}
{Clarke}, L., {Shapley}, A.~E., {Sanders}, R.~L., {et~al.} 2024, \apj, 977, 133

\bibitem[{{Cooper} {et~al.}(2024){Cooper}, {Brammer}, {Heintz}, {Toft},
  {Casey}, {Setton}, {de Graaff}, {Boogaard}, {Cleri}, {Gillman},
  {Gottumukkala}, {Greene}, {Gullberg}, {Hirschmann}, {Hviding}, {Lambrides},
  {Leja}, {Long}, {Manning}, {Maseda}, {McConachie}, {McKinney}, {Narayanan},
  {Price}, {Strait}, {Weibel}, \& {Williams}}]{Cooper2024}
{Cooper}, O.~R., {Brammer}, G., {Heintz}, K.~E., {et~al.} 2024, arXiv e-prints,
  arXiv:2410.08387

\bibitem[{{Curti} {et~al.}(2023){Curti}, {D'Eugenio}, {Carniani}, {Maiolino},
  {Sandles}, {Witstok}, {Baker}, {Bennett}, {Piotrowska}, {Tacchella},
  {Charlot}, {Nakajima}, {Maheson}, {Mannucci}, {Amiri}, {Arribas}, {Belfiore},
  {Bonaventura}, {Bunker}, {Chevallard}, {Cresci}, {Curtis-Lake},
  {Hayden-Pawson}, {Jones}, {Kumari}, {Laseter}, {Looser}, {Marconi}, {Maseda},
  {Scholtz}, {Smit}, {{\"U}bler}, \& {Wallace}}]{Curti2023}
{Curti}, M., {D'Eugenio}, F., {Carniani}, S., {et~al.} 2023, \mnras, 518, 425

\bibitem[{{Curti} {et~al.}(2024){Curti}, {Maiolino}, {Curtis-Lake},
  {Chevallard}, {Carniani}, {D'Eugenio}, {Looser}, {Scholtz}, {Charlot},
  {Cameron}, {{\"U}bler}, {Witstok}, {Boyett}, {Laseter}, {Sandles}, {Arribas},
  {Bunker}, {Giardino}, {Maseda}, {Rawle}, {Rodr{\'\i}guez Del Pino}, {Smit},
  {Willott}, {Eisenstein}, {Hausen}, {Johnson}, {Rieke}, {Robertson},
  {Tacchella}, {Williams}, {Willmer}, {Baker}, {Bhatawdekar}, {Egami},
  {Helton}, {Ji}, {Kumari}, {Perna}, {Shivaei}, \& {Sun}}]{Curti2024}
{Curti}, M., {Maiolino}, R., {Curtis-Lake}, E., {et~al.} 2024, \aap, 684, A75

\bibitem[{{Curtis-Lake} {et~al.}(2023){Curtis-Lake}, {Carniani}, {Cameron},
  {Charlot}, {Jakobsen}, {Maiolino}, {Bunker}, {Witstok}, {Smit}, {Chevallard},
  {Willott}, {Ferruit}, {Arribas}, {Bonaventura}, {Curti}, {D'Eugenio},
  {Franx}, {Giardino}, {Looser}, {L{\"u}tzgendorf}, {Maseda}, {Rawle}, {Rix},
  {Rodr{\'\i}guez del Pino}, {{\"U}bler}, {Sirianni}, {Dressler}, {Egami},
  {Eisenstein}, {Endsley}, {Hainline}, {Hausen}, {Johnson}, {Rieke},
  {Robertson}, {Shivaei}, {Stark}, {Tacchella}, {Williams}, {Willmer},
  {Bhatawdekar}, {Bowler}, {Boyett}, {Chen}, {de Graaff}, {Helton}, {Hviding},
  {Jones}, {Kumari}, {Lyu}, {Nelson}, {Perna}, {Sandles}, {Saxena}, {Suess},
  {Sun}, {Topping}, {Wallace}, \& {Whitler}}]{CurtisLake2023}
{Curtis-Lake}, E., {Carniani}, S., {Cameron}, A., {et~al.} 2023, Nature
  Astronomy, 7, 622

\bibitem[{{Davies}(2024)}]{snowblind}
{Davies}, J. 2024, {snowblind}

\bibitem[{{de Graaff} {et~al.}(2024){de Graaff}, {Rix}, {Carniani}, {Suess},
  {Charlot}, {Curtis-Lake}, {Arribas}, {Baker}, {Boyett}, {Bunker}, {Cameron},
  {Chevallard}, {Curti}, {Eisenstein}, {Franx}, {Hainline}, {Hausen}, {Ji},
  {Johnson}, {Jones}, {Maiolino}, {Maseda}, {Nelson}, {Parlanti}, {Rawle},
  {Robertson}, {Tacchella}, {{\"U}bler}, {Williams}, {Willmer}, \&
  {Willott}}]{deGraaff2024a}
{de Graaff}, A., {Rix}, H.-W., {Carniani}, S., {et~al.} 2024, \aap, 684, A87

\bibitem[{{de Graaff} {et~al.}(2025){de Graaff}, {Setton}, {Brammer}, {Cutler},
  {Suess}, {Labb{\'e}}, {Leja}, {Weibel}, {Maseda}, {Whitaker}, {Bezanson},
  {Boogaard}, {Cleri}, {De Lucia}, {Franx}, {Greene}, {Hirschmann}, {Matthee},
  {McConachie}, {Naidu}, {Oesch}, {Price}, {Rix}, {Valentino}, {Wang}, \&
  {Williams}}]{deGraaff2024c}
{de Graaff}, A., {Setton}, D.~J., {Brammer}, G., {et~al.} 2025, Nature
  Astronomy, 9, 280

\bibitem[{{D'Eugenio} {et~al.}(2025){D'Eugenio}, {Cameron}, {Scholtz},
  {Carniani}, {Willott}, {Curtis-Lake}, {Bunker}, {Parlanti}, {Maiolino},
  {Willmer}, {Jakobsen}, {Robertson}, {Johnson}, {Tacchella}, {Cargile},
  {Rawle}, {Arribas}, {Chevallard}, {Curti}, {Egami}, {Eisenstein}, {Kumari},
  {Looser}, {Rieke}, {Rodr{\'\i}guez Del Pino}, {Saxena}, {{\"U}bler},
  {Venturi}, {Witstok}, {Baker}, {Bhatawdekar}, {Bonaventura}, {Boyett},
  {Charlot}, {Danhaive}, {Hainline}, {Hausen}, {Helton}, {Ji}, {Ji}, {Jones},
  {Juod{\v{z}}balis}, {Maseda}, {P{\'e}rez-Gonz{\'a}lez}, {Perna},
  {Pusk{\'a}s}, {Shivaei}, {Silcock}, {Simmonds}, {Smit}, {Sun}, {Villanueva},
  {Williams}, \& {Zhu}}]{JADES_DR3}
{D'Eugenio}, F., {Cameron}, A.~J., {Scholtz}, J., {et~al.} 2025, \apjs, 277, 4

\bibitem[{Donnan {et~al.}(2022)Donnan, McLeod, Dunlop, McLure, Carnall, Begley,
  Cullen, Hamadouche, Bowler, Magee, McCracken, Milvang-Jensen, Moneti, \&
  Targett}]{Donnan_2022}
Donnan, C.~T., McLeod, D.~J., Dunlop, J.~S., {et~al.} 2022, Monthly Notices of
  the Royal Astronomical Society, 518, 6011–6040

\bibitem[{{Donnan} {et~al.}(2024){Donnan}, {McLure}, {Dunlop}, {McLeod},
  {Magee}, {Arellano-C{\'o}rdova}, {Barrufet}, {Begley}, {Bowler}, {Carnall},
  {Cullen}, {Ellis}, {Fontana}, {Illingworth}, {Grogin}, {Hamadouche},
  {Koekemoer}, {Liu}, {Mason}, {Santini}, \& {Stanton}}]{Donnan2024}
{Donnan}, C.~T., {McLure}, R.~J., {Dunlop}, J.~S., {et~al.} 2024, \mnras, 533,
  3222

\bibitem[{{Eisenstein} {et~al.}(2023){Eisenstein}, {Willott}, {Alberts},
  {Arribas}, {Bonaventura}, {Bunker}, {Cameron}, {Carniani}, {Charlot},
  {Curtis-Lake}, {D'Eugenio}, {Endsley}, {Ferruit}, {Giardino}, {Hainline},
  {Hausen}, {Jakobsen}, {Johnson}, {Maiolino}, {Rieke}, {Rieke}, {Rix},
  {Robertson}, {Stark}, {Tacchella}, {Williams}, {Willmer}, {Baker}, {Baum},
  {Bhatawdekar}, {Boyett}, {Chen}, {Chevallard}, {Circosta}, {Curti},
  {Danhaive}, {DeCoursey}, {de Graaff}, {Dressler}, {Egami}, {Helton},
  {Hviding}, {Ji}, {Jones}, {Kumari}, {L{\"u}tzgendorf}, {Laseter}, {Looser},
  {Lyu}, {Maseda}, {Nelson}, {Parlanti}, {Perna}, {Pusk{\'a}s}, {Rawle},
  {Rodr{\'\i}guez Del Pino}, {Sandles}, {Saxena}, {Scholtz}, {Sharpe},
  {Shivaei}, {Silcock}, {Simmonds}, {Skarbinski}, {Smit}, {Stone}, {Suess},
  {Sun}, {Tang}, {Topping}, {{\"U}bler}, {Villanueva}, {Wallace}, {Whitler},
  {Witstok}, \& {Woodrum}}]{Eisenstein2023}
{Eisenstein}, D.~J., {Willott}, C., {Alberts}, S., {et~al.} 2023, arXiv
  e-prints, arXiv:2306.02465

\bibitem[{{Endsley} {et~al.}(2023){Endsley}, {Stark}, {Whitler}, {Topping},
  {Chen}, {Plat}, {Chisholm}, \& {Charlot}}]{Endsley2023}
{Endsley}, R., {Stark}, D.~P., {Whitler}, L., {et~al.} 2023, \mnras, 524, 2312

\bibitem[{{Eyles} {et~al.}(2005){Eyles}, {Bunker}, {Stanway}, {Lacy}, {Ellis},
  \& {Doherty}}]{Eyles2005}
{Eyles}, L.~P., {Bunker}, A.~J., {Stanway}, E.~R., {et~al.} 2005, \mnras, 364,
  443

\bibitem[{{Ferruit} {et~al.}(2022){Ferruit}, {Jakobsen}, {Giardino}, {Rawle},
  {Alves de Oliveira}, {Arribas}, {Beck}, {Birkmann}, {B{\"o}ker}, {Bunker},
  {Charlot}, {de Marchi}, {Franx}, {Henry}, {Karakla}, {Kassin}, {Kumari},
  {L{\'o}pez-Caniego}, {L{\"u}tzgendorf}, {Maiolino}, {Manjavacas}, {Marston},
  {Moseley}, {Muzerolle}, {Pirzkal}, {Rauscher}, {Rix}, {Sabbi}, {Sirianni},
  {te Plate}, {Valenti}, {Willott}, \& {Zeidler}}]{Ferruit2022}
{Ferruit}, P., {Jakobsen}, P., {Giardino}, G., {et~al.} 2022, \aap, 661, A81

\bibitem[{{Finkelstein} {et~al.}(2025){Finkelstein}, {Bagley}, {Arrabal Haro},
  {Dickinson}, {Ferguson}, {Kartaltepe}, {Kocevski}, {Koekemoer}, {Lotz},
  {Papovich}, {Perez-Gonzalez}, {Pirzkal}, {Somerville}, {Trump}, {Yang},
  {Yung}, {Fontana}, {Grazian}, {Grogin}, {Kewley}, {Kirkpatrick}, {Larson},
  {Pentericci}, {Ravindranath}, {Wilkins}, {Almaini}, {Amorin}, {Barro},
  {Bhatawdekar}, {Bisigello}, {Brooks}, {Buitrago}, {Calabro}, {Castellano},
  {Cheng}, {Cleri}, {Cole}, {Cooper}, {Cooper}, {Costantin}, {Cox}, {Croton},
  {Daddi}, {Davis}, {Dekel}, {Elbaz}, {Fernandez}, {Fujimoto}, {Gandolfi},
  {Gardner}, {Gawiser}, {Giavalisco}, {Gomez-Guijarro}, {Guo}, {Gupta},
  {Hathi}, {Harish}, {Henry}, {Hirschmann}, {Hu}, {Hutchison}, {Iyer},
  {Jaskot}, {Jha}, {Jung}, {Kokorev}, {Kurczynski}, {Leung}, {Llerena}, {Long},
  {Lucas}, {Lu}, {McGrath}, {McIntosh}, {Merlin}, {Morales}, {Napolitano},
  {Pacucci}, {Pandya}, {Rafelski}, {Rodighiero}, {Rose}, {Santini}, {Seille},
  {Simons}, {Shen}, {Straughn}, {Tacchella}, {Vanderhoof}, {Vega-Ferrero},
  {Weiner}, {Willmer}, {Zhu}, {Bell}, {Wuyts}, {Holwerda}, {Wang}, {Wang}, \&
  {Zavala}}]{Finkelstein2025}
{Finkelstein}, S.~L., {Bagley}, M.~B., {Arrabal Haro}, P., {et~al.} 2025, arXiv
  e-prints, arXiv:2501.04085

\bibitem[{Finkelstein {et~al.}(2023)Finkelstein, Bagley, Ferguson, Wilkins,
  Kartaltepe, Papovich, Yung, Arrabal~Haro, Behroozi, Dickinson, Kocevski,
  Koekemoer, Larson, Le~Bail, Morales, Pérez-González, Burgarella, Davé,
  Hirschmann, Somerville, Wuyts, Bromm, Casey, Fontana, Fujimoto, Gardner,
  Giavalisco, Grazian, Grogin, Hathi, Hutchison, Jha, Jogee, Kewley,
  Kirkpatrick, Long, Lotz, Pentericci, Pierel, Pirzkal, Ravindranath, Ryan,
  Trump, Yang, Bhatawdekar, Bisigello, Buat, Calabrò, Castellano, Cleri,
  Cooper, Croton, Daddi, Dekel, Elbaz, Franco, Gawiser, Holwerda,
  Huertas-Company, Jaskot, Leung, Lucas, Mobasher, Pandya, Tacchella, Weiner,
  \& Zavala}]{Finkelstein_2023}
Finkelstein, S.~L., Bagley, M.~B., Ferguson, H.~C., {et~al.} 2023, The
  Astrophysical Journal Letters, 946, L13

\bibitem[{{Foreman-Mackey} {et~al.}(2013){Foreman-Mackey}, {Hogg}, {Lang}, \&
  {Goodman}}]{emcee}
{Foreman-Mackey}, D., {Hogg}, D.~W., {Lang}, D., \& {Goodman}, J. 2013, \pasp,
  125, 306

\bibitem[{{Franco} {et~al.}(2018){Franco}, {Elbaz}, {B{\'e}thermin},
  {Magnelli}, {Schreiber}, {Ciesla}, {Dickinson}, {Nagar}, {Silverman},
  {Daddi}, {Alexander}, {Wang}, {Pannella}, {Le Floc'h}, {Pope}, {Giavalisco},
  {Maury}, {Bournaud}, {Chary}, {Demarco}, {Ferguson}, {Finkelstein}, {Inami},
  {Iono}, {Juneau}, {Lagache}, {Leiton}, {Lin}, {Magdis}, {Messias},
  {Motohara}, {Mullaney}, {Okumura}, {Papovich}, {Pforr}, {Rujopakarn},
  {Sargent}, {Shu}, \& {Zhou}}]{Franco2018}
{Franco}, M., {Elbaz}, D., {B{\'e}thermin}, M., {et~al.} 2018, \aap, 620, A152

\bibitem[{{Fruchter} \& {Hook}(2002)}]{Drizzle}
{Fruchter}, A.~S. \& {Hook}, R.~N. 2002, \pasp, 114, 144

\bibitem[{{Fudamoto} {et~al.}(2022){Fudamoto}, {Inoue}, \&
  {Sugahara}}]{Fudamoto2022}
{Fudamoto}, Y., {Inoue}, A.~K., \& {Sugahara}, Y. 2022, \apjl, 938, L24

\bibitem[{{Fujimoto} {et~al.}(2023){Fujimoto}, {Arrabal Haro}, {Dickinson},
  {Finkelstein}, {Kartaltepe}, {Larson}, {Burgarella}, {Bagley}, {Behroozi},
  {Chworowsky}, {Hirschmann}, {Trump}, {Wilkins}, {Yung}, {Koekemoer},
  {Papovich}, {Pirzkal}, {Ferguson}, {Fontana}, {Grogin}, {Grazian}, {Kewley},
  {Kocevski}, {Lotz}, {Pentericci}, {Ravindranath}, {Somerville}, {Wilkins},
  {Amor{\'\i}n}, {Backhaus}, {Calabr{\`o}}, {Casey}, {Cooper}, {Fern{\'a}ndez},
  {Franco}, {Giavalisco}, {Hathi}, {Harish}, {Hutchison}, {Iyer}, {Jung},
  {Lucas}, \& {Zavala}}]{Fujimoto2023}
{Fujimoto}, S., {Arrabal Haro}, P., {Dickinson}, M., {et~al.} 2023, \apjl, 949,
  L25

\bibitem[{{Fujimoto} {et~al.}(2024){Fujimoto}, {Wang}, {Weaver}, {Kokorev},
  {Atek}, {Bezanson}, {Labbe}, {Brammer}, {Greene}, {Chemerynska}, {Dayal}, {de
  Graaff}, {Furtak}, {Oesch}, {Setton}, {Price}, {Miller}, {Williams},
  {Whitaker}, {Zitrin}, {Cutler}, {Leja}, {Pan}, {Coe}, {van Dokkum},
  {Feldmann}, {Fudamoto}, {Goulding}, {Khullar}, {Marchesini}, {Maseda},
  {Nanayakkara}, {Nelson}, {Smit}, {Stefanon}, \& {Weibel}}]{Fujimoto2023b}
{Fujimoto}, S., {Wang}, B., {Weaver}, J.~R., {et~al.} 2024, \apj, 977, 250

\bibitem[{{Furtak} {et~al.}(2023){Furtak}, {Zitrin}, {Plat}, {Fujimoto},
  {Wang}, {Nelson}, {Labb{\'e}}, {Bezanson}, {Brammer}, {van Dokkum},
  {Endsley}, {Glazebrook}, {Greene}, {Leja}, {Price}, {Smit}, {Stark},
  {Weaver}, {Whitaker}, {Atek}, {Chevallard}, {Curtis-Lake}, {Dayal}, {Feltre},
  {Franx}, {Fudamoto}, {Marchesini}, {Mowla}, {Pan}, {Suess},
  {Vidal-Garc{\'\i}a}, \& {Williams}}]{Furtak2023}
{Furtak}, L.~J., {Zitrin}, A., {Plat}, A., {et~al.} 2023, \apj, 952, 142

\bibitem[{{Gardner} {et~al.}(2023){Gardner}, {Mather}, {Abbott}, {Abell},
  {Abernathy}, {Abney}, {Abraham}, {Abraham}, {Abul-Huda}, {Acton}, {Adams},
  {Adams}, {Adler}, {Adriaensen}, {Aguilar}, {Ahmed}, {Ahmed}, {Ahmed},
  {Albat}, \& {Albert}}]{Gardner2023}
{Gardner}, J.~P., {Mather}, J.~C., {Abbott}, R., {et~al.} 2023, \pasp, 135,
  068001

\bibitem[{{Giavalisco}(2002)}]{Giavalisco2002}
{Giavalisco}, M. 2002, \araa, 40, 579

\bibitem[{{Glazebrook} {et~al.}(2024){Glazebrook}, {Nanayakkara}, {Schreiber},
  {Lagos}, {Kawinwanichakij}, {Jacobs}, {Chittenden}, {Brammer}, {Kacprzak},
  {Labbe}, {Marchesini}, {Marsan}, {Oesch}, {Papovich}, {Remus}, {Tran},
  {Esdaile}, \& {Chandro-Gomez}}]{Glazebrook2024}
{Glazebrook}, K., {Nanayakkara}, T., {Schreiber}, C., {et~al.} 2024, \nat, 628,
  277

\bibitem[{{Gordon} {et~al.}(2022){Gordon}, {Bohlin}, {Sloan}, {Rieke}, {Volk},
  {Boyer}, {Muzerolle}, {Schlawin}, {Deustua}, {Hines}, {Kraemer}, {Mullally},
  \& {Su}}]{Gordon2022}
{Gordon}, K.~D., {Bohlin}, R., {Sloan}, G.~C., {et~al.} 2022, \aj, 163, 267

\bibitem[{{Gottumukkala} {et~al.}(2024){Gottumukkala}, {Barrufet}, {Oesch},
  {Weibel}, {Allen}, {Alcalde Pampliega}, {Nelson}, {Williams}, {Brammer},
  {Fudamoto}, {Gonz{\'a}lez}, {Heintz}, {Illingworth}, {Magee}, {Naidu},
  {Shuntov}, {Stefanon}, {Toft}, {Valentino}, \& {Xiao}}]{Gottumukkala2024}
{Gottumukkala}, R., {Barrufet}, L., {Oesch}, P.~A., {et~al.} 2024, \mnras, 530,
  966

\bibitem[{{Greene} {et~al.}(2024){Greene}, {Labbe}, {Goulding}, {Furtak},
  {Chemerynska}, {Kokorev}, {Dayal}, {Volonteri}, {Williams}, {Wang}, {Setton},
  {Burgasser}, {Bezanson}, {Atek}, {Brammer}, {Cutler}, {Feldmann}, {Fujimoto},
  {Glazebrook}, {de Graaff}, {Khullar}, {Leja}, {Marchesini}, {Maseda},
  {Matthee}, {Miller}, {Naidu}, {Nanayakkara}, {Oesch}, {Pan}, {Papovich},
  {Price}, {van Dokkum}, {Weaver}, {Whitaker}, \& {Zitrin}}]{Greene2024}
{Greene}, J.~E., {Labbe}, I., {Goulding}, A.~D., {et~al.} 2024, \apj, 964, 39

\bibitem[{{Grogin} {et~al.}(2011){Grogin}, {Kocevski}, {Faber}, {Ferguson},
  {Koekemoer}, {Riess}, {Acquaviva}, {Alexander}, {Almaini}, {Ashby}, {Barden},
  {Bell}, {Bournaud}, {Brown}, {Caputi}, {Casertano}, {Cassata}, {Castellano},
  {Challis}, {Chary}, {Cheung}, {Cirasuolo}, {Conselice}, {Roshan Cooray},
  {Croton}, {Daddi}, {Dahlen}, {Dav{\'e}}, {de Mello}, {Dekel}, {Dickinson},
  {Dolch}, {Donley}, {Dunlop}, {Dutton}, {Elbaz}, {Fazio}, {Filippenko},
  {Finkelstein}, {Fontana}, {Gardner}, {Garnavich}, {Gawiser}, {Giavalisco},
  {Grazian}, {Guo}, {Hathi}, {H{\"a}ussler}, {Hopkins}, {Huang}, {Huang},
  {Jha}, {Kartaltepe}, {Kirshner}, {Koo}, {Lai}, {Lee}, {Li}, {Lotz}, {Lucas},
  {Madau}, {McCarthy}, {McGrath}, {McIntosh}, {McLure}, {Mobasher},
  {Moustakas}, {Mozena}, {Nandra}, {Newman}, {Niemi}, {Noeske}, {Papovich},
  {Pentericci}, {Pope}, {Primack}, {Rajan}, {Ravindranath}, {Reddy}, {Renzini},
  {Rix}, {Robaina}, {Rodney}, {Rosario}, {Rosati}, {Salimbeni}, {Scarlata},
  {Siana}, {Simard}, {Smidt}, {Somerville}, {Spinrad}, {Straughn}, {Strolger},
  {Telford}, {Teplitz}, {Trump}, {van der Wel}, {Villforth}, {Wechsler},
  {Weiner}, {Wiklind}, {Wild}, {Wilson}, {Wuyts}, {Yan}, \& {Yun}}]{Grogin2011}
{Grogin}, N.~A., {Kocevski}, D.~D., {Faber}, S.~M., {et~al.} 2011, \apjs, 197,
  35

\bibitem[{{Hainline} {et~al.}(2024){Hainline}, {Helton}, {Johnson}, {Sun},
  {Topping}, {Leisenring}, {Baker}, {Eisenstein}, {Hausen}, {Hviding}, {Lyu},
  {Robertson}, {Tacchella}, {Williams}, {Willmer}, \& {Roellig}}]{Hainline2024}
{Hainline}, K.~N., {Helton}, J.~M., {Johnson}, B.~D., {et~al.} 2024, \apj, 964,
  66

\bibitem[{{Harikane} {et~al.}(2023){Harikane}, {Zhang}, {Nakajima}, {Ouchi},
  {Isobe}, {Ono}, {Hatano}, {Xu}, \& {Umeda}}]{Harikane2023}
{Harikane}, Y., {Zhang}, Y., {Nakajima}, K., {et~al.} 2023, \apj, 959, 39

\bibitem[{{Harvey} {et~al.}(2025){Harvey}, {Conselice}, {Adams}, {Austin},
  {Juod{\v{z}}balis}, {Trussler}, {Li}, {Ormerod}, {Ferreira}, {Lovell},
  {Duan}, {Westcott}, {Harris}, {Bhatawdekar}, {Coe}, {Cohen}, {Caruana},
  {Cheng}, {Driver}, {Frye}, {Furtak}, {Grogin}, {Hathi}, {Holwerda}, {Jansen},
  {Koekemoer}, {Marshall}, {Nonino}, {Vijayan}, {Wilkins}, {Windhorst},
  {Willmer}, {Yan}, \& {Zitrin}}]{Harvey2024}
{Harvey}, T., {Conselice}, C.~J., {Adams}, N.~J., {et~al.} 2025, \apj, 978, 89

\bibitem[{{Heintz} {et~al.}(2025){Heintz}, {Brammer}, {Watson}, {Oesch},
  {Keating}, {Hayes}, {Abdurro'uf}, {Arellano-C{\'o}rdova}, {Carnall},
  {Christiansen}, {Cullen}, {Dav{\'e}}, {Dayal}, {Ferrara}, {Finlator},
  {Fynbo}, {Flury}, {Gelli}, {Gillman}, {Gottumukkala}, {Gould}, {Greve},
  {Hardin}, {Hsiao}, {Hutter}, {Jakobsson}, {Killi}, {Khosravaninezhad},
  {Laursen}, {Lee}, {Magdis}, {Matthee}, {Naidu}, {Narayanan}, {Pollock},
  {Prescott}, {Rusakov}, {Shuntov}, {Sneppen}, {Smit}, {Tanvir}, {Terp},
  {Toft}, {Valentino}, {Vijayan}, {Weaver}, {Wise}, \& {Witstok}}]{Heintz2024}
{Heintz}, K.~E., {Brammer}, G.~B., {Watson}, D., {et~al.} 2025, \aap, 693, A60

\bibitem[{{Herard-Demanche} {et~al.}(2025){Herard-Demanche}, {Bouwens},
  {Oesch}, {Naidu}, {Decarli}, {Nelson}, {Brammer}, {Weibel}, {Xiao},
  {Stefanon}, {Walter}, {Matthee}, {Meyer}, {Wuyts}, {Reddy}, {Rowland}, {van
  Leeuwen}, {Haro}, {Dannerbauer}, {Shapley}, {Chisholm}, {van Dokkum},
  {Labbe}, {Illingworth}, {Schaerer}, \& {Shivaei}}]{Herard2023}
{Herard-Demanche}, T., {Bouwens}, R.~J., {Oesch}, P.~A., {et~al.} 2025, \mnras,
  537, 788

\bibitem[{{Hodge} \& {da Cunha}(2020)}]{Hodge2020}
{Hodge}, J.~A. \& {da Cunha}, E. 2020, Royal Society Open Science, 7, 200556

\bibitem[{{Holwerda} {et~al.}(2024){Holwerda}, {Hsu}, {Hathi}, {Bisigello}, {de
  la Vega}, {Haro}, {Bagley}, {Dickinson}, {Finkelstein}, {Kartaltepe},
  {Koekemoer}, {Papovich}, {Pirzkal}, {Cook}, {Robertson}, {Casey}, {Aganze},
  {P{\'e}rez-Gonz{\'a}lez}, {Lucas}, {Jogee}, {Wilkins}, {Burgarella}, \&
  {Kirkpatrick}}]{Holwerda2024}
{Holwerda}, B.~W., {Hsu}, C.-C., {Hathi}, N., {et~al.} 2024, \mnras, 529, 1067

\bibitem[{{Horne}(1986)}]{Horne86}
{Horne}, K. 1986, \pasp, 98, 609

\bibitem[{{Jakobsen} {et~al.}(2022){Jakobsen}, {Ferruit}, {Alves de Oliveira},
  {Arribas}, {Bagnasco}, {Barho}, {Beck}, {Birkmann}, {B{\"o}ker}, {Bunker},
  {Charlot}, {de Jong}, {de Marchi}, {Ehrenwinkler}, {Falcolini}, {Fels},
  {Franx}, {Franz}, {Funke}, {Giardino}, {Gnata}, {Holota}, {Honnen}, {Jensen},
  {Jentsch}, {Johnson}, {Jollet}, {Karl}, {Kling}, {K{\"o}hler}, {Kolm},
  {Kumari}, {Lander}, {Lemke}, {L{\'o}pez-Caniego}, {L{\"u}tzgendorf},
  {Maiolino}, {Manjavacas}, {Marston}, {Maschmann}, {Maurer}, {Messerschmidt},
  {Moseley}, {Mosner}, {Mott}, {Muzerolle}, {Pirzkal}, {Pittet}, {Plitzke},
  {Posselt}, {Rapp}, {Rauscher}, {Rawle}, {Rix}, {R{\"o}del}, {Rumler},
  {Sabbi}, {Salvignol}, {Schmid}, {Sirianni}, {Smith}, {Strada}, {te Plate},
  {Valenti}, {Wettemann}, {Wiehe}, {Wiesmayer}, {Willott}, {Wright}, {Zeidler},
  \& {Zincke}}]{Jakobsen2022}
{Jakobsen}, P., {Ferruit}, P., {Alves de Oliveira}, C., {et~al.} 2022, \aap,
  661, A80

\bibitem[{{Kartaltepe} {et~al.}(2023){Kartaltepe}, {Rose}, {Vanderhoof},
  {McGrath}, {Costantin}, {Cox}, {Yung}, {Kocevski}, {Wuyts}, {Ferguson},
  {Bagley}, {Finkelstein}, {Amor{\'\i}n}, {Andrews}, {Arrabal Haro},
  {Backhaus}, {Behroozi}, {Bisigello}, {Calabr{\`o}}, {Casey}, {Coogan},
  {Cooper}, {Croton}, {de la Vega}, {Dickinson}, {Fontana}, {Franco},
  {Grazian}, {Grogin}, {Hathi}, {Holwerda}, {Huertas-Company}, {Iyer}, {Jogee},
  {Jung}, {Kewley}, {Kirkpatrick}, {Koekemoer}, {Liu}, {Lotz}, {Lucas},
  {Newman}, {Pacifici}, {Pandya}, {Papovich}, {Pentericci},
  {P{\'e}rez-Gonz{\'a}lez}, {Petersen}, {Pirzkal}, {Rafelski}, {Ravindranath},
  {Simons}, {Snyder}, {Somerville}, {Stanway}, {Straughn}, {Tacchella},
  {Trump}, {Vega-Ferrero}, {Wilkins}, {Yang}, \& {Zavala}}]{Kartaltepe2023}
{Kartaltepe}, J.~S., {Rose}, C., {Vanderhoof}, B.~N., {et~al.} 2023, \apjl,
  946, L15

\bibitem[{{Katz} {et~al.}(2024){Katz}, {Cameron}, {Saxena}, {Barrufet},
  {Choustikov}, {Cleri}, {de Graaff}, {Ellis}, {Fosbury}, {Heintz}, {Maseda},
  {Matthee}, {McConchie}, \& {Oesch}}]{Katz2024}
{Katz}, H., {Cameron}, A.~J., {Saxena}, A., {et~al.} 2024, arXiv e-prints,
  arXiv:2408.03189

\bibitem[{{Killi} {et~al.}(2023){Killi}, {Watson}, {Brammer}, {McPartland},
  {Antwi-Danso}, {Newshore}, {Coe}, {Allen}, {Fynbo}, {Gould}, {Heintz},
  {Rusakov}, \& {Vejlgaard}}]{Killi2023}
{Killi}, M., {Watson}, D., {Brammer}, G., {et~al.} 2023, arXiv e-prints,
  arXiv:2312.03065

\bibitem[{{Kocevski} {et~al.}(2023){Kocevski}, {Onoue}, {Inayoshi}, {Trump},
  {Arrabal Haro}, {Grazian}, {Dickinson}, {Finkelstein}, {Kartaltepe},
  {Hirschmann}, {Aird}, {Holwerda}, {Fujimoto}, {Juneau}, {Amor{\'\i}n},
  {Backhaus}, {Bagley}, {Barro}, {Bell}, {Bisigello}, {Calabr{\`o}}, {Cleri},
  {Cooper}, {Ding}, {Grogin}, {Ho}, {Hutchison}, {Inoue}, {Jiang}, {Jones},
  {Koekemoer}, {Li}, {Li}, {McGrath}, {Molina}, {Papovich},
  {P{\'e}rez-Gonz{\'a}lez}, {Pirzkal}, {Wilkins}, {Yang}, \&
  {Yung}}]{Kocevski2023}
{Kocevski}, D.~D., {Onoue}, M., {Inayoshi}, K., {et~al.} 2023, \apjl, 954, L4

\bibitem[{{Koekemoer} {et~al.}(2011){Koekemoer}, {Faber}, {Ferguson}, {Grogin},
  {Kocevski}, {Koo}, {Lai}, {Lotz}, {Lucas}, {McGrath}, {Ogaz}, {Rajan},
  {Riess}, {Rodney}, {Strolger}, {Casertano}, {Castellano}, {Dahlen},
  {Dickinson}, {Dolch}, {Fontana}, {Giavalisco}, {Grazian}, {Guo}, {Hathi},
  {Huang}, {van der Wel}, {Yan}, {Acquaviva}, {Alexander}, {Almaini}, {Ashby},
  {Barden}, {Bell}, {Bournaud}, {Brown}, {Caputi}, {Cassata}, {Challis},
  {Chary}, {Cheung}, {Cirasuolo}, {Conselice}, {Roshan Cooray}, {Croton},
  {Daddi}, {Dav{\'e}}, {de Mello}, {de Ravel}, {Dekel}, {Donley}, {Dunlop},
  {Dutton}, {Elbaz}, {Fazio}, {Filippenko}, {Finkelstein}, {Frazer}, {Gardner},
  {Garnavich}, {Gawiser}, {Gruetzbauch}, {Hartley}, {H{\"a}ussler},
  {Herrington}, {Hopkins}, {Huang}, {Jha}, {Johnson}, {Kartaltepe},
  {Khostovan}, {Kirshner}, {Lani}, {Lee}, {Li}, {Madau}, {McCarthy},
  {McIntosh}, {McLure}, {McPartland}, {Mobasher}, {Moreira}, {Mortlock},
  {Moustakas}, {Mozena}, {Nandra}, {Newman}, {Nielsen}, {Niemi}, {Noeske},
  {Papovich}, {Pentericci}, {Pope}, {Primack}, {Ravindranath}, {Reddy},
  {Renzini}, {Rix}, {Robaina}, {Rosario}, {Rosati}, {Salimbeni}, {Scarlata},
  {Siana}, {Simard}, {Smidt}, {Snyder}, {Somerville}, {Spinrad}, {Straughn},
  {Telford}, {Teplitz}, {Trump}, {Vargas}, {Villforth}, {Wagner}, {Wandro},
  {Wechsler}, {Weiner}, {Wiklind}, {Wild}, {Wilson}, {Wuyts}, \&
  {Yun}}]{Koekemoer2011}
{Koekemoer}, A.~M., {Faber}, S.~M., {Ferguson}, H.~C., {et~al.} 2011, \apjs,
  197, 36

\bibitem[{{Labb{\'e}} {et~al.}(2010){Labb{\'e}}, {Gonz{\'a}lez}, {Bouwens},
  {Illingworth}, {Franx}, {Trenti}, {Oesch}, {van Dokkum}, {Stiavelli},
  {Carollo}, {Kriek}, \& {Magee}}]{Labbe2010}
{Labb{\'e}}, I., {Gonz{\'a}lez}, V., {Bouwens}, R.~J., {et~al.} 2010, \apjl,
  716, L103

\bibitem[{{Labbe} {et~al.}(2025){Labbe}, {Greene}, {Bezanson}, {Fujimoto},
  {Furtak}, {Goulding}, {Matthee}, {Naidu}, {Oesch}, {Atek}, {Brammer},
  {Chemerynska}, {Coe}, {Cutler}, {Dayal}, {Feldmann}, {Franx}, {Glazebrook},
  {Leja}, {Maseda}, {Marchesini}, {Nanayakkara}, {Nelson}, {Pan}, {Papovich},
  {Price}, {Suess}, {Wang}, {Weaver}, {Whitaker}, {Williams}, \&
  {Zitrin}}]{Labbe2023b}
{Labbe}, I., {Greene}, J.~E., {Bezanson}, R., {et~al.} 2025, \apj, 978, 92

\bibitem[{{Labb{\'e}} {et~al.}(2023){Labb{\'e}}, {van Dokkum}, {Nelson},
  {Bezanson}, {Suess}, {Leja}, {Brammer}, {Whitaker}, {Mathews}, {Stefanon}, \&
  {Wang}}]{Labbe2023}
{Labb{\'e}}, I., {van Dokkum}, P., {Nelson}, E., {et~al.} 2023, \nat, 616, 266

\bibitem[{{Larson} {et~al.}(2023){Larson}, {Finkelstein}, {Kocevski},
  {Hutchison}, {Trump}, {Arrabal Haro}, {Bromm}, {Cleri}, {Dickinson},
  {Fujimoto}, {Kartaltepe}, {Koekemoer}, {Papovich}, {Pirzkal}, {Tacchella},
  {Zavala}, {Bagley}, {Behroozi}, {Champagne}, {Cole}, {Jung}, {Morales},
  {Yang}, {Zhang}, {Zitrin}, {Amor{\'\i}n}, {Burgarella}, {Casey}, {Ch{\'a}vez
  Ortiz}, {Cox}, {Chworowsky}, {Fontana}, {Gawiser}, {Grazian}, {Grogin},
  {Harish}, {Hathi}, {Hirschmann}, {Holwerda}, {Juneau}, {Leung}, {Lucas},
  {McGrath}, {P{\'e}rez-Gonz{\'a}lez}, {Rigby}, {Seill{\'e}}, {Simons}, {de La
  Vega}, {Weiner}, {Wilkins}, {Yung}, \& {Ceers Team}}]{Larson2023}
{Larson}, R.~L., {Finkelstein}, S.~L., {Kocevski}, D.~D., {et~al.} 2023, \apjl,
  953, L29

\bibitem[{{Li} {et~al.}(2024){Li}, {Leja}, {Johnson}, {Tacchella}, \&
  {Naidu}}]{Li2024}
{Li}, Y., {Leja}, J., {Johnson}, B.~D., {Tacchella}, S., \& {Naidu}, R.~P.
  2024, \apjl, 969, L5

\bibitem[{{Long} {et~al.}(2024){Long}, {Antwi-Danso}, {Lambrides}, {Lovell},
  {de la Vega}, {Valentino}, {Zavala}, {Casey}, {Wilkins}, {Yung}, {Arrabal
  Haro}, {Bagley}, {Bisigello}, {Chworowsky}, {Cooper}, {Cooper}, {Cooray},
  {Croton}, {Dickinson}, {Finkelstein}, {Franco}, {Gould}, {Hirschmann},
  {Hutchison}, {Kartaltepe}, {Kocevski}, {Koekemoer}, {Lucas}, {McKinney},
  {Nere}, {Papovich}, {P{\'e}rez-Gonz{\'a}lez}, {Pirzkal}, \&
  {Santini}}]{Long2024}
{Long}, A.~S., {Antwi-Danso}, J., {Lambrides}, E.~L., {et~al.} 2024, \apj, 970,
  68

\bibitem[{Madau {et~al.}(1996)Madau, Ferguson, Dickinson, Giavalisco, Steidel,
  \& Fruchter}]{Madau_1996}
Madau, P., Ferguson, H.~C., Dickinson, M.~E., {et~al.} 1996, Monthly Notices of
  the Royal Astronomical Society, 283, 1388–1404

\bibitem[{{Manning} {et~al.}(2022){Manning}, {Casey}, {Zavala}, {Magdis},
  {Drew}, {Champagne}, {Aravena}, {B{\'e}thermin}, {Clements}, {Finkelstein},
  {Fujimoto}, {Hayward}, {Hodge}, {Ilbert}, {Kartaltepe}, {Knudsen},
  {Koekemoer}, {Man}, {Sanders}, {Sheth}, {Spilker}, {Staguhn}, {Talia},
  {Treister}, \& {Yun}}]{Manning2022}
{Manning}, S.~M., {Casey}, C.~M., {Zavala}, J.~A., {et~al.} 2022, \apj, 925, 23

\bibitem[{{Maseda} {et~al.}(2024){Maseda}, {de Graaff}, {Franx}, {Rix},
  {Carniani}, {Laseter}, {Dudzevi{\v{c}}i{\={u}}t{\.{e}}}, {Rawle}, {Parlanti},
  {Arribas}, {Bunker}, {Cameron}, {Charlot}, {Curti}, {D'Eugenio}, {Jones},
  {Kumari}, {Maiolino}, {{\"U}bler}, {Saxena}, {Smit}, {Willott}, \&
  {Witstok}}]{Maseda2024}
{Maseda}, M.~V., {de Graaff}, A., {Franx}, M., {et~al.} 2024, \aap, 689, A73

\bibitem[{{Maseda} {et~al.}(2023){Maseda}, {Lewis}, {Matthee}, {Hennawi},
  {Boogaard}, {Feltre}, {Nanayakkara}, {Bacon}, {Barger}, {Brinchmann},
  {Franx}, {Hashimoto}, {Inami}, {Kusakabe}, {Leclercq}, {Rowland}, {Taylor},
  {Tremonti}, {Urrutia}, {Schaye}, {Simmonds}, \& {Vitte}}]{Maseda2023}
{Maseda}, M.~V., {Lewis}, Z., {Matthee}, J., {et~al.} 2023, \apj, 956, 11

\bibitem[{{Matthee} {et~al.}(2024){Matthee}, {Naidu}, {Brammer}, {Chisholm},
  {Eilers}, {Goulding}, {Greene}, {Kashino}, {Labbe}, {Lilly}, {Mackenzie},
  {Oesch}, {Weibel}, {Wuyts}, {Xiao}, {Bordoloi}, {Bouwens}, {van Dokkum},
  {Illingworth}, {Kramarenko}, {Maseda}, {Mason}, {Meyer}, {Nelson}, {Reddy},
  {Shivaei}, {Simcoe}, \& {Yue}}]{Matthee2024}
{Matthee}, J., {Naidu}, R.~P., {Brammer}, G., {et~al.} 2024, \apj, 963, 129

\bibitem[{{Muzzin} {et~al.}(2013){Muzzin}, {Marchesini}, {Stefanon}, {Franx},
  {McCracken}, {Milvang-Jensen}, {Dunlop}, {Fynbo}, {Brammer}, {Labb{\'e}}, \&
  {van Dokkum}}]{Muzzin2013}
{Muzzin}, A., {Marchesini}, D., {Stefanon}, M., {et~al.} 2013, \apj, 777, 18

\bibitem[{{Naidu} {et~al.}(2022){Naidu}, {Oesch}, {van Dokkum}, {Nelson},
  {Suess}, {Brammer}, {Whitaker}, {Illingworth}, {Bouwens}, {Tacchella},
  {Matthee}, {Allen}, {Bezanson}, {Conroy}, {Labbe}, {Leja}, {Leonova},
  {Magee}, {Price}, {Setton}, {Strait}, {Stefanon}, {Toft}, {Weaver}, \&
  {Weibel}}]{Naidu2022}
{Naidu}, R.~P., {Oesch}, P.~A., {van Dokkum}, P., {et~al.} 2022, \apjl, 940,
  L14

\bibitem[{{Nakajima} {et~al.}(2023){Nakajima}, {Ouchi}, {Isobe}, {Harikane},
  {Zhang}, {Ono}, {Umeda}, \& {Oguri}}]{Nakajima2023}
{Nakajima}, K., {Ouchi}, M., {Isobe}, Y., {et~al.} 2023, \apjs, 269, 33

\bibitem[{{Nelson} {et~al.}(2023){Nelson}, {Suess}, {Bezanson}, {Price}, {van
  Dokkum}, {Leja}, {Wang}, {Whitaker}, {Labb{\'e}}, {Barrufet}, {Brammer},
  {Eisenstein}, {Gibson}, {Hartley}, {Johnson}, {Heintz}, {Mathews}, {Miller},
  {Oesch}, {Sandles}, {Setton}, {Speagle}, {Tacchella}, {Tadaki}, {{\"U}bler},
  \& {Weaver}}]{Nelson2023}
{Nelson}, E.~J., {Suess}, K.~A., {Bezanson}, R., {et~al.} 2023, \apjl, 948, L18

\bibitem[{{Oke} \& {Gunn}(1983)}]{ABmags}
{Oke}, J.~B. \& {Gunn}, J.~E. 1983, \apj, 266, 713

\bibitem[{{Ormerod} {et~al.}(2024){Ormerod}, {Conselice}, {Adams}, {Harvey},
  {Austin}, {Trussler}, {Ferreira}, {Caruana}, {Lucatelli}, {Li}, \&
  {Roper}}]{Ormerod2024}
{Ormerod}, K., {Conselice}, C.~J., {Adams}, N.~J., {et~al.} 2024, \mnras, 527,
  6110

\bibitem[{{P{\'e}rez-Gonz{\'a}lez} {et~al.}(2023){P{\'e}rez-Gonz{\'a}lez},
  {Barro}, {Annunziatella}, {Costantin}, {Garc{\'\i}a-Argum{\'a}nez},
  {McGrath}, {M{\'e}rida}, {Zavala}, {Arrabal Haro}, {Bagley}, {Backhaus},
  {Behroozi}, {Bell}, {Bisigello}, {Buat}, {Calabr{\`o}}, {Casey}, {Cleri},
  {Coogan}, {Cooper}, {Cooray}, {Dekel}, {Dickinson}, {Elbaz}, {Ferguson},
  {Finkelstein}, {Fontana}, {Franco}, {Gardner}, {Giavalisco},
  {G{\'o}mez-Guijarro}, {Grazian}, {Grogin}, {Guo}, {Huertas-Company}, {Jogee},
  {Kartaltepe}, {Kewley}, {Kirkpatrick}, {Kocevski}, {Koekemoer}, {Long},
  {Lotz}, {Lucas}, {Papovich}, {Pirzkal}, {Ravindranath}, {Somerville},
  {Tacchella}, {Trump}, {Wang}, {Wilkins}, {Wuyts}, {Yang}, \&
  {Yung}}]{PerezGonzalez2023}
{P{\'e}rez-Gonz{\'a}lez}, P.~G., {Barro}, G., {Annunziatella}, M., {et~al.}
  2023, \apjl, 946, L16

\bibitem[{{Price} {et~al.}(2025){Price}, {Suess}, {Williams}, {Bezanson},
  {Khullar}, {Nelson}, {Wang}, {Weaver}, {Fujimoto}, {Kokorev}, {Greene},
  {Brammer}, {Cutler}, {Dayal}, {Furtak}, {Labbe}, {Leja}, {Miller},
  {Nanayakkara}, {Pan}, \& {Whitaker}}]{Price2023}
{Price}, S.~H., {Suess}, K.~A., {Williams}, C.~C., {et~al.} 2025, \apj, 980, 11

\bibitem[{{Rieke} {et~al.}(2023){Rieke}, {Kelly}, {Misselt}, {Stansberry},
  {Boyer}, {Beatty}, {Egami}, {Florian}, {Greene}, {Hainline}, {Leisenring},
  {Roellig}, {Schlawin}, {Sun}, {Tinnin}, {Williams}, {Willmer}, {Wilson},
  {Clark}, {Rohrbach}, {Brooks}, {Canipe}, {Correnti}, {DiFelice}, {Gennaro},
  {Girard}, {Hartig}, {Hilbert}, {Koekemoer}, {Nikolov}, {Pirzkal}, {Rest},
  {Robberto}, {Sunnquist}, {Telfer}, {Wu}, {Ferry}, {Lewis}, {Baum},
  {Beichman}, {Doyon}, {Dressler}, {Eisenstein}, {Ferrarese}, {Hodapp},
  {Horner}, {Jaffe}, {Johnstone}, {Krist}, {Martin}, {McCarthy}, {Meyer},
  {Rieke}, {Trauger}, \& {Young}}]{Rieke2023}
{Rieke}, M.~J., {Kelly}, D.~M., {Misselt}, K., {et~al.} 2023, \pasp, 135,
  028001

\bibitem[{{Rigby} {et~al.}(2023){Rigby}, {Perrin}, {McElwain}, {Kimble},
  {Friedman}, {Lallo}, {Doyon}, {Feinberg}, {Ferruit}, {Glasse}, {Rieke},
  {Rieke}, {Wright}, {Willott}, {Colon}, {Milam}, {Neff}, {Stark}, {Valenti},
  {Abell}, {Abney}, {Abul-Huda}, {Acton}, {Adams}, {Adler}, {Aguilar}, {Ahmed},
  {Albert}, {Alberts}, {Aldridge}, {Allen}, {Altenburg},
  {{\'A}lvarez-M{\'a}rquez}, {Alves de Oliveira}, {Andersen}, {Anderson},
  {Anderson}, {Argyriou}, {Armstrong}, {Arribas}, {Artigau}, {Arvai},
  {Atkinson}, {Bacon}, {Bair}, {Banks}, {Barrientes}, {Barringer}, {Bartosik},
  {Bast}, {Baudoz}, {Beatty}, {Bechtold}, {Beck}, {Bergeron}, {Bergkoetter},
  {Bhatawdekar}, {Birkmann}, {Blazek}, {Blome}, {Boccaletti}, {B{\"o}ker},
  {Boia}, {Bonaventura}, {Bond}, {Bosley}, {Boucarut}, {Bourque}, {Bouwman},
  {Bower}, {Bowers}, {Boyer}, {Bradley}, {Brady}, {Braun}, {Breda},
  {Bresnahan}, {Bright}, {Britt}, {Bromenschenkel}, {Brooks}, {Brooks},
  {Brown}, {Brown}, {Brown}, {Bunker}, {Burger}, {Bushouse}, {Cale}, {Cameron},
  {Cameron}, {Canipe}, {Caplinger}, {Caputo}, {Cara}, {Carey}, {Carniani},
  {Carrasquilla}, {Carruthers}, {Case}, {Catherine}, {Chance}, {Chapman},
  {Charlot}, {Charlow}, {Chayer}, {Chen}, {Cherinka}, {Chichester}, {Chilton},
  {Chonis}, {Clampin}, {Clark}, {Clark}, {Coe}, {Coleman}, {Comber}, {Comeau},
  {Connolly}, {Cooper}, {Cooper}, {Coppock}, {Correnti}, {Cossou}, {Coulais},
  {Coyle}, {Cracraft}, {Curti}, {Cuturic}, {Davis}, {Davis}, {Dean}, {DeLisa},
  {deMeester}, {Dencheva}, {Dencheva}, {DePasquale}, {Deschenes}, {Hunor
  Detre}, {Diaz}, {Dicken}, {DiFelice}, {Dillman}, {Dixon}, {Doggett},
  {Donaldson}, {Douglas}, {DuPrie}, {Dupuis}, {Durning}, {Easmin}, {Eck},
  {Edeani}, {Egami}, {Ehrenwinkler}, {Eisenhamer}, {Eisenhower}, {Elie},
  {Elliott}, {Elliott}, {Ellis}, {Engesser}, {Espinoza}, {Etienne}, {Etxaluze},
  {Falini}, {Feeney}, {Ferry}, {Filippazzo}, {Fincham}, {Fix}, {Flagey},
  {Florian}, {Flynn}, {Fontanella}, {Ford}, {Forshay}, {Fox}, {Franz}, {Fu},
  {Fullerton}, {Galkin}, {Galyer}, {Garc{\'\i}a Mar{\'\i}n}, {Gardner},
  {Gardner}, {Garland}, {Garrett}, {Gasman}, {Gaspar}, {Gaudreau}, {Gauthier},
  {Geers}, {Geithner}, {Gennaro}, {Giardino}, {Girard}, {Giuliano},
  {Glassmire}, {Glauser}, {Glazer}, {Godfrey}, {Golimowski}, {Gollnitz},
  {Gong}, {Gonzaga}, {Gordon}, {Gordon}, {Goudfrooij}, {Greene}, {Greenhouse},
  {Grimaldi}, {Groebner}, {Grundy}, {Guillard}, {Gutman}, {Ha}, {Haderlein},
  {Hagedorn}, {Hainline}, {Haley}, {Hami}, {Hamilton}, {Hammel}, {Hansen},
  {Harkins}, {Harr}, {Hart}, {Hart}, {Hartig}, {Hashimoto}, {Haskins},
  {Hathaway}, {Havey}, {Hayden}, {Hecht}, {Heller-Boyer}, {Henriques}, {Henry},
  {Hermann}, {Hernandez}, {Hesman}, {Hicks}, {Hilbert}, {Hines}, {Hoffman},
  {Holfeltz}, {Holler}, {Hoppa}, {Hott}, {Howard}, {Howard}, {Hunter},
  {Hunter}, {Hurst}, {Husemann}, {Hustak}, {Ilinca Ignat}, {Illingworth},
  {Irish}, {Jackson}, {Jahromi}, {Jakobsen}, {James}, {James}, {Januszewski},
  {Jenkins}, {Jirdeh}, {Johnson}, {Johnson}, {Jones}, {Jones}, {Jones},
  {Jones}, {Jordan}, {Jordan}, {Jurczyk}, {Jurling}, {Kaleida}, {Kalmanson},
  {Kammerer}, {Kang}, {Kao}, {Karakla}, {Kavanagh}, {Kelly}, {Kendrew},
  {Kennedy}, {Kenny}, {Keski-kuha}, {Keyes}, {Kidwell}, {Kinzel}, {Kirk},
  {Kirkpatrick}, {Kirshenblat}, {Klaassen}, {Knapp}, {Knight}, {Knollenberg},
  {Koehler}, {Koekemoer}, {Kovacs}, {Kulp}, {Kumari}, {Kyprianou}, {La Massa},
  {Labador}, {Labiano}, {Lagage}, {Lajoie}, {Lallo}, {Lam}, {Lamb}, {Lambros},
  {Lampenfield}, {Langston}, {Larson}, {Law}, {Lawrence}, {Lee}, {Leisenring},
  {Lepo}, {Leveille}, {Levenson}, {Levine}, {Levy}, {Lewis}, {Lewis},
  {Libralato}, {Lightsey}, {Link}, {Liu}, {Lo}, {Lockwood}, {Logue}, {Long},
  {Long}, {Loomis}, {Lopez-Caniego}, {Lorenzo Alvarez}, {Love-Pruitt}, {Lucy},
  {Luetzgendorf}, {Maghami}, {Maiolino}, {Major}, {Malla}, {Malumuth},
  {Manjavacas}, {Mannfolk}, {Marrione}, {Marston}, {Martel}, {Maschmann},
  {Masci}, {Masciarelli}, {Maszkiewicz}, {Mather}, {McKenzie}, {McLean},
  {McMaster}, {Melbourne}, {Mel{\'e}ndez}, {Menzel}, {Merz}, {Meyett}, {Meza},
  {Miskey}, {Misselt}, {Moller}, {Morrison}, {Morse}, {Moseley}, {Mosier},
  {Mountain}, {Mueckay}, {Mueller}, {Mullally}, {Murphy}, {Murray}, {Murray},
  {Mustelier}, {Muzerolle}, {Mycroft}, {Myers}, {Myrick}, {Nanavati}, {Nance},
  {Nayak}, {Naylor}, {Nelan}, {Nickson}, {Nielson}, {Nieto-Santisteban},
  {Nikolov}, {Noriega-Crespo}, {O'Shaughnessy}, {O'Sullivan}, {Ochs}, {Ogle},
  {Oleszczuk}, {Olmsted}, {Osborne}, {Ottens}, {Owens}, {Pacifici}, {Pagan},
  {Page}, {Park}, {Parrish}, {Patapis}, {Paul}, {Pauly}, {Pavlovsky}, {Pedder},
  {Peek}, {Pena-Guerrero}, {Penanen}, {Perez}, {Perna}, {Perriello},
  {Phillips}, {Pietraszkiewicz}, {Pinaud}, {Pirzkal}, {Pitman}, {Piwowar},
  {Platais}, {Player}, {Plesha}, {Pollizi}, {Polster}, {Pontoppidan},
  {Porterfield}, {Proffitt}, {Pueyo}, {Pulliam}, {Quirt}, {Quispe Neira},
  {Ramos Alarcon}, {Ramsay}, {Rapp}, {Rapp}, {Rauscher}, {Ravindranath},
  {Rawle}, {Regan}, {Reichard}, {Reis}, {Ressler}, {Rest}, {Reynolds}, {Rhue},
  {Richon}, {Rickman}, {Ridgaway}, {Ritchie}, {Rix}, {Robberto}, {Robinson},
  {Robinson}, {Robinson}, {Rock}, {Rodriguez}, {Rodriguez Del Pino}, {Roellig},
  {Rohrbach}, {Roman}, {Romelfanger}, {Rose}, {Roteliuk}, {Roth}, {Rothwell},
  {Rowlands}, {Roy}, {Royer}, {Royle}, {Rui}, {Rumler}, {Runnels}, {Russ},
  {Rustamkulov}, {Ryden}, {Ryer}, {Sabata}, {Sabatke}, {Sabbi}, {Samuelson},
  {Sapp}, {Sappington}, {Sargent}, {Sauer}, {Scheithauer}, {Schlawin},
  {Schlitz}, {Schmitz}, {Schneider}, {Schreiber}, {Schulze}, {Schwab}, {Scott},
  {Sembach}, {Shanahan}, {Shaughnessy}, {Shaw}, {Shawger}, {Shay}, {Sheehan},
  {Shen}, {Sherman}, {Shiao}, {Shih}, {Shivaei}, {Sienkiewicz}, {Sing},
  {Sirianni}, {Sivaramakrishnan}, {Skipper}, {Sloan}, {Slocum}, {Slowinski},
  {Smith}, {Smith}, {Smith}, {Smith}, {Snyder}, {Soh}, {Sohn}, {Soto},
  {Spencer}, {Stallcup}, {Stansberry}, {Starr}, {Starr}, {Stewart},
  {Stiavelli}, {Straughn}, {Strickland}, {Stys}, {Summers}, {Sun}, {Sunnquist},
  {Swade}, {Swam}, {Swaters}, {Swoish}, {Taylor}, {Taylor}, {Te Plate}, {Tea},
  {Teague}, {Telfer}, {Temim}, {Thatte}, {Thompson}, {Thompson}, {Thomson},
  {Tikkanen}, {Tippet}, {Todd}, {Toolan}, {Tran}, {Trejo}, {Truong},
  {Tsukamoto}, {Tustain}, {Tyra}, {Ubeda}, {Underwood}, {Uzzo}, {Van Campen},
  {Vandal}, {Vandenbussche}, {Vila}, {Volk}, {Wahlgren}, {Waldman}, {Walker},
  {Wander}, {Warfield}, {Warner}, {Wasiak}, {Watkins}, {Weaver}, {Weilert},
  {Weiser}, {Weiss}, {Weissman}, {Welty}, {West}, {Wheate}, {Wheatley},
  {Wheeler}, {White}, {Whiteaker}, {Whitehouse}, {Whiteleather}, {Whitman},
  {Williams}, {Willmer}, {Willoughby}, {Wilson}, {Wirth}, {Wislowski}, {Wolf},
  {Wolfe}, {Wolff}, {Workman}, {Wright}, {Wu}, {Wu}, {Wymer}, {Yates},
  {Yeager}, {Yeates}, {Yerger}, {Yoon}, {Young}, {Yu}, {Zak}, {Zeidler},
  {Zhou}, {Zielinski}, {Zincke}, \& {Zonak}}]{RigbyJWST}
{Rigby}, J., {Perrin}, M., {McElwain}, M., {et~al.} 2023, \pasp, 135, 048001

\bibitem[{{Rinaldi} {et~al.}(2025){Rinaldi}, {Navarro-Carrera}, {Caputi},
  {Iani}, {{\"O}stlin}, {Colina}, {Alberts}, {{\'A}lvarez-M{\'a}rquez},
  {Annunziatella}, {Boogaard}, {Costantin}, {Hjorth}, {Langeroodi}, {Melinder},
  {Moutard}, \& {Walter}}]{Rinaldi2024}
{Rinaldi}, P., {Navarro-Carrera}, R., {Caputi}, K.~I., {et~al.} 2025, \apj,
  981, 161

\bibitem[{{Roberts-Borsani} {et~al.}(2023){Roberts-Borsani}, {Treu}, {Chen},
  {Morishita}, {Vanzella}, {Zitrin}, {Bergamini}, {Castellano}, {Fontana},
  {Glazebrook}, {Grillo}, {Kelly}, {Merlin}, {Nanayakkara}, {Paris}, {Rosati},
  {Yang}, {Acebron}, {Bonchi}, {Boyett}, {Brada{\v{c}}}, {Brammer},
  {Broadhurst}, {Calabr{\'o}}, {Diego}, {Dressler}, {Furtak}, {Filippenko},
  {Henry}, {Koekemoer}, {Leethochawalit}, {Malkan}, {Mason}, {Mercurio},
  {Metha}, {Pentericci}, {Pierel}, {Rieck}, {Roy}, {Santini}, {Strait},
  {Strausbaugh}, {Trenti}, {Vulcani}, {Wang}, {Wang}, \&
  {Windhorst}}]{RobertsBorsani2023}
{Roberts-Borsani}, G., {Treu}, T., {Chen}, W., {et~al.} 2023, \nat, 618, 480

\bibitem[{{Sanders} {et~al.}(2023){Sanders}, {Shapley}, {Topping}, {Reddy}, \&
  {Brammer}}]{Sanders2023}
{Sanders}, R.~L., {Shapley}, A.~E., {Topping}, M.~W., {Reddy}, N.~A., \&
  {Brammer}, G.~B. 2023, \apj, 955, 54

\bibitem[{{Shapley} {et~al.}(2023){Shapley}, {Sanders}, {Reddy}, {Topping}, \&
  {Brammer}}]{Shapley2023}
{Shapley}, A.~E., {Sanders}, R.~L., {Reddy}, N.~A., {Topping}, M.~W., \&
  {Brammer}, G.~B. 2023, \apj, 954, 157

\bibitem[{{Skelton} {et~al.}(2014){Skelton}, {Whitaker}, {Momcheva}, {Brammer},
  {van Dokkum}, {Labb{\'e}}, {Franx}, {van der Wel}, {Bezanson}, {Da Cunha},
  {Fumagalli}, {F{\"o}rster Schreiber}, {Kriek}, {Leja}, {Lundgren}, {Magee},
  {Marchesini}, {Maseda}, {Nelson}, {Oesch}, {Pacifici}, {Patel}, {Price},
  {Rix}, {Tal}, {Wake}, \& {Wuyts}}]{Skelton2014}
{Skelton}, R.~E., {Whitaker}, K.~E., {Momcheva}, I.~G., {et~al.} 2014, \apjs,
  214, 24

\bibitem[{{Speagle} {et~al.}(2014){Speagle}, {Steinhardt}, {Capak}, \&
  {Silverman}}]{Speagle2014}
{Speagle}, J.~S., {Steinhardt}, C.~L., {Capak}, P.~L., \& {Silverman}, J.~D.
  2014, \apjs, 214, 15

\bibitem[{Steidel {et~al.}(2003)Steidel, Adelberger, Shapley, Pettini,
  Dickinson, \& Giavalisco}]{Steidel_2003}
Steidel, C.~C., Adelberger, K.~L., Shapley, A.~E., {et~al.} 2003, The
  Astrophysical Journal, 592, 728–754

\bibitem[{Steidel {et~al.}(1996)Steidel, Giavalisco, Pettini, Dickinson, \&
  Adelberger}]{Steidel_1996}
Steidel, C.~C., Giavalisco, M., Pettini, M., Dickinson, M., \& Adelberger,
  K.~L. 1996, The Astrophysical Journal, 462, L17–L21

\bibitem[{{Sun} {et~al.}(2024{\natexlab{a}}){Sun}, {Helton}, {Egami},
  {Hainline}, {Rieke}, {Willmer}, {Eisenstein}, {Johnson}, {Rieke},
  {Robertson}, {Tacchella}, {Alberts}, {Baker}, {Bhatawdekar}, {Boyett},
  {Bunker}, {Charlot}, {Chen}, {Chevallard}, {Curtis-Lake}, {Danhaive},
  {DeCoursey}, {Ji}, {Lyu}, {Maiolino}, {Rujopakarn}, {Sandles}, {Shivaei},
  {{\"U}bler}, {Willott}, \& {Witstok}}]{Sun2024}
{Sun}, F., {Helton}, J.~M., {Egami}, E., {et~al.} 2024{\natexlab{a}}, \apj,
  961, 69

\bibitem[{{Sun} {et~al.}(2024{\natexlab{b}}){Sun}, {Ho}, {Zhuang}, {Ma},
  {Chen}, \& {Li}}]{Sun2024:morph}
{Sun}, W., {Ho}, L.~C., {Zhuang}, M.-Y., {et~al.} 2024{\natexlab{b}}, \apj,
  960, 104

\bibitem[{{Tacchella} {et~al.}(2024){Tacchella}, {McClymont}, {Scholtz},
  {Maiolino}, {Ji}, {Villanueva}, {Charlot}, {D'Eugenio}, {Helton}, {Williams},
  {Witstok}, {Bhatawdekar}, {Carniani}, {Chevallard}, {Curti}, {Hainline},
  {Ji}, {Johnson}, {Leja}, {Li}, {Maseda}, {Pusk{\'a}s}, {Rieke}, {Robertson},
  {Shivaei}, {Silcock}, {Simmonds}, {{\"U}bler}, {Willmer}, \&
  {Willott}}]{Tacchella2024}
{Tacchella}, S., {McClymont}, W., {Scholtz}, J., {et~al.} 2024, arXiv e-prints,
  arXiv:2404.02194

\bibitem[{{Treu} {et~al.}(2022){Treu}, {Roberts-Borsani}, {Bradac}, {Brammer},
  {Fontana}, {Henry}, {Mason}, {Morishita}, {Pentericci}, {Wang}, {Acebron},
  {Bagley}, {Bergamini}, {Belfiori}, {Bonchi}, {Boyett}, {Boutsia},
  {Calabr{\'o}}, {Caminha}, {Castellano}, {Dressler}, {Glazebrook}, {Grillo},
  {Jacobs}, {Jones}, {Kelly}, {Leethochawalit}, {Malkan}, {Marchesini},
  {Mascia}, {Mercurio}, {Merlin}, {Nanayakkara}, {Nonino}, {Paris},
  {Poggianti}, {Rosati}, {Santini}, {Scarlata}, {Shipley}, {Strait}, {Trenti},
  {Tubthong}, {Vanzella}, {Vulcani}, \& {Yang}}]{Treu2022}
{Treu}, T., {Roberts-Borsani}, G., {Bradac}, M., {et~al.} 2022, \apj, 935, 110

\bibitem[{{Valentino} {et~al.}(2023){Valentino}, {Brammer}, {Gould}, {Kokorev},
  {Fujimoto}, {Jespersen}, {Vijayan}, {Weaver}, {Ito}, {Tanaka}, {Ilbert},
  {Magdis}, {Whitaker}, {Faisst}, {Gallazzi}, {Gillman}, {Gim{\'e}nez-Arteaga},
  {G{\'o}mez-Guijarro}, {Kubo}, {Heintz}, {Hirschmann}, {Oesch}, {Onodera},
  {Rizzo}, {Lee}, {Strait}, \& {Toft}}]{Valentino2023}
{Valentino}, F., {Brammer}, G., {Gould}, K. M.~L., {et~al.} 2023, \apj, 947, 20

\bibitem[{{van Dokkum} {et~al.}(2011){van Dokkum}, {Brammer}, {Fumagalli},
  {Nelson}, {Franx}, {Rix}, {Kriek}, {Skelton}, {Patel}, {Schmidt}, {Bezanson},
  {Bian}, {da Cunha}, {Erb}, {Fan}, {F{\"o}rster Schreiber}, {Illingworth},
  {Labb{\'e}}, {Lundgren}, {Magee}, {Marchesini}, {McCarthy}, {Muzzin},
  {Quadri}, {Steidel}, {Tal}, {Wake}, {Whitaker}, \&
  {Williams}}]{vanDokkum2011}
{van Dokkum}, P.~G., {Brammer}, G., {Fumagalli}, M., {et~al.} 2011, \apjl, 743,
  L15

\bibitem[{{Wang} {et~al.}(2024{\natexlab{a}}){Wang}, {de Graaff}, {Davies},
  {Greene}, {Leja}, {Goulding}, {Williams}, {Brammer}, {Suess}, {Weibel},
  {Bezanson}, {Boogaard}, {Cleri}, {Hirschmann}, {Katz}, {Labbe}, {Maseda},
  {Matthee}, {McConachie}, {Naidu}, {Oesch}, {Rix}, {Setton}, \&
  {Whitaker}}]{Wang2024a}
{Wang}, B., {de Graaff}, A., {Davies}, R.~L., {et~al.} 2024{\natexlab{a}},
  arXiv e-prints, arXiv:2403.02304

\bibitem[{{Wang} {et~al.}(2023){Wang}, {Fujimoto}, {Labb{\'e}}, {Furtak},
  {Miller}, {Setton}, {Zitrin}, {Atek}, {Bezanson}, {Brammer}, {Leja}, {Oesch},
  {Price}, {Chemerynska}, {Cutler}, {Dayal}, {van Dokkum}, {Goulding},
  {Greene}, {Fudamoto}, {Khullar}, {Kokorev}, {Marchesini}, {Pan}, {Weaver},
  {Whitaker}, \& {Williams}}]{Wang2023}
{Wang}, B., {Fujimoto}, S., {Labb{\'e}}, I., {et~al.} 2023, \apjl, 957, L34

\bibitem[{{Wang} {et~al.}(2024{\natexlab{b}}){Wang}, {Leja}, {de Graaff},
  {Brammer}, {Weibel}, {van Dokkum}, {Baggen}, {Suess}, {Greene}, {Bezanson},
  {Cleri}, {Hirschmann}, {Labb{\'e}}, {Matthee}, {McConachie}, {Naidu},
  {Nelson}, {Oesch}, {Setton}, \& {Williams}}]{Wang2024b}
{Wang}, B., {Leja}, J., {de Graaff}, A., {et~al.} 2024{\natexlab{b}}, \apjl,
  969, L13

\bibitem[{{Wang} {et~al.}(2019){Wang}, {Schreiber}, {Elbaz}, {Yoshimura},
  {Kohno}, {Shu}, {Yamaguchi}, {Pannella}, {Franco}, {Huang}, {Lim}, \&
  {Wang}}]{Wang2019}
{Wang}, T., {Schreiber}, C., {Elbaz}, D., {et~al.} 2019, \nat, 572, 211

\bibitem[{{Wang} {et~al.}(2024{\natexlab{c}}){Wang}, {Sun}, {Zhou}, {Xu},
  {Cheng}, {Li}, {Chen}, {Mo}, {Dekel}, {Zheng}, {Cai}, {Yang}, {Dai}, {Elbaz},
  \& {Huang}}]{Wang2024_massfunction}
{Wang}, T., {Sun}, H., {Zhou}, L., {et~al.} 2024{\natexlab{c}}, arXiv e-prints,
  arXiv:2403.02399

\bibitem[{{Weaver} {et~al.}(2023){Weaver}, {Davidzon}, {Toft}, {Ilbert},
  {McCracken}, {Gould}, {Jespersen}, {Steinhardt}, {Lagos}, {Capak}, {Casey},
  {Chartab}, {Faisst}, {Hayward}, {Kartaltepe}, {Kauffmann}, {Koekemoer},
  {Kokorev}, {Laigle}, {Liu}, {Long}, {Magdis}, {McPartland}, {Milvang-Jensen},
  {Mobasher}, {Moneti}, {Peng}, {Sanders}, {Shuntov}, {Sneppen}, {Valentino},
  {Zalesky}, \& {Zamorani}}]{Weaver2023}
{Weaver}, J.~R., {Davidzon}, I., {Toft}, S., {et~al.} 2023, \aap, 677, A184

\bibitem[{{Weibel} {et~al.}(2024{\natexlab{a}}){Weibel}, {de Graaff}, {Setton},
  {Miller}, {Oesch}, {Brammer}, {Lagos}, {Whitaker}, {Williams}, {Baggen},
  {Bezanson}, {Boogaard}, {Cleri}, {Greene}, {Hirschmann}, {Hviding},
  {Kuruvanthodi}, {Labb{\'e}}, {Leja}, {Maseda}, {Matthee}, {McConachie},
  {Naidu}, {Roberts-Borsani}, {Schaerer}, {Suess}, {Valentino}, {van Dokkum},
  \& {Wang}}]{Weibel2024b}
{Weibel}, A., {de Graaff}, A., {Setton}, D.~J., {et~al.} 2024{\natexlab{a}},
  arXiv e-prints, arXiv:2409.03829

\bibitem[{{Weibel} {et~al.}(2024{\natexlab{b}}){Weibel}, {Oesch}, {Barrufet},
  {Gottumukkala}, {Ellis}, {Santini}, {Weaver}, {Allen}, {Bouwens}, {Bowler},
  {Brammer}, {Carnall}, {Cullen}, {Dayal}, {Dickinson}, {Donnan}, {Dunlop},
  {Giavalisco}, {Grogin}, {Illingworth}, {Koekemoer}, {Labbe}, {Marchesini},
  {McLeod}, {McLure}, {Naidu}, {P{\'e}rez-Gonz{\'a}lez}, {Shuntov}, {Stefanon},
  {Toft}, \& {Xiao}}]{Weibel2024}
{Weibel}, A., {Oesch}, P.~A., {Barrufet}, L., {et~al.} 2024{\natexlab{b}},
  \mnras, 533, 1808

\bibitem[{{Whitaker} {et~al.}(2011){Whitaker}, {Labb{\'e}}, {van Dokkum},
  {Brammer}, {Kriek}, {Marchesini}, {Quadri}, {Franx}, {Muzzin}, {Williams},
  {Bezanson}, {Illingworth}, {Lee}, {Lundgren}, {Nelson}, {Rudnick}, {Tal}, \&
  {Wake}}]{Whitaker2011}
{Whitaker}, K.~E., {Labb{\'e}}, I., {van Dokkum}, P.~G., {et~al.} 2011, \apj,
  735, 86

\bibitem[{{Williams} {et~al.}(2024){Williams}, {Alberts}, {Ji}, {Hainline},
  {Lyu}, {Rieke}, {Endsley}, {Suess}, {Sun}, {Johnson}, {Florian}, {Shivaei},
  {Rujopakarn}, {Baker}, {Bhatawdekar}, {Boyett}, {Bunker}, {Cameron},
  {Carniani}, {Charlot}, {Curtis-Lake}, {DeCoursey}, {de Graaff}, {Egami},
  {Eisenstein}, {Gibson}, {Hausen}, {Helton}, {Maiolino}, {Maseda}, {Nelson},
  {P{\'e}rez-Gonz{\'a}lez}, {Rieke}, {Robertson}, {Saxena}, {Tacchella},
  {Willmer}, \& {Willott}}]{Williams2024}
{Williams}, C.~C., {Alberts}, S., {Ji}, Z., {et~al.} 2024, \apj, 968, 34

\bibitem[{{Williams} {et~al.}(2019){Williams}, {Labbe}, {Spilker}, {Stefanon},
  {Leja}, {Whitaker}, {Bezanson}, {Narayanan}, {Oesch}, \&
  {Weiner}}]{Williams2019}
{Williams}, C.~C., {Labbe}, I., {Spilker}, J., {et~al.} 2019, \apj, 884, 154

\bibitem[{{Williams} {et~al.}(2025){Williams}, {Oesch}, {Weibel}, {Brammer},
  {Cloonan}, {Whitaker}, {Barrufet}, {Bezanson}, {Bowler}, {Dayal}, {Franx},
  {Greene}, {Hutter}, {Ji}, {Labb{\'e}}, {Manning}, {Maseda}, \&
  {Xiao}}]{Williams2024:panoramic}
{Williams}, C.~C., {Oesch}, P.~A., {Weibel}, A., {et~al.} 2025, \apj, 979, 140

\bibitem[{{Wright} {et~al.}(2023){Wright}, {Rieke}, {Glasse}, {Ressler},
  {Garc{\'\i}a Mar{\'\i}n}, {Aguilar}, {Alberts}, {{\'A}lvarez-M{\'a}rquez},
  {Argyriou}, {Banks}, {Baudoz}, {Boccaletti}, {Bouchet}, {Bouwman}, {Brandl},
  {Breda}, {Bright}, {Cale}, {Colina}, {Cossou}, {Coulais}, {Cracraft}, {De
  Meester}, {Dicken}, {Engesser}, {Etxaluze}, {Fox}, {Friedman}, {Fu},
  {Gasman}, {G{\'a}sp{\'a}r}, {Gastaud}, {Geers}, {Glauser}, {Gordon},
  {Greene}, {Greve}, {Grundy}, {G{\"u}del}, {Guillard}, {Haderlein},
  {Hashimoto}, {Henning}, {Hines}, {Holler}, {Detre}, {Jahromi}, {James},
  {Jones}, {Justtanont}, {Kavanagh}, {Kendrew}, {Klaassen}, {Krause},
  {Labiano}, {Lagage}, {Lambros}, {Larson}, {Law}, {Lee}, {Libralato}, {Lorenzo
  Alverez}, {Meixner}, {Morrison}, {Mueller}, {Murray}, {Mycroft}, {Myers},
  {Nayak}, {Naylor}, {Nickson}, {Noriega-Crespo}, {{\"O}stlin}, {O'Sullivan},
  {Ottens}, {Patapis}, {Penanen}, {Pietraszkiewicz}, {Ray}, {Regan},
  {Roteliuk}, {Royer}, {Samara-Ratna}, {Samuelson}, {Sargent}, {Scheithauer},
  {Schneider}, {Schreiber}, {Shaughnessy}, {Sheehan}, {Shivaei}, {Sloan},
  {Tamas}, {Teague}, {Temim}, {Tikkanen}, {Tustain}, {van Dishoeck},
  {Vandenbussche}, {Weilert}, {Whitehouse}, \& {Wolff}}]{Wright2023}
{Wright}, G.~S., {Rieke}, G.~H., {Glasse}, A., {et~al.} 2023, \pasp, 135,
  048003

\bibitem[{{Xiao} {et~al.}(2024){Xiao}, {Oesch}, {Elbaz}, {Bing}, {Nelson},
  {Weibel}, {Illingworth}, {van Dokkum}, {Naidu}, {Daddi}, {Bouwens},
  {Matthee}, {Wuyts}, {Chisholm}, {Brammer}, {Dickinson}, {Magnelli}, {Leroy},
  {Schaerer}, {Herard-Demanche}, {Lim}, {Barrufet}, {Endsley}, {Fudamoto},
  {G{\'o}mez-Guijarro}, {Gottumukkala}, {Labb{\'e}}, {Magee}, {Marchesini},
  {Maseda}, {Qin}, {Reddy}, {Shapley}, {Shivaei}, {Shuntov}, {Stefanon},
  {Whitaker}, \& {Wyithe}}]{Xiao2023}
{Xiao}, M., {Oesch}, P.~A., {Elbaz}, D., {et~al.} 2024, \nat, 635, 311

\bibitem[{{Xu} {et~al.}(2023){Xu}, {Ouchi}, {Nakajima}, {Harikane}, {Isobe},
  {Ono}, {Umeda}, \& {Zhang}}]{Xu2023}
{Xu}, Y., {Ouchi}, M., {Nakajima}, K., {et~al.} 2023, arXiv e-prints,
  arXiv:2310.06614

\end{thebibliography}

\appendix

\section{Sky spectra}\label{sec:sky_appendix}

\begin{figure*}[!t]
    \centering
    \includegraphics[width=0.9\linewidth]{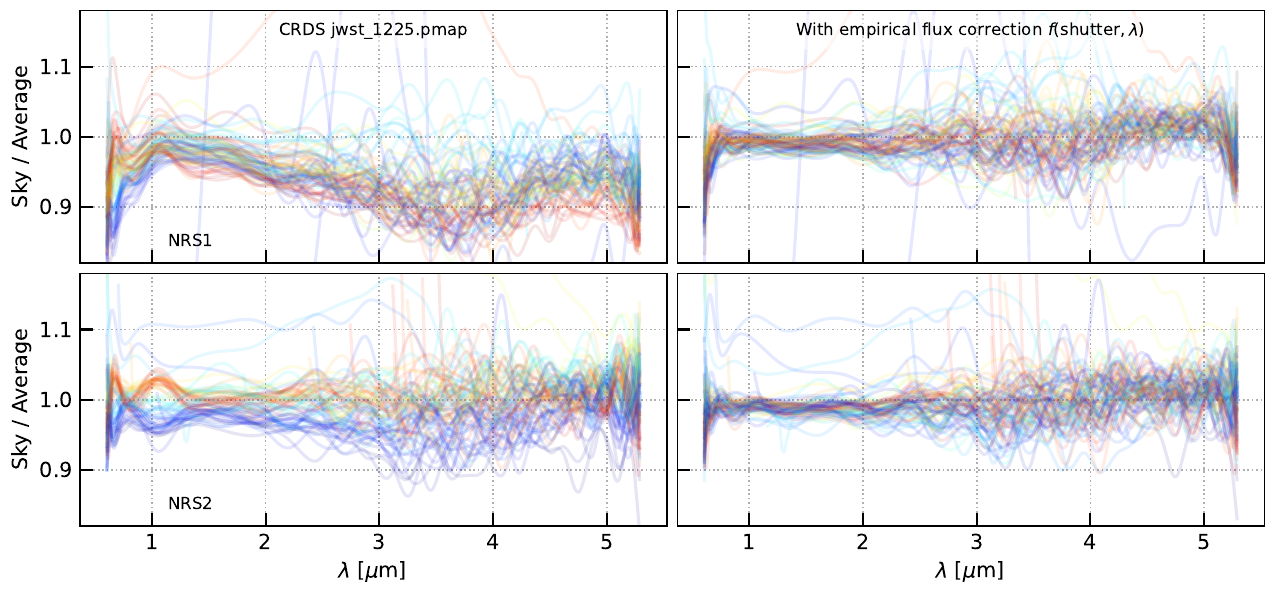}
    \caption{Position / detector normalisation derived from sky spectra.  The lines in the left panels show the ratio of the background spectra measured from individual slitlets relative to the average spectrum of all slitlets, with the line colours indicating the ``y'' position of a slitlet within the field of view. Clearly, there are wavelength- and y-dependent systematics. The right panels show the ratios after applying the position- and wavelength-dependent flux scale correction.
    }
    \label{fig:sky_flux_correciton}
\end{figure*}

Given the excellent intrinsic quality of the NIRSpec detectors, the sensitivity of observations of faint sources is generally limited by shot noise from sky photons. Furthermore, the intensity of the sky background in NIRSpec PRISM observations is often similar to, if not orders of magnitude brighter than the faint astronomical sources of interest and \textit{systematic} errors on the sky removal can be a dominant component of the error budget in the analysis of such sources.

\subsection{Calibration corrections derived from empty slitlets}\label{sec:barshadow}

We take advantage of the relatively bright background to use it as a uniform illumination source to refine two aspects of the PRISM calibration: vignetting from the MSA bars (``bar shadow'') and the field dependence of the absolute flux calibration.  For this exercise we extract the 2D spectra of a large number of empty slitlets from program GO-2750 (PI: Arrabal-Haro), which has more and deeper exposures than the RUBIES observational setup. 

We first estimate the 1D sky spectrum of each slitlet by fitting a high-order cubic spline to the intensities of pixels near the expected centres of the open shutters across $N$ exposures including that slitlet (typically $N=9$ for GO-2750).  Assuming that the shape and normalisation of the sky spectrum should be the same for every slitlet across the detector, these spectra can be used as a highly-multiplexed observation of an (infinitely) extended ``standard'' source.  Using the nominal calibrations from the Calibration Reference Data System (CRDS, \texttt{jwst\_1225.pmap}) we find significant variation in the shape of the sky spectra between and within the two NIRCam detectors (Fig.~\ref{fig:sky_flux_correciton}, left panels).  We derive a shutter- and wavelength-dependent photometric correction of the PRISM spectra by fitting a quadratic 2D polynomial to the spline coefficients of the sky spectra as a function of the MSA shutter row and column indices.  This correction reduces the spatial systematics to $\lesssim 1\%$ at $\lambda < 2~\mu\mathrm{m}$ where the sky is bright and $\lesssim 5\%$ at $\lambda = 3.5~\mu\mathrm{m}$ where the sky is faintest (Fig.~\ref{fig:sky_flux_correciton}, right panels).

\begin{figure*}[!t]
    \centering
    \includegraphics[width=\linewidth]{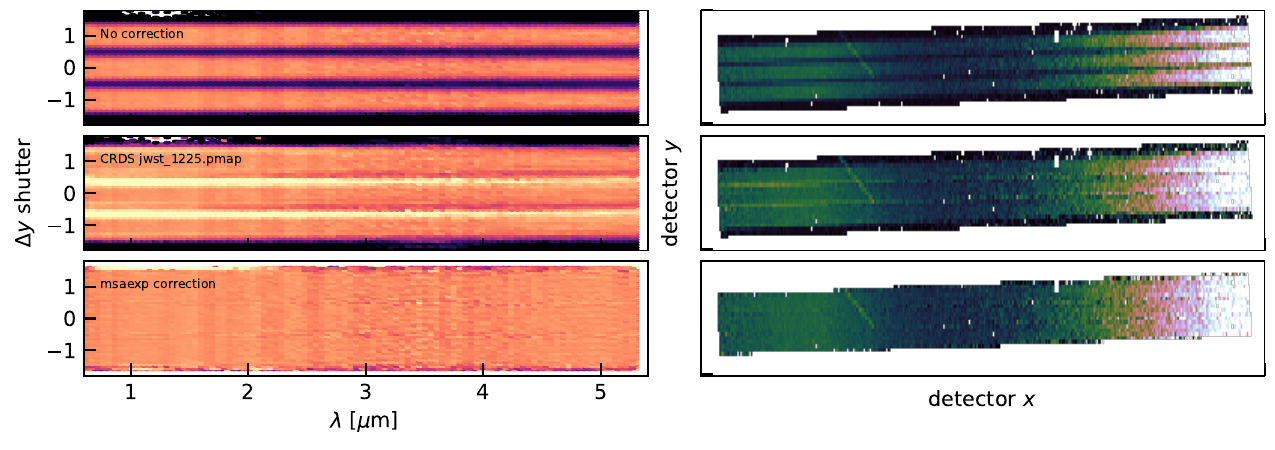}
    \caption{PRISM bar shadow for 3-shutter slitlets.  The left panels show the sky-normalised average spectra for 168 empty background slitlets extracted from program GO-275O.  The vertical axis is the rectified ``shutter'' coordinate frame.  The right panels show the 2D spectrum of a single exposure/slitlet in the original detector coordinate frame.  The top panels show the spectra without any bar shadow correction.  The centre panels show the correction using the CRDS reference files, and the bottom panels show the wavelength-dependent correction using \texttt{msaexp}.}
    \label{fig:barshadow}
\end{figure*}

With the same empty sky spectra corrected for the spatial variations and normalised by the average 1D sky spectrum , we measure the average 2D cross-dispersion profile across the entire detector.  With no bar shadow correction applied, the vignetting by the MSA bars is readily apparent (Fig.~\ref{fig:barshadow}, top panels). With slitlets observed across the entire MSA, the GO-2750 sky spectra finely sample the ``$\Delta y$ shutter'' cross-dispersion coordinate at all wavelengths.  The intensity at the centre of the open shutters is roughly twice that under the bars, i.e., a correction for this effect involves multiplying the vignetted pixels by a factor as large as 2.  We find that the currently-available CRDS bar-shadow calibration does not adequately correct for the bar vignetting: a bright \textit{excess} near the bar centres is consistent with a small shift of the cross-dispersion shutter coordinate relative to the profile in the calibration file (Fig.~\ref{fig:barshadow}, middle panels).  Again we note that the bar shadow correction residuals that result from multiplying the bright sky by factors as large as 2 can be many times larger than the intensity of the faint astronomical sources of interest. We derive a purely empirical bar shadow correction from these profiles by again using flexible cubic splines to approximate the cross-dispersion profiles in wavelength bins across the PRISM bandpass, where the wavelength dependence results from diffraction effects of the wavelength-dependent PSF. While our correction makes the 2D sky very flat by design (Fig.~\ref{fig:barshadow}, bottom panels), we note that such a correction (just as with the \texttt{jwst} pipeline implementation) is only strictly valid for very extended sources.  The correction for compact sources near the shutter edges will be very uncertain and is beyond the scope of the work here. 

\subsection{Master sky background removal} \label{sec:skybackground}

With an improved bar shadow correction in hand that produces flat 2D spectra, we are in position to develop  a strategy for performing a global sky removal from the primary target spectra without relying on taking image differences from the nod offsets. The benefits of a master sky removal are 1) better overall sensitivity (the noise difference doubles the variance) and 2) eliminating ``self-subtraction'' of spatial structures with sizes of order of the 0\farcs5 nod offset.

\begin{figure}
    \centering
    \includegraphics[width=\linewidth]{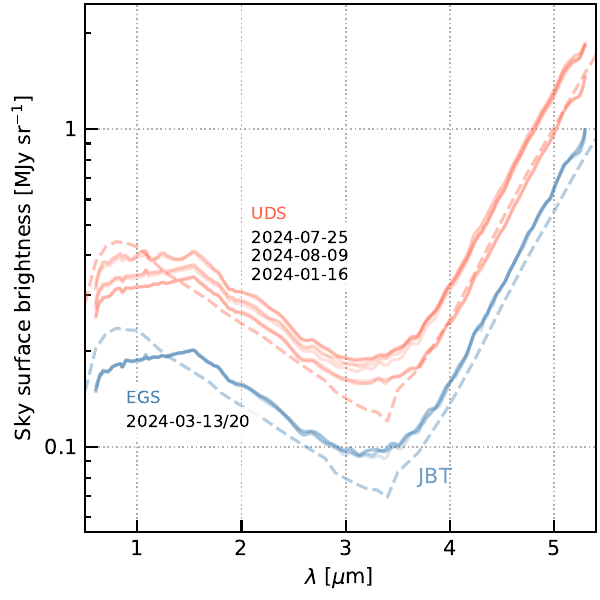}
    \caption{Average sky background surface brightness in the RUBIES visits measured from empty regions of the science and filler sky slitlets.  The background predicted by the ``JWST Backgrounds Tool'' (JBT) for the UDS 2024-01-16 and EGS 2024-03-20 visits are shown in the dashed curves.}
    \label{fig:sky_background}
\end{figure}

We first estimate the average 1D sky spectrum of each RUBIES mask using both the empty sky slitlets included in the mask design and relatively empty portions of slitlets of faint sources.  The sky spectrum is \textit{fit} as the combination of the Solar spectrum with a modified slope resulting from reflected zodiacal light and cubic splines to approximate the long-wavelength thermal emission from zodiacal dust. Both the shape and intensity of the sky spectra differ on timescales as short as a few weeks (Fig.~\ref{fig:sky_background}).  The magnitude of the variation is roughly consistent with the predictions of the JWST Backgrounds Tool (JBT, Rigby \& Pontoppidon), though the spectral shape is somewhat different, especially at blue wavelengths dominated by the reflected zodiacal light.  Similar 1D sky spectra for many additional public PRISM datasets are included in the \href{https://github.com/gbrammer/msaexp/tree/main/msaexp/data/msa_sky}{msaexp repository}.

\begin{figure}
    \centering
    \includegraphics[width=\linewidth]{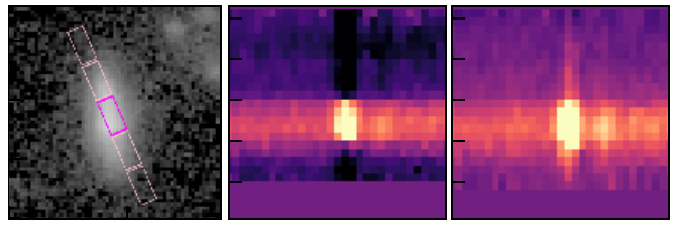}
    \caption{\textit{Left:} F444W cutout and shutter footprints for RUBIES UDS-42150 ($z=3.191$).  \textit{Centre:} 2D spectrum around H$\alpha$ emission with nod-offset background removal.  \textit{Right:} 2D spectrum with global sky background removal. }
    \label{fig:global_comparison}
\end{figure}

For the global sky removal, we subtract the 1D master sky from the bar-shadow-corrected 2D spectrum.  In \texttt{msaexp} we can remove the master sky without modification or optionally fit the master sky to relatively empty portions of the source slitlets allowing for small normalization and shape corrections that might not be correctly accounted for by the spatially-dependent flux calibration of the PRISM spectra described above.   An comparison of the image-difference and master sky background removal approaches for a large extended source is shown in Fig.~\ref{fig:global_comparison}. We only perform this global sky removal for the PRISM spectra, and use the image-difference background removal for the G395M grating spectra, which frequently have overlaps that compromise a global background determination even if the sky spectrum was perfectly known.

\end{document}